\documentstyle[twoside,psfig]{article}
\catcode`\@=11
\long\def\@makefntext#1{
\protect\noindent \hbox to 3.2pt {\hskip-.9pt
$^{{\eightrm\@thefnmark}}$\hfil}#1\hfill}       %CAN BE USED

\def\@makefnmark{\hbox to 0pt{$^{\@thefnmark}$\hss}}    %ORIGINAL

\def\ps@myheadings{\let\@mkboth\@gobbletwo
\def\@oddhead{\hbox{}
\rightmark\hfil\eightrm\thepage}
\def\@oddfoot{}\def\@evenhead{\eightrm\thepage\hfil
\leftmark\hbox{}}\def\@evenfoot{}
\def\sectionmark##1{}\def\subsectionmark##1{}}

\oddsidemargin=\evensidemargin
\newcounter{sectionc}\newcounter{subsectionc}\newcounter{subsubsectionc}
\renewcommand{\section}[1] {\vspace{12pt}\addtocounter{sectionc}{1}
\setcounter{subsectionc}{0}\setcounter{subsubsectionc}{0}\noindent
    {\tenbf\thesectionc. #1}\par\vspace{5pt}}
\renewcommand{\subsection}[1] {\vspace{12pt}\addtocounter{subsectionc}{1}
    \setcounter{subsubsectionc}{0}\noindent
    {\bf\thesectionc.\thesubsectionc. {\kern1pt \bfit #1}}\par\vspace{5pt}}
\renewcommand{\subsubsection}[1] {\vspace{12pt}\addtocounter{subsubsectionc}{1}
    \noindent{\tenrm\thesectionc.\thesubsectionc.\thesubsubsectionc.
    {\kern1pt \tenit #1}}\par\vspace{5pt}}
\newcommand{\nonumsection}[1] {\vspace{12pt}\noindent{\tenbf #1}
    \par\vspace{5pt}}

\newcounter{appendixc}
\newcounter{subappendixc}[appendixc]
\newcounter{subsubappendixc}[subappendixc]
\renewcommand{\thesubappendixc}{\Alph{appendixc}.\arabic{subappendixc}}
\renewcommand{\thesubsubappendixc}
    {\Alph{appendixc}.\arabic{subappendixc}.\arabic{subsubappendixc}}

\renewcommand{\appendix}[1] {\vspace{12pt}
        \refstepcounter{appendixc}
        \setcounter{figure}{0}
        \setcounter{table}{0}
        \setcounter{lemma}{0}
        \setcounter{theorem}{0}
        \setcounter{corollary}{0}
        \setcounter{definition}{0}
        \setcounter{equation}{0}
        \renewcommand{\thefigure}{\Alph{appendixc}.\arabic{figure}}
        \renewcommand{\thetable}{\Alph{appendixc}.\arabic{table}}
        \renewcommand{\theappendixc}{\Alph{appendixc}}
        \renewcommand{\thelemma}{\Alph{appendixc}.\arabic{lemma}}
        \renewcommand{\thetheorem}{\Alph{appendixc}.\arabic{theorem}}
        \renewcommand{\thedefinition}{\Alph{appendixc}.\arabic{definition}}
        \renewcommand{\thecorollary}{\Alph{appendixc}.\arabic{corollary}}
        \renewcommand{\theequation}{\Alph{appendixc}.\arabic{equation}}
        \noindent{\tenbf Appendix \theappendixc #1}\par\vspace{5pt}}
\newcommand{\subappendix}[1] {\vspace{12pt}
        \refstepcounter{subappendixc}
        \noindent{\bf Appendix \thesubappendixc. {\kern1pt \bfit #1}}
    \par\vspace{5pt}}
\newcommand{\subsubappendix}[1] {\vspace{12pt}
        \refstepcounter{subsubappendixc}
        \noindent{\rm Appendix \thesubsubappendixc. {\kern1pt \tenit #1}}
    \par\vspace{5pt}}

\newcommand{\textlineskip}{\baselineskip=13pt}
\newcommand{\smalllineskip}{\baselineskip=10pt}

\def\eightcirc{
\begin{picture}(0,0)
\put(4.4,1.8){\circle{6.5}}
\end{picture}}
\def\eightcopyright{\eightcirc\kern2.7pt\hbox{\eightrm c}}

\newcommand{\copyrightheading}[1]
    {\vspace*{-2.5cm}\smalllineskip{\flushleft
    {\footnotesize International Journal of Modern E #1}\\
    {\footnotesize $\eightcopyright$\, World Scientific Publishing
     Company}\\
     }}

\newcommand{\publisher}[2]{{\begin{center}\footnotesize\smalllineskip
    Received #1\\
    Revised #2
    \end{center}
    }}

\def\abstracts#1#2#3{{
    \centering{\begin{minipage}{4.5in}\footnotesize\baselineskip=10pt
    \parindent=0pt #1\par
    \parindent=15pt #2\par
    \parindent=15pt #3
    \end{minipage}}\par}}

\def\keywords#1{{
    \centering{\begin{minipage}{4.5in}\footnotesize\baselineskip=10pt
    {\footnotesize\it Keywords}\/: #1
     \end{minipage}}\par}}

\newcommand{\bibit}{\nineit}
\newcommand{\bibbf}{\ninebf}
\renewenvironment{thebibliography}[1]
    {\frenchspacing
     \ninerm\baselineskip=11pt
     \begin{list}{\arabic{enumi}.}
        {\usecounter{enumi}\setlength{\parsep}{0pt}
     \setlength{\leftmargin 12.7pt}{\rightmargin 0pt} %FOR 1--9 ITEMS
         \setlength{\itemsep}{0pt} \settowidth
    {\labelwidth}{#1.}\sloppy}}{\end{list}}

\newcounter{itemlistc}
\newcounter{romanlistc}
\newcounter{alphlistc}
\newcounter{arabiclistc}

\newcommand{\fcaption}[1]{
        \refstepcounter{figure}
        \setbox\@tempboxa = \hbox{\footnotesize Fig.~\thefigure. #1}
        \ifdim \wd\@tempboxa > 5in
           {\begin{center}
        \parbox{5in}{\footnotesize\smalllineskip Fig.~\thefigure. #1}
            \end{center}}
        \else
             {\begin{center}
             {\footnotesize Fig.~\thefigure. #1}
              \end{center}}
        \fi}

\newcommand{\tcaption}[1]{
        \refstepcounter{table}
        \setbox\@tempboxa = \hbox{\footnotesize Table~\thetable. #1}
        \ifdim \wd\@tempboxa > 5in
           {\begin{center}
        \parbox{5in}{\footnotesize\smalllineskip Table~\thetable. #1}
            \end{center}}
        \else
             {\begin{center}
             {\footnotesize Table~\thetable. #1}
              \end{center}}
        \fi}

\def\@citex[#1]#2{\if@filesw\immediate\write\@auxout
    {\string\citation{#2}}\fi
\def\@citea{}\@cite{\@for\@citeb:=#2\do
    {\@citea\def\@citea{,}\@ifundefined
    {b@\@citeb}{{\bf ?}\@warning
    {Citation `\@citeb' on page \thepage \space undefined}}
    {\csname b@\@citeb\endcsname}}}{#1}}

\newif\if@cghi
\def\cite{\@cghitrue\@ifnextchar [{\@tempswatrue
    \@citex}{\@tempswafalse\@citex[]}}
\def\citelow{\@cghifalse\@ifnextchar [{\@tempswatrue
    \@citex}{\@tempswafalse\@citex[]}}
\def\@cite#1#2{{$\null^{#1}$\if@tempswa\typeout
    {IJCGA warning: optional citation argument
    ignored: `#2'} \fi}}

\def\pmb#1{\setbox0=\hbox{#1}
    \kern-.025em\copy0\kern-\wd0
    \kern.05em\copy0\kern-\wd0
    \kern-.025em\raise.0433em\box0}

\def\fnt#1#2{\footnotetext{\kern-.3em
    {$^{\mbox{\scriptsize #1}}$}{#2}}}

\def\ps@myheadings{%
    \def\@evenhead{\slshape\leftmark\hfil}%       %EVEN PAGE
    \def\@oddhead{\hfil{\slshape\rightmark}}%     %ODD PAGE
    \let\@mkboth\@gobbletwo
    \let\sectionmark\@gobble
    \let\subsectionmark\@gobble
    }
\font\tenrm=cmr10
\font\tenit=cmti10
\font\tenbf=cmbx10
\font\bfit=cmbxti10 at 10pt
\font\ninerm=cmr9
\font\nineit=cmti9
\font\ninebf=cmbx9
\font\eightrm=cmr8

\def\qed{\hbox{${\vcenter{\vbox{            %HOLLOW SQUARE
   \hrule height 0.4pt\hbox{\vrule width 0.4pt height 6pt
   \kern5pt\vrule width 0.4pt}\hrule height 0.4pt}}}$}}

\begin{document}
\setlength{\textheight}{7.7truein}  %for 2nd page onwards

\thispagestyle{empty}

\markboth{\protect{\footnotesize\it RMF + BCS description of
drip-line nuclei }}{\protect{\footnotesize\it  Yadav, Kaushik and
Toki }}

\normalsize\textlineskip

\setcounter{page}{1}

\copyrightheading{} %{Vol.~0, No.~0 (2002) 000--000}

\vspace*{0.88truein}

\centerline{\bf DESCRIPTION OF DRIP-LINE NUCLEI} \centerline{\bf
WITHIN RELATIVISTIC MEAN FIELD PLUS BCS APPROACH }
%\baselineskip=13pt
%\centerline{\bf MANUSCRIPTS USING COMPUTER SOFTWARE\footnote{For
%the title, try not to use more than three lines. Typeset the title
%in 10-pt Times Roman, uppercase and boldface.}}
%\vspace*{0.37truein}
\vspace*{0.4truein} \centerline{\footnotesize H. L.
Yadav$^{1,2}$\footnote{email: hlyadav@sancharnet.in (H. L. Yadav)}
, M. Kaushik$^{1}$, H. Toki$^{2,3,}$\footnote{email:
toki@rcnp.osaka-u.ac.jp (H. Toki)}} \vspace*{12pt}
\centerline{\footnotesize\it $^{1}$ Department of Physics,
Rajasthan  University, Jaipur 302004, India} \vspace*{12pt}
\baselineskip=12pt \centerline{\footnotesize\it $^{2}$Research
Center for Nuclear Physics (RCNP), Osaka University, 10-1,
Mihogaoka} \baselineskip=10pt \centerline{\footnotesize\it
Ibaraki, Osaka 567-0047, Japan} \vspace*{12pt}
\baselineskip=12pt \centerline{\footnotesize\it $^{3}$The
Institute of Physical and Chemical Research (RIKEN), 2-1 Hirosawa}
\baselineskip=10pt \centerline{\footnotesize\it Wako, Saitama
351-0198, Japan} \vspace*{12pt}
\vspace*{0.228truein}

\publisher{(received date)}{(revised date)}

%\vspace*{0.21truein}
\vspace*{0.23truein} \abstracts{Recently it has been demonstrated,
considering Ni and Ca isotopes as prototypes,  that the
relativistic mean-field plus BCS (RMF+BCS) approach wherein the
single particle continuum corresponding to the RMF is replaced by
a set of discrete positive energy states for the calculation of
pairing energy provides a good approximation to the full
relativistic Hartree-Bogoliubov (RHB) description of the ground
state properties of  the drip-line neutron rich nuclei. The
applicability of RMF+BCS approach even for the drip-line nuclei is
essentially due to the fact that the main contribution to the
pairing correlations for the neutron rich nuclei is provided by
the low-lying resonant states, in addition to the contributions
coming from the states close to the Fermi surface. In order to
show the general validity of this approach we present here the
results of our detailed calculations for the ground state
properties of the chains of isotopes of $\rm {O, Ca, Ni, Zr, Sn}$
and $\rm {Pb}$ nuclei. The TMA force parameter set has been used
for the effective mean-field Lagrangian with the nonlinear terms
for the sigma and omega mesons. Further, to check the validity of
our treatment for different mean-field descriptions, calculations
have also been carried out for the NL-SH force parameterization
usually employed for the description of drip-line nuclei.
Comprehensive results for the two neutron separation energy, rms
radii, single particle pairing gaps and pairing energies etc. are
presented. Especially, the $\rm{Ca}$ isotopes are found to exhibit
distinct features near the neutron drip line whereby it is found
that further addition of neutrons causes a rapid increase in the
neutron rms radius with almost no increase in the binding energy,
indicating the occurrence of halos. It is mainly caused by the
pairing correlations and results in the existence of bound states
of extremely neutron rich exotic nuclei. Similar characteristics
though less pronounced, are also exhibited by the neutron rich
$\rm{Zr}$ isotopes. A comparison of these results with the
available experimental data and with the recent continuum
relativistic Hartree-Bogoliubov (RCHB) calculations amply
demonstrates the validity and usefulness of this fast RMF+BCS
approach for the description of nuclei including those near the
drip-lines. }{}{} \vspace*{10pt}

 \vspace*{2pt}

%\textlineskip          %) USE THIS MEASUREMENT WHEN THERE IS
%\vspace*{12pt}          %) NO SECTION HEADING

\baselineskip=13pt          %) ACTUAL LEADING
\normalsize                 %) USE THIS MEASUREMENT WHEN THERE IS
%\section{The Main Text}        %) A SECTION HEADING
%\vspace*{-0.5pt}
\section{Introduction}
\noindent Production of radioactive beams have facilitated the
nuclear structure studies away from the line of $\beta$-stability,
especially for the neutron rich nuclei. The structure of these
exotic nuclei with unusually large $\mid N - Z\mid$ value is
characterized by several interesting features. It exhibits
extremely small separation energy of the outermost nucleons and
the Fermi level lies close to the single particle continuum. In
the case of neutron drip-line nuclei the neutron density
distribution shows a much extended tail with  a diffused neutron
skin\cite{tanihata}. In some cases it even leads to the phenomenon
of neutron halo made of several neutrons outside a core with
separation energy of the order of $100$ keV or less. Due to the
weak binding and large spatial dimension of the outermost
nucleons, the role of continuum states and their coupling to the
bound states become exceedingly important, especially for the
pairing energy contribution to the total binding energy of the
system. Theoretical investigations of such nuclei have been
carried out extensively within the framework of mean field
theories such as Hartree-Fock-Bogoliubov (HFB),
HF+BCS\cite{dobac}$^{-}$\cite{grasso} and their relativistic
counterparts\cite{walecka}$^{-}$\cite{estal}. A detailed
comparative study of the HFB approach with those of the HFB based
on the box boundary conditions, and the HF+BCS+Resonant continuum
approach has been carried out by Grasso {\it et al}.\cite{grasso}
providing insight to the validity of different approaches for the
treatment of drip-line nuclei. Earlier Sandulescu {\it et
al}.\cite{sand2} have also studied the effect of resonant
continuum on the properties of neutron rich nuclei within the
HF+BCS approximation. The interesting result of these
investigation is that only a few low energy resonant states,
especially those near the Fermi surface influence in an
appreciable way  the pairing properties of nuclei far from the
$\beta$-stability. Indeed, comparison between the results given by
the resonant continuum HF+BCS\cite{sand2} and the continuum HFB
calculations\cite{grasso} shows that a few low-lying resonances
give practically the full effect of the continuum on pairing
related properties. This finding has proved to be of immense
significance because one can eventually make systematic studies of
a large number of nuclei by using a simpler HF+BCS approximation.

Recently the relativistic mean field (RMF) theory has been
extensively used for the study of unstable nuclei
\cite{toki,hirata,suga}. The advantage of the RMF approach is that
it provides the spin-orbit interaction in the entire mass region,
which is consistent with the nucleon density\cite{walecka,pgr}.
This indeed could be very important for the study of unstable
nuclei near the drip line, since the single particle properties
near the threshold change largely as compared to the case of
deeply  bound levels in the nuclear potential. In addition to
this, the pairing properties are also important for nuclei near
the drip line where we have to take into account the coupling of
bound pairs with the pairs in the continuum.  In order to take
into account the pairing correlations together with a realistic
mean field, the framework of RHB approach is commonly
used\cite{meng,lala}. In this connection, the finding above for
the non-relativistic frameworks has turned out to be very
important for the systematic work of unstable nuclei in the
relativistic approach. This has been demonstrated recently by
Yadav et al.\cite{yadav} for the case of $^{48-98}\rm{Ni}$
isotopes. Indeed the RMF+BCS scheme\cite{yadav} wherein the single
particle continuum corresponding to the RMF is replaced by a set
of discrete positive energy states yields results which are found
to be in close agreement with the experimental data and with those
of recent continuum relativistic Hartree-Bogoliubov (RCHB) and
other similar mean-field calculations\cite{meng,meng2}. In fact
the applicability of RMF+BCS approach even for the drip-line
nuclei is essentially due to the fact that the main contribution
to the pairing correlations for the neutron rich nuclei is
provided by the low-lying resonant states, in addition to the
contributions coming from the states close to the Fermi surface.

 With the success of the RMF+BCS approach for the prototype
calculations of $\rm{Ni}$ and $\rm{Ca}$ isotopes\cite{yadav}, it
is natural to investigate its validity for other nuclei in
different regions of the periodic table. Also it would be
interesting to check the results with different popular RMF force
parameterizations. With this in view we have carried out detailed
calculations for the chains of isotopes of $\rm{O, Ca, Zr, Sn}$
and $\rm{Pb}$ nuclei using the TMA and the NLSH force
parameterizations. Excepting $\rm{Zr}$, the other nuclei
considered in this studies are proton magic nuclei. The results of
these calculations for the two neutron separation energy, neutron,
proton, and matter rms radii, and single particle pairing gaps
etc., and their comparison with the available experimental data
and with the RCHB results are presented here to demonstrate the
general validity of our RMF+BCS approach.

\section{Theoretical Formulation and Model}

\noindent Our RMF calculations have been carried out using the
model Lagrangian density with nonlinear terms both for the
${\sigma}$ and ${\omega}$ mesons as described in detail in
Ref.~\cite{suga}, which is given by
\begin{eqnarray}
       {\cal L}& = &{\bar\psi} [\imath \gamma^{\mu}\partial_{\mu}
                  - M]\psi\nonumber\\
                  &&+ \frac{1}{2}\, \partial_{\mu}\sigma\partial^{\mu}\sigma
                - \frac{1}{2}m_{\sigma}^{2}\sigma^2- \frac{1}{3}g_{2}\sigma
                  ^{3} - \frac{1}{4}g_{3}\sigma^{4} -g_{\sigma}
                 {\bar\psi}  \sigma  \psi\nonumber\\
                &&-\frac{1}{4}H_{\mu \nu}H^{\mu \nu} + \frac{1}{2}m_{\omega}
                   ^{2}\omega_{\mu}\omega^{\mu} + \frac{1}{4} c_{3}
                  (\omega_{\mu} \omega^{\mu})^{2}
                   - g_{\omega}{\bar\psi} \gamma^{\mu}\psi
                  \omega_{\mu}\nonumber\\
               &&-\frac{1}{4}G_{\mu \nu}^{a}G^{a\mu \nu}
                  + \frac{1}{2}m_{\rho}
                   ^{2}\rho_{\mu}^{a}\rho^{a\mu}
                   - g_{\rho}{\bar\psi} \gamma_{\mu}\tau^{a}\psi
                  \rho^{\mu a}\nonumber\nonumber\\
                &&-\frac{1}{4}F_{\mu \nu}F^{\mu \nu}
                  - e{\bar\psi} \gamma_{\mu} \frac{(1-\tau_{3})}
                  {2} A^{\mu} \psi\,\,,%\nonumber\
\end{eqnarray}
where the field tensors $H$, $G$ and $F$ for the vector fields are
defined by
\begin{eqnarray}
                 H_{\mu \nu} &=& \partial_{\mu} \omega_{\nu} -
                       \partial_{\nu} \omega_{\mu}\nonumber\\
                 G_{\mu \nu}^{a} &=& \partial_{\mu} \rho_{\nu}^{a} -
                       \partial_{\nu} \rho_{\mu}^{a}
                     -2 g_{\rho}\,\epsilon^{abc} \rho_{\mu}^{b}
                    \rho_{\nu}^{c} \nonumber\\
                  F_{\mu \nu} &=& \partial_{\mu} A_{\nu} -
                       \partial_{\nu} A_{\mu}\,\,,\nonumber\
\end{eqnarray}
and other symbols have their usual meaning.

The set of parameters appearing in the effective Lagrangian (1)
have been obtained in an extensive study which provides a good
description for the ground state of nuclei and that of the nuclear
matter properties\cite{suga}. This set, termed as TMA, has an
$A$-dependence and covers the light as well as heavy nuclei from
$^{16}\rm{O}$ to $^{208}\rm{Pb}$. Table 1 lists the TMA set of
parameters along with the results for the calculated bulk
properties of nuclear matter. As mentioned earlier we have also
carried the RMF+BCS calculations using the NL-SH force parameters
\cite{sharma}  in order to compare our results with those obtained
in the RHB calculations \cite{meng2} using this force
parameterization. The NL-SH parameters are also listed in Table 1
together with the corresponding nuclear matter properties.

Based on the single-particle spectrum calculated by the RMF
described above, we perform a state dependent BCS
calculations\cite{lane,ring2}. As we already mentioned, the
continuum is replaced by a set of positive energy states generated
by enclosing the nucleus in a spherical box. Thus the gap
equations have the standard form for all the single particle
states, i.e.
\begin{eqnarray}
     \Delta_{j_1}& =&\,-\frac{1}{2}\frac{1}{\sqrt{2j_1+1}}
     \sum_{j_2}\frac{\left<{({j_1}^2)\,0^+\,|V|\,({j_2}^2)\,0^+}\right>}
      {\sqrt{\big(\varepsilon_{j_2}\,-\,\lambda \big)
       ^2\,+\,{\Delta_{j_2}^2}}}\,\,\sqrt{2j_2+1}\,\,\, \Delta_{j_2}\,\,,
\end{eqnarray}\\
where $\varepsilon_{j_2}$ are the single particle energies, and
$\lambda$ is the Fermi energy, whereas the particle number
condition is given by $\sum_j \, (2j+1) v^2_{j}\,=\,{\rm N}$. In
the calculations we use for the pairing interaction a delta force,
i.e., $V=-V_0 \delta(r)$ with the same strength $V_0$ for both
protons and neutrons. The value of the interaction strength $V_0 =
350\,$ MeV fm$^3$ was determined by obtaining a best fit to the
binding energy of $\rm{Ni}$ isotopes. We use the same value of
$V_0 = 350\,$ for our present studies of isotopes of other nuclei
as well. Apart from its simplicity, the applicability and
justification of using such a $\delta$-function form of
interaction has been recently discussed in  Refs.~{2} and ~{4},
whereby it has been shown in the context of HFB calculations that
the use of a delta force in a finite space simulates the effect of
finite range interaction in a phenomenological manner. The pairing
matrix element for the $\delta$-function force is given by\
\begin{eqnarray}
\left<{({j_1}^2)\,0^+\,|V|\,({j_2}^2)\,0^+}\right>& =&\,-\,\frac{V_0}{8\pi}
       \sqrt{(2j_1+1)(2j_2+1)}\,\,I_R\,\,,
\end{eqnarray}
where $I_R$ is the radial integral having the form
\begin{eqnarray}
   I_R& =&\,\int\,dr \frac{1}{r^2}\,\left(G^\star_{j_ 1}\, G_{j_2}\,+\,
     F^\star_{j_ 1}\, F_{j_2}\right)^2
\end{eqnarray}
Here $G_{\alpha}$ and $F_{\alpha}$ denote the radial wave
functions for the upper and lower components, respectively, of the
nucleon wave function expressed as
\begin{equation}\psi_\alpha={1 \over r} \,\, \left({i \,\,\, G_\alpha \,\,\,
 {\mathcal Y}_{j_\alpha l_\alpha m_\alpha}
\atop{F_\alpha \, {\sigma} \cdot \hat{r}\,
\, {\mathcal Y}_{j_\alpha l_\alpha m_\alpha}}} \right)\,\,,
%\qquad\qquad \rm{for}\qquad
%\hat{r}={{\bf r} \over r}
%\label
\end{equation}\\
and satisfy the normalization condition\
 \begin{eqnarray}
         \int dr\, {\{|G_{\alpha}|^2\,+\,|F_{\alpha}|^2}\}\,=\,1
 \end{eqnarray}\
In Eq. (5) the symbol ${\mathcal Y}_{jlm}$ has been used for the
standard spinor spherical harmonics with the phase $i^l$. The
coupled field equations obtained from the Lagrangian density in
(1) are finally reduced to a set of simple radial
equations\cite{pgr} which are solved self consistently along with
the equations  for the state dependent pairing gap $\Delta_{j}$
and the total particle number $\rm N$ for a given nucleus.

\section{Results and Discussion}

The present RMF+BCS calculations are restricted to the spherical
shape and have been carried out for the entire chain of isotopes
of proton magic nuclei $\rm{O, Ca, Ni}$, $\rm{Sn}$ and $\rm{Pb}$
as well as those of isotopes of $\rm{Zr}$ with proton sub-magic
number Z= 40. We note that earlier non-relativistic as well as
relativistic calculations indicate that $\rm{Zr}$ nuclei with $A >
122$ are spherical \cite{meng3}. It is found that the isotopes of
$\rm{Ca}$ and $\rm{Zr}$, to a large extent, exhibit similar
characteristics near the drip line which is pushed out to
unusually heavy neutron rich isotopes. Similarly the isotopes of
$\rm{Ni, Sn, Pb }$ and that of $\rm{O}$ yield results with common
features. Thus in order to save space, we have made the
presentation of results in the following manner. First a detailed
description of the $\rm{Ca}$ and $\rm{Ni}$ isotopes is given as
representative cases of the nuclei considered in the present
investigation. This is then followed by a  description of the
results for the rest of the nuclei $\rm{O, Zr, Sn}$ and $\rm{Pb}$.

Again, from amongst the chain of isotopes of these nuclei, we
consider only a few of the selected neutron rich isotopes to
describe in detail their single particle spectra and resonant
states, and their pairing gap energies etc. These chosen isotopes
are considered here as representative cases in order to
demonstrate the main properties of neutron rich isotopes for a
particular element. The total pairing energy contribution to the
binding energy of the system plays a crucial role in the
understanding of exotic nuclei and has been described next. This
is followed by a detailed description of the results of two
neutron separation energies, proton and neutron radii, and the
density distributions for the chains of isotopes of different
nuclei mentioned above.

Furthermore, the structure of single particle spectra near the
Fermi level, and its variation with further addition of neutrons
have been usefully employed to explain the results for the
two-neutron drip line, and also for the possibility of occurrence
of neutron halos as well as for the asymptotic radial dependence
of the neutron density distributions. These involve both the bound
as well as positive energy states in the continuum. These states
play an important role as scattering of particles from bound to
continuum states near the Fermi level and vice versa due to the
the pairing interaction involves mainly these very states. The
last few occupied states near the Fermi level also provide an
understanding of the radii of the loosely bound exotic nuclei. The
neutron rich nuclei in which the last filled single particle state
near the Fermi level is of low angular momentum ($s_{1/2}$ or
$p_{1/2}$ state), especially the $l=0$ state, can have large radii
due to large spatial extension of the $s_{1/2}$ state which has no
centrifugal barrier. In our calculations such a situation is seen
to occur in the neutron rich $\rm{Ca}$ isotopes wherein the last
single particle state involved is the $3s_{1/2}$ leading to halo
like phenomenon already well known in light nuclei. Similar
effect, though less pronounced, is found in the neutron rich
$\rm{Zr}$ isotope due to the $3p_{1/2}$ and $3sp_{3/2}$ states
near the Fermi level. In contrast, the single particle structures
in the neutron rich $\rm{Ni}$, $\rm{Sn}$ and $\rm{Pb}$ isotopes do
not favor the formation of such halos as has been described later.

As stated earlier our RMF+BCS calculations have been performed
with two different force parameterizations, TMA and NL-SH, in
order to check if the results have any dependence on different
mean field descriptions. Details of our calculations show that the
two interactions employed here produce very similar results and,
therefore, for the detailed description of single particle and
resonant states mentioned above, we have presented results
obtained with only the TMA force.

\subsection{$\rm{Ca}$ Isotopes}

Neutron rich members of $\rm{Ca}$ isotopes constitute interesting
example of loosely bound system. In order to describe the
contribution of various single particle states to the pairing
energy, we plot in fig. 1 the calculated RMF potential, a sum of
scalar and vector potentials, for the neutron rich nucleus
$^{62}\rm{Ca}$ along with the spectrum for the bound neutron
single particle states. This is a typical example of the neutron
rich nucleus amongst the $\rm{Ca}$ isotopes. The figure also shows
the positive energy state corresponding to the first low-lying
resonance $1g_{9/2}$, and other positive energy states, for
example, $3s_{1/2}$, $2d_{5/2}$, $2d_{3/2}$ and $1g_{7/2}$ close
to the Fermi surface which play significant role for the binding
of neutron rich isotopes, ranging from $^{62}\rm{Ca}$ to
$^{72}\rm{Ca}$, through their contributions to the total pairing
energy . In contrast to other states in the box which correspond
to the non-resonant continuum, the position of the resonant
$1g_{9/2}$ state is not much affected by changing the box radius
around $R=30$ fm. For the purpose of illustration we have also
depicted in fig. 1 the total mean field potential for the neutron
$1g_{9/2}$ state, obtained by adding the centrifugal potential
energy.

\vskip 0.15in
\psfig{figure=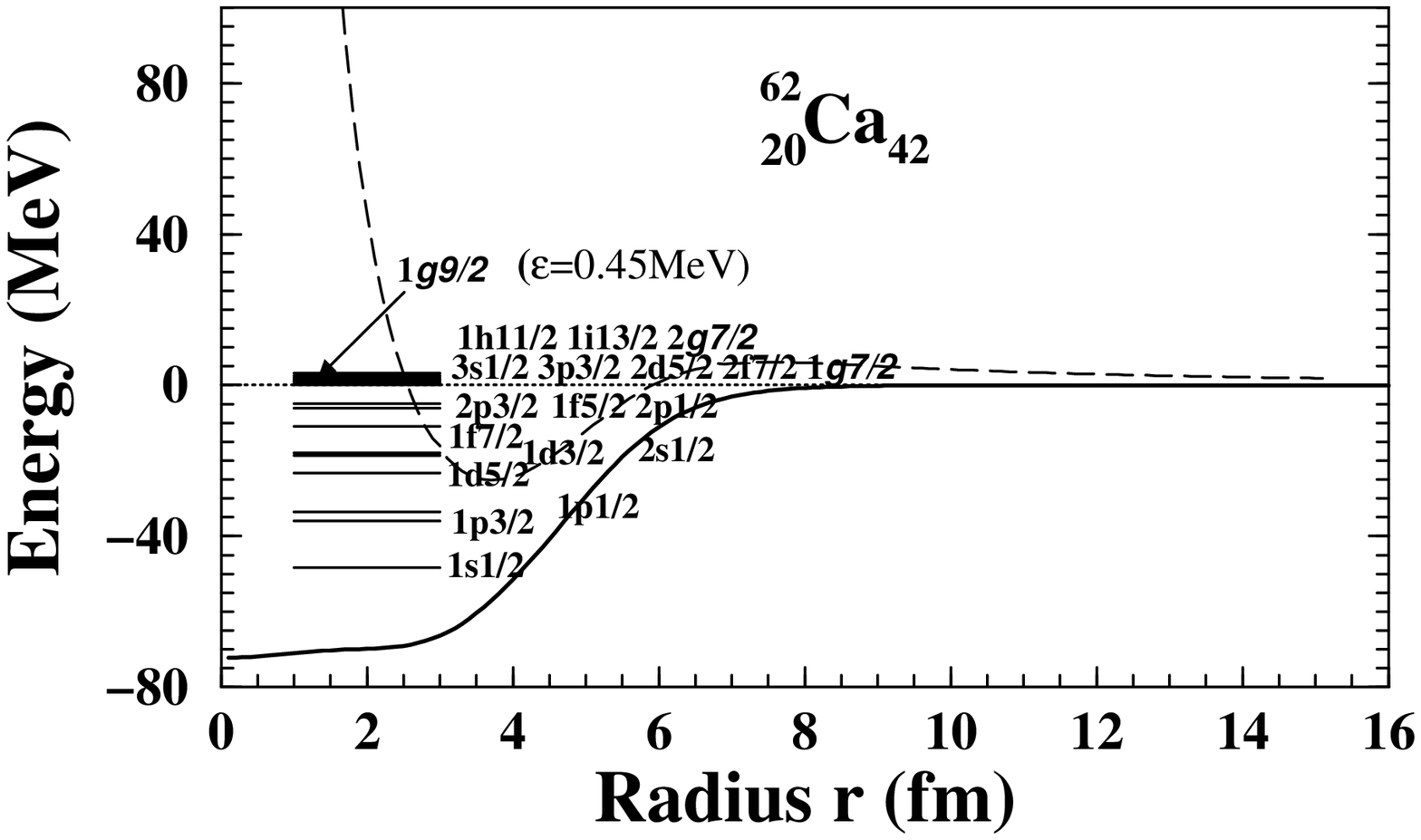,width=13cm,height=8cm} \vskip
0.001in {\noindent \small {{\bf Fig. 1.} The RMF potential energy
(sum of the scalar and vector potentials), for the nucleus
$^{62}\rm{Ca}$ as a function of radius is shown by the solid line.
The long dashed line represents the sum of RMF potential energy
and the centrifugal barrier energy for the neutron resonant state
$1g_{9/2}$. The figure also shows the energy spectrum of some
important neutron single particle states along with the resonant
$1g_{9/2}$ state at 0.45 MeV. A few high lying continuum states
like $1h_{11/2}$, $1i_{13/2}$ etc. are also indicated. \mbox{} }}
\vspace{ 0.5cm}

It is evident from the figure that the effective total potential
for the $1g_{9/2}$ state has an appreciable barrier for the
trapping of waves to form a quasi-bound or resonant state. Such a
meta-stable state remains mainly confined to the region of the
potential well and the wave function exhibits characteristics
similar to that of a bound state. This is clearly observed in fig.
2 which depicts the radial wave functions of some of the neutron
single particle states lying close to the Fermi surface, the
neutron Fermi energy being $\lambda_n\, =\,-0.204$ MeV. These
include the bound $1f_{5/2}$ and $2p_{1/2}$, and the continuum
$3s_{1/2}$ and $2d_{5/2}$ states in addition to the state
corresponding to the resonant $1g_{9/2}$.

The wave function for the $1g_{9/2}$ state plotted in fig. 2 is
clearly seen to be confined within a radial range of about 8 fm
and has a decaying component outside this region, characterizing a
resonant state. In contrast, the main part of the wave function
for the non-resonant states, e.g. $2d_{5/2}$, is seen to be spread
over outside the potential region, though a small part is also
contained inside the potential range. This type of state thus has
a poorer overlap with the bound states near the Fermi surface
leading to small value for the pairing gap $\Delta_{2d_{5/2}}$.
Further, the positive energy states lying much higher from the
Fermi level, for example, $1h_{11/2}$, $1i_{13/2}$ etc.  have a
negligible contribution to the total pairing energy of the system.

\vskip 0.1in \psfig{figure=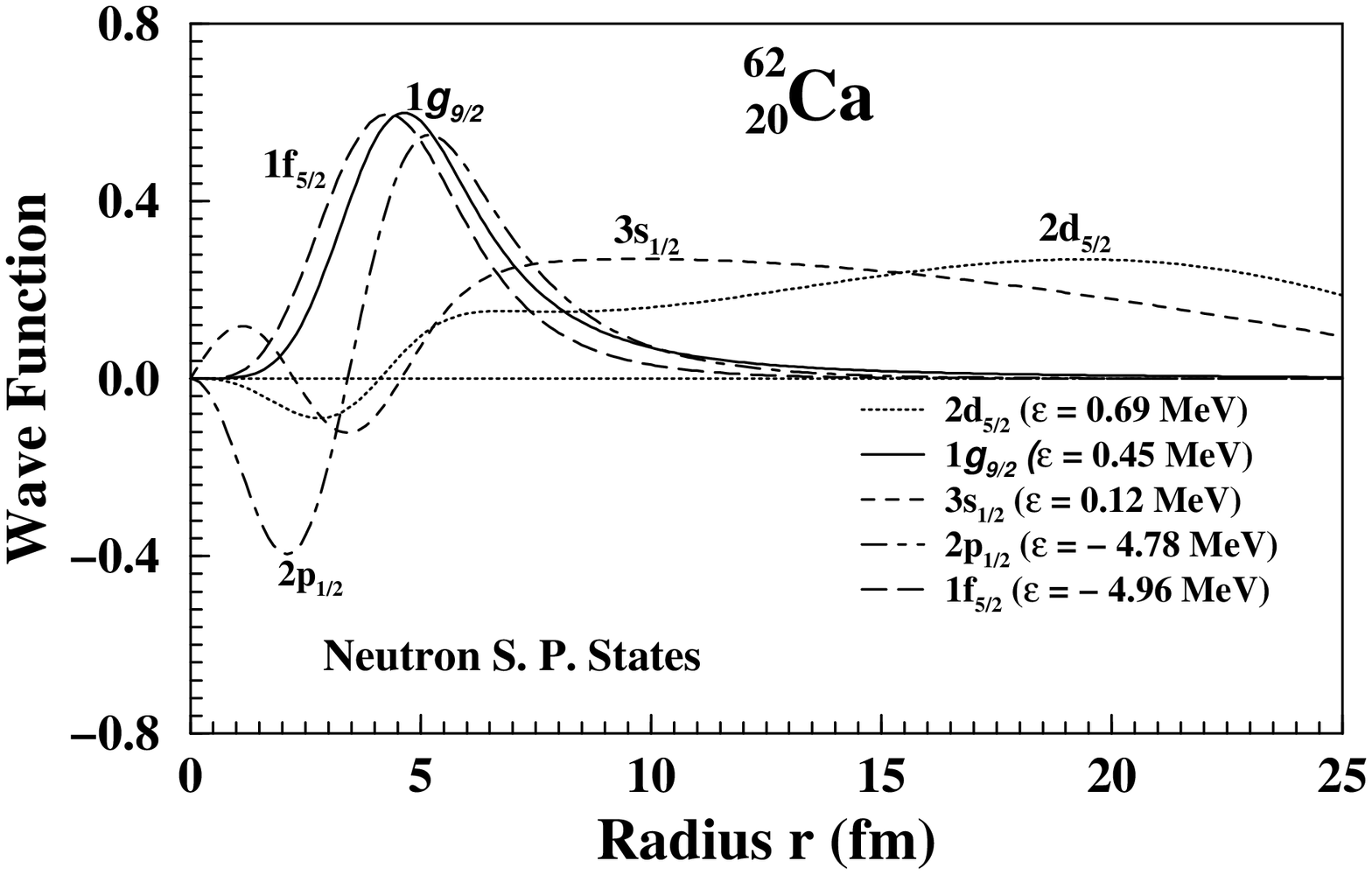,width=13cm} \vskip
0.1in {\noindent \small {{\bf Fig. 2.} Radial wave functions of a
few representative neutron single particle states with energy
close to the Fermi surface for the nucleus $^{62}\rm{Ca}$. The
solid line shows the resonant $1g_{9/2}$ state at energy 0.45 MeV,
while the bound $1f_{5/2}$ state at -4.96 MeV, and $2p_{1/2}$
state at -4.78 MeV are shown by long dashed line, and dashed
-dotted lines, respectively. The $3s_{1/2}$ and $2d_{5/2}$ states
with positive energies 0.12 MeV and 0.69 MeV have been depicted by
small dashed and dotted lines, respectively. \mbox{} }} \vspace{
0.5cm}

 These features can be seen from fig. 3 which depicts the
calculated pairing gap energy $\Delta_j$ for the neutron states in
the nucleus $^{62}\rm{Ca}$. However, we have not shown in the
figure the single particle states having very small $\Delta_j$
values as these do not contribute significantly to the total
pairing energy. One observes indeed in fig. 3 that the gap energy
for the $1g_{9/2}$ state has a value close to $1$ MeV which is
quantitatively similar to that of bound states $1f_{7/2}$ and
$2p_{3/2}$ etc. Also, fig. 3 shows that the pairing gap values for
the non-resonant states like $3s_{1/2}$ and $2d_{5/2}$ in
continuum have much smaller gap energy. However, as the number of
neutrons increases while approaching the neutron drip line nucleus
$^{72}\rm{Ca}$, the single particle states $3s_{1/2}$, $1g_{9/2}$,
$2d_{5/2}$ and $2d_{3/2}$ which lie near the Fermi level gradually
come down close to zero energy, and subsequently the $1g_{9/2}$
and $3s_{1/2}$ states even become bound states. This helps in
accommodating more and more neutrons which are just bound. In
fact, the occupancy of the $3s_{1/2}$ state in these extremely
neutron rich isotopes causes the halo formation as will be
discussed later.

\vskip 0.1in \psfig{figure=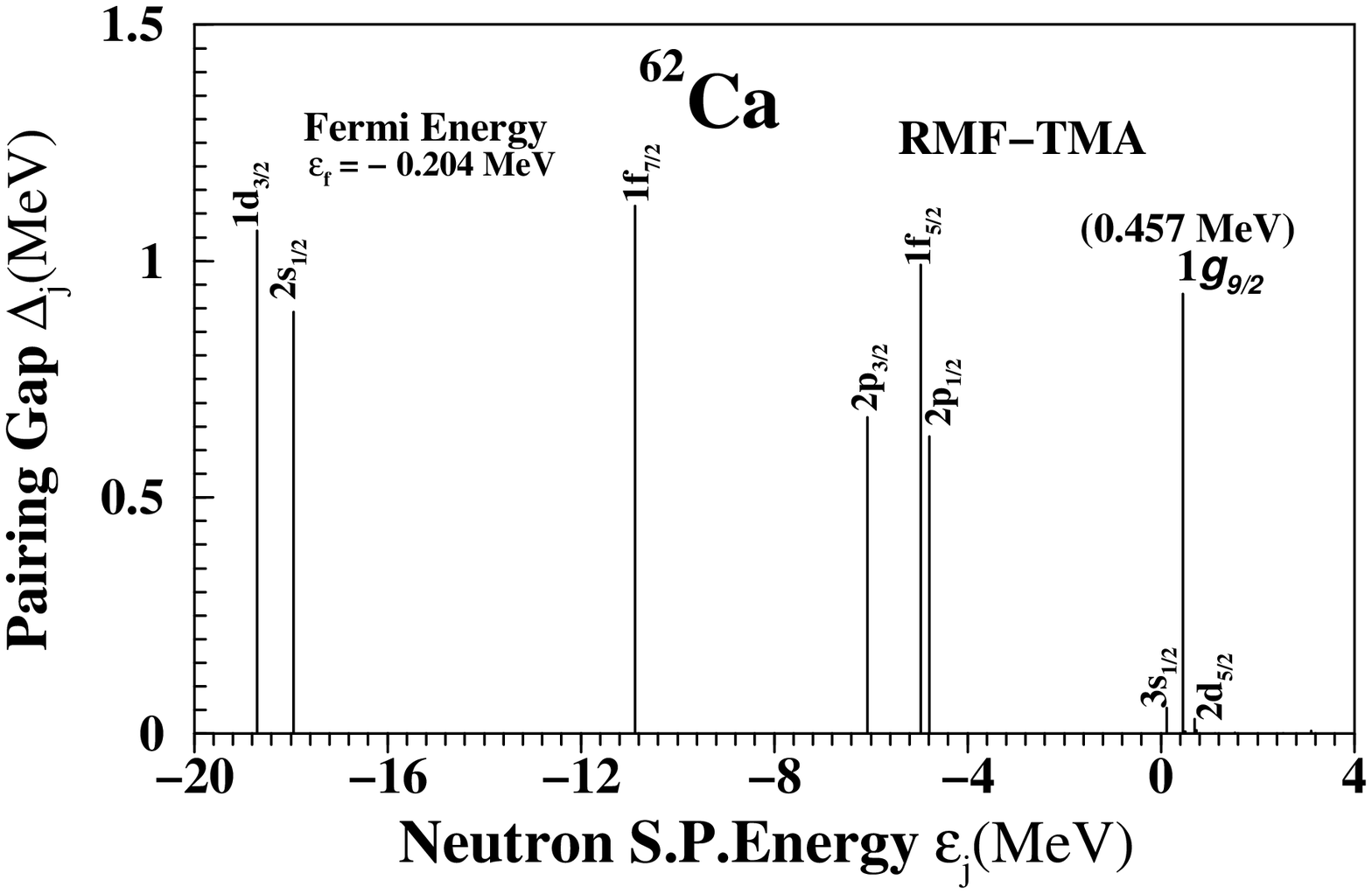,width=13cm} \vskip
0.15in {\noindent \small {{\bf Fig. 3.} Pairing gap energy
$\Delta_{j}$ of neutron single particle states with energy close
to the Fermi surface for the nucleus $^{62}\rm{Ca}$. The resonant
$1g_{9/2}$ state at energy 0.457 MeV has the gap energy of about
$1$ MeV which is close to that of bound states like $1f_{5/2}$,
$1f_{7/2}$, $1d_{3/2}$ etc. \mbox{} }} \vspace{0.5cm}

The contribution of pairing energy plays an important role for the
stability of the neutron rich nuclei and consequently in deciding
the position of the neutron and proton drip lines. It is noted
again that the RMF+BCS calculations carried out with two different
sets of force parameters, the TMA and NL-SH, yield almost similar
results also for the pairing energies for the isotopes of various
nuclei considered in the present investigation. In fig. 4. this
has been shown for the case of $\rm{Ca}$ isotopes. The differences
in the two results can be attributed to the difference in the
detailed structure of single particle energies  obtained with the
TMA and NL-SH forces. It is seen from fig. 4 that the pairing
energy vanishes for the neutron numbers $N = 14, 20, 28$ and $40$
indicating the shell closures. It is observed that the usual shell
closure at $N=50$ is absent for the neutron rich $\rm{Ca}$
isotopes and at $N=40$ a new shell closure appears. Also, for
$N=52$ the pairing energy value is rather small. This
reorganization of single particle energies with large values of
N/Z ratio (for the neutron rich $\rm{Ca}$ isotopes N/Z $\geq 2$)
has its origin in the deviation of the strength of spin-orbit
splitting from the conventional shell model results for nuclei
with not so large $N/Z$ ratio. Further, from the figure one sees
the appearance of a new shell closure at the neutron number $N=
14$. This represents the case of nuclei having large number of
protons as compared to that of neutrons.

\vskip 0.15in \psfig{figure=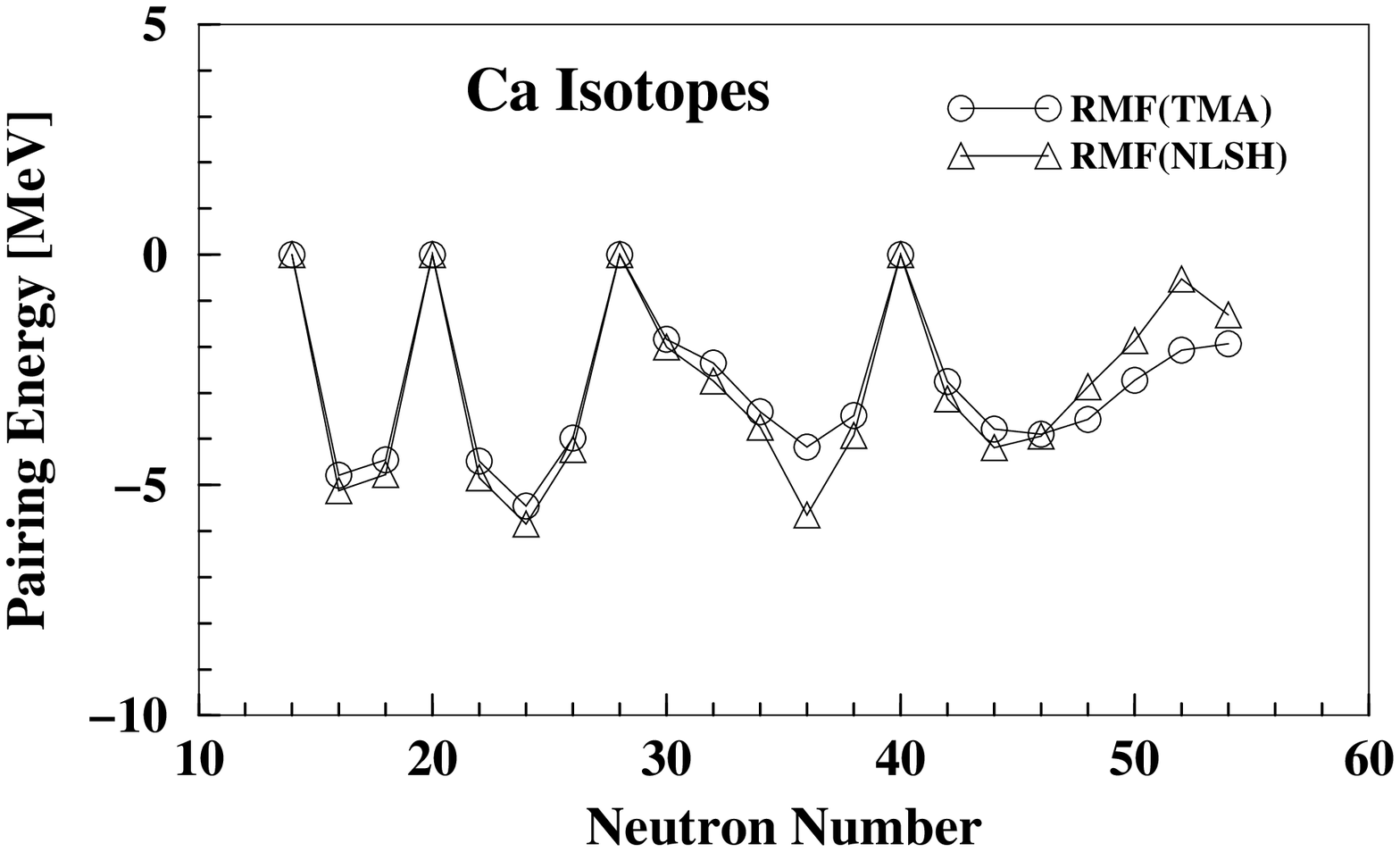,width=13 cm}
\vskip 0.15in {\noindent \small {{\bf Fig. 4.} Present RMF results
for the pairing energy  for the $\rm{Ca}$ isotopes obtained with
the TMA (open circles) are compared with those obtained using the
NL-SH (open triangles) force parameters.\mbox{} }}\vspace{0.5cm}

It is also seen from the figure that the pairing energies for the
nuclei falling in between two closed shells, as expected,
increases from zero and have maximum values in the middle of the
two closed shells. For the neutron rich $\rm{Ca}$ isotopes there
is a tendency of slow rate of change in the pairing energy while
one moves from one isotope to another. The shell structure as
revealed by the pairing energies are also exhibited in the
variation of two neutron separation energies $S_{2n}$ as shown in
fig. 5 for the $\rm{Ca}$ isotopes. An abrupt increase in the
$S_{2n}$ values for the isotopes next to magic numbers is
evidently seen in Fig. 5. Thus, appearance of new magic numbers as
well as the disappearance of conventional magic numbers for nuclei
with extreme isospin values are expected to be quite general
feature related to the reorganization of the single particle
states due to changed characteristics of the spin-orbit splitting
heather to known from stable nuclei with normal isospin values.

The two neutron separation energies provide a crucial means to
check the validity of model calculations as extensive experimental
data are available for long chains of isotopes of the nuclei
considered in the present investigation. It is important to
emphasize that such extensive experimental data are not available
for the rms radii and charge densities especially for the neutron
rich isotopes.  In the lower panel of Fig. 5 we present the
results of two neutron separation energy and their comparison with
the RCHB, and also with available experimental data, for the
even-even $^{34-76}\rm{Ca}$ isotopes . As stated earlier, in order
to check the validity of our treatment for different mean field
descriptions, calculations have been done for two different RMF
parameterizations: the TMA\cite{suga} force discussed above and
the NL-SH\cite{sharma} force often used for the drip-line nuclei.
The upper panel of fig. 5 depicts the difference between the
experimental and calculated values as well as the difference
between our RMF+BCS predictions and those obtained from the RCHB
approach\cite{meng2}.

\vskip 0.15in \psfig{figure=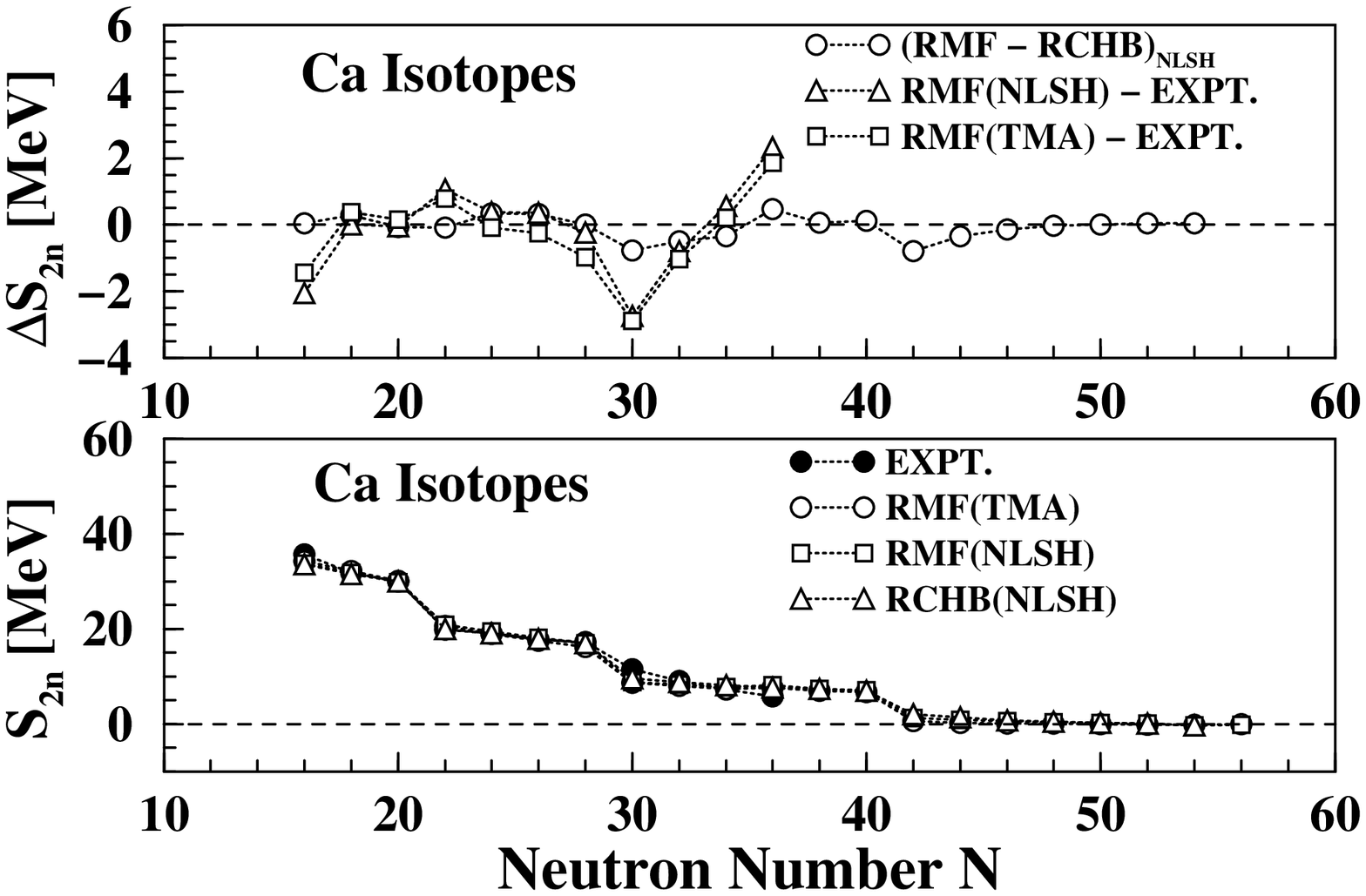,width=13 cm} \vskip
0.15in {\noindent \small {{\bf Fig. 5.} Lower panel: The present
RMF results for the two neutron separation energy  for the
$\rm{Ca}$ isotopes obtained with the TMA (open circles) and the
NL-SH (open squares) force parameters are compared with the
continuum relativistic Hartree-Bogoliubov (RCHB) calculations of
Ref.~26 carried out with the NL-SH force (open triangles). The two
neutron drip line is predicted to occur at $N=52$ corresponding to
$^{72}\rm{Ca}$. This part of the plot also depicts the available
experimental data\cite{audi} (solid circles) for the purpose of
comparison. Upper panel: It depicts the difference in the RMF+BCS
results  and the RCHB results of Ref.~26 for the two neutron
separation energy obtained for the NL-SH force. the plot also
shows the difference of the calculated results with respect to the
available experimental data\cite{audi}.\mbox{} }} \vspace{0.5cm}

The interesting results for the $\rm Ca$ isotopes is that the
isotopes beyond A$=$66 have two neutron separation energy close to
zero. It is seen from the figure that the two forces yield similar
results. The small differences, especially for the isotopes with
N$ > 42$, are due to difference in the energy spacing between the
neutron $1g_{9/2}$, and the neutron $3s_{1/2}$ and $2d_{5/2}$
single particle states in the two RMF descriptions. Our
calculations for the TMA description predicts that the heaviest
stable isotope against two neutron emission corresponds to A$=$70
(N$=$50) whereas for the the NL-SH force it corresponds to A$=$72
(N$=$52). As mentioned above, since the isotopes beyond A$=$66 are
just bound this difference in the two mean-field description is
not of much significance and should not be taken too seriously.
The isotopes with mass number $70<A<76$ for the TMA, and those
with $72<A<76$ for the NL-SH case are similarly found to be just
unbound with negative separation energy very close to zero.
Accordingly in Fig. 5 we have shown the results up to N$=$56 to
emphasize this point. Also, for our purpose we shall neglect this
small difference in the drip line mentioned above.

Fig. 5 also shows the results of RCHB calculations of Ref.~26
carried out with the NL-SH force parameterization. A comparison
with these results shows that the RMF+BCS description is almost
similar to that of the RCHB results. From fig. 5 it is evident
that our RMF+BCS results obtained with the NL-SH force, in
contrast to those for the TMA force, are closer to the RCHB
results which are also obtained using the NL-SH force. The upper
panel of the figure explicitly shows the difference in the
separation energy obtained in the RCHB and the RMF+BCS
calculations for the NL-SH force. Indeed the difference is quite
small. This figure also depicts the difference of calculated
values with respect to the available experimental data for the
separation energy. The maximum difference is of the order of less
than 2 MeV. However, it should be emphasized that the difference
with respect to the experimental data for both the RMF+BCS and the
RCHB calculations are of similar nature as is easily seen in the
figure.

 The rms radii for the proton and neutron , $r_{p,n}\,
=\,(\langle r^2_{p(n)}\rangle\,)^{1/2} $ calculated from the
respective density distributions are obtained from\

\begin{eqnarray}
         \langle r^2_{p(n)}\rangle\,=\,\frac{\int \rho_{p(n)}\, r^2
         d\tau}{\int \rho_{p(n)}\,d\tau}
\end{eqnarray}\

The experimental data for the rms charge radii are used to deduce
the nuclear rms proton radii using the relation
$r_c^2\,=\,r_p^2\,+\,0.64 \, fm^2$ for the purpose of comparison.
In the lower panel of fig. 6a we have shown the results for the
neutron and proton rms radii of the $\rm{Ca}$ isotopes obtained
with the NL-SH force. For the purpose of comparison the figure
also depicts the results of RCHB calculations\cite{meng2} carried
out with the same NL-SH force. A comparison of the RMF+BCS results
with that for the RCHB  shows that the two approaches yield almost
similar values for the neutron and proton  rms radii $r_n$ and
$r_p$.

\vskip 0.15in \psfig{figure=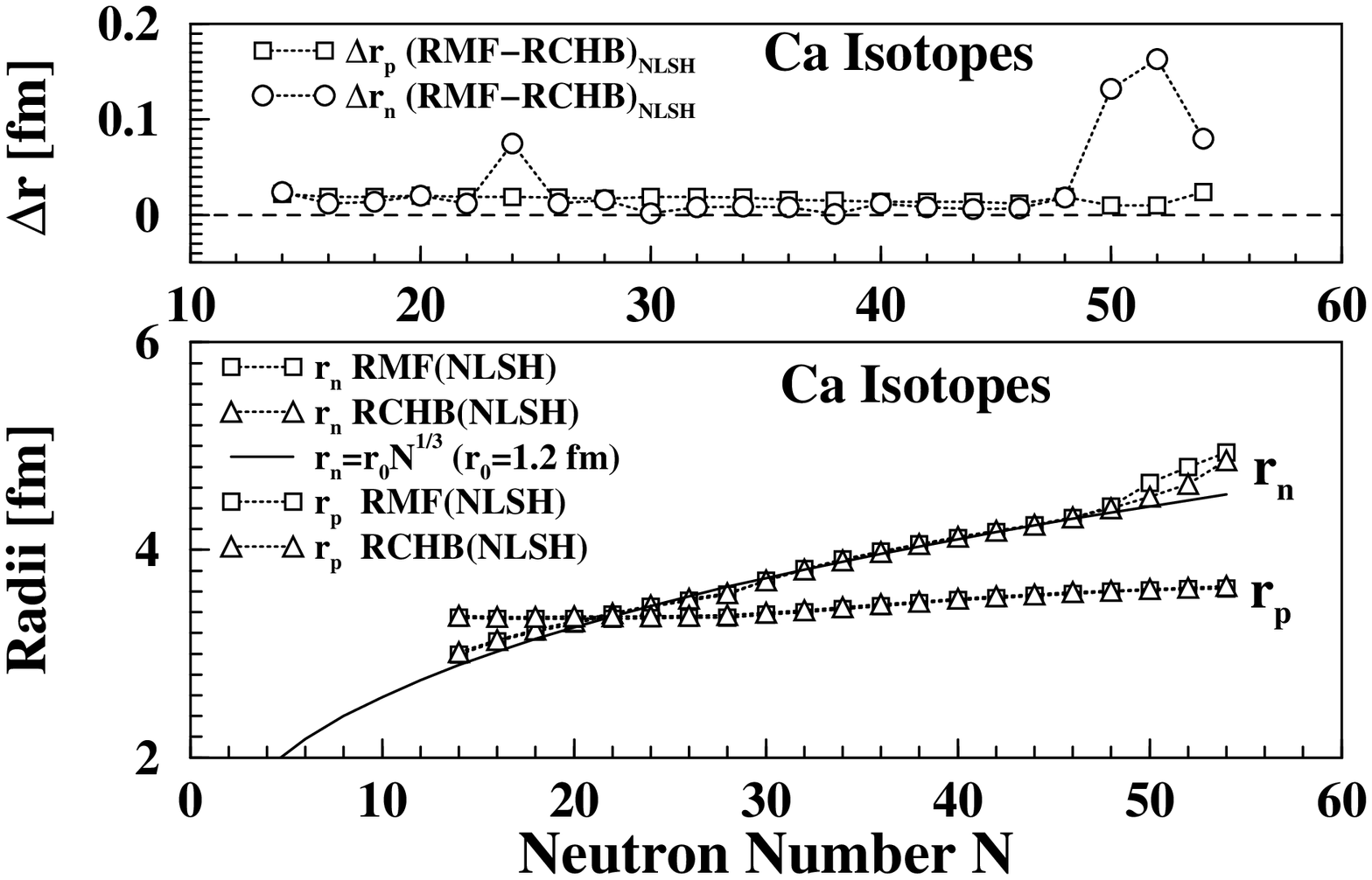,width=13 cm} \vskip
0.15in {\noindent \small {{\bf Fig. 6a.} The present RMF results
for the rms radii of neutron $r_n$ and proton $r_p$(lower panel)
for the $\rm{Ca}$ isotopes obtained with the NL-SH (open squares)
force parameters are compared with the continuum relativistic
Hartree-Bogoliubov (RCHB) calculations of Ref.~26 carried out with
the NL-SH force (open triangles). The upper panel depicts the
difference between the results obtained from  RMF+BCS and the RCHB
approaches for the proton, and the neutron rms radii using the
NL-SH force.\mbox{} }}

We have not plotted in fig. 6a the results obtained with the TMA
force as it is found  that the calculated values for these radii
using two different forces, TMA and the NL-SH, are almost similar
to each other. Instead the TMA results are shown separately in
fig. 6b along with the available experimental data to keep the
figure uncluttered. In fact experimental data for the proton and
neutron rms radii are available only for a few stable $\rm Ca$
isotopes as is seen from fig. 6b. It is evident that the measured
proton radii $r_p$ for the isotopes $^{40-48}\rm Ca$ are in
excellent agreement with our RMF+BCS results. Similarly the
neutron radii $r_n$ for the $^{40,42,44,48}\rm Ca$ isotopes are
found to compare reasonably well, though for the $^{40,42}\rm Ca$
isotopes the calculated values are slightly lower than the
measured ones as is seen in the figure.

\vskip 0.15in \psfig{figure=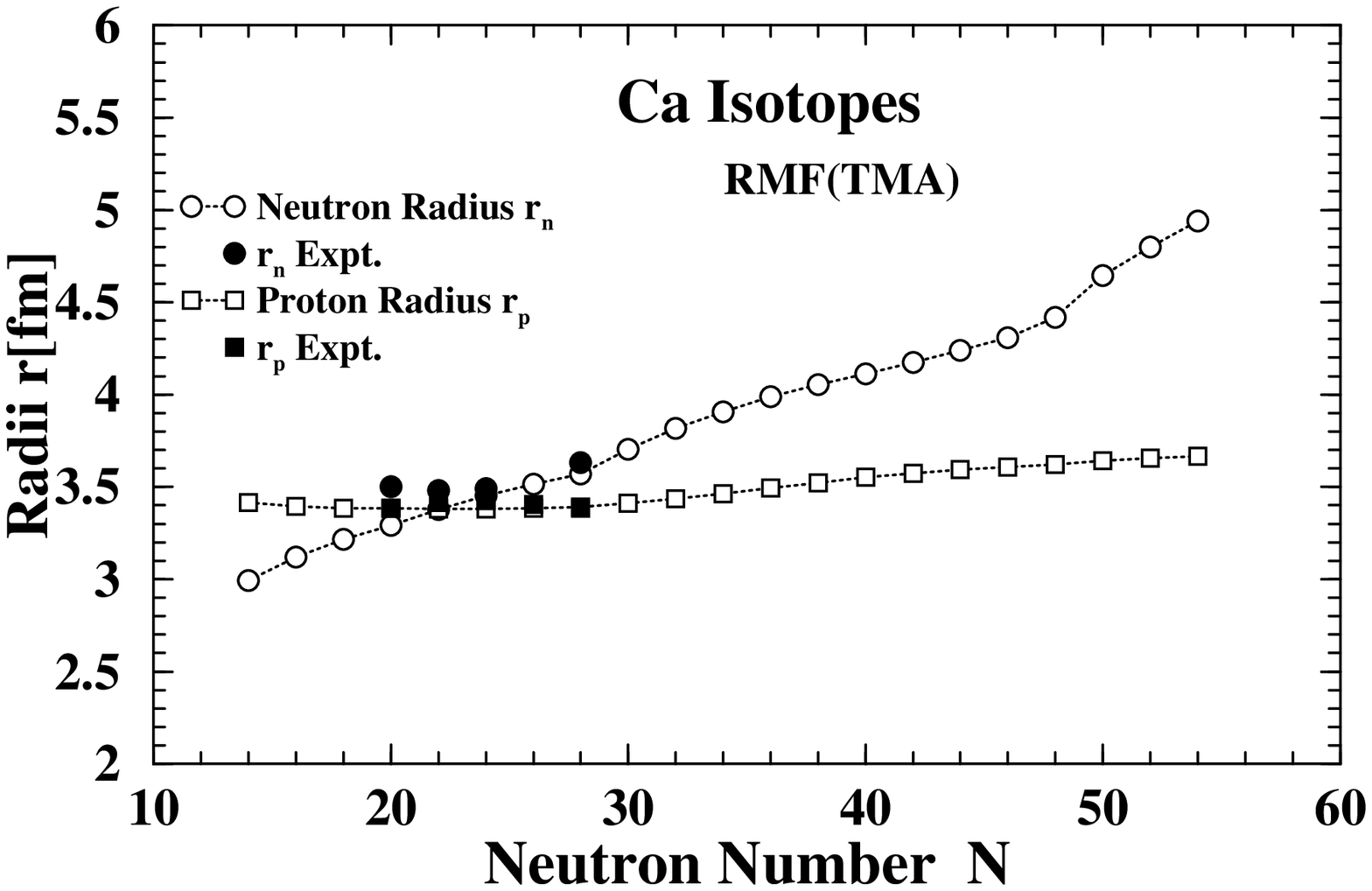,width=13 cm}
\vskip 0.15in {\noindent \small {{\bf Fig. 6b.} The present RMF
results for the rms radii of neutron distribution $r_n$ (open
circles), and that of proton distribution $r_p$ (open squares) for
the $\rm{Ca}$ isotopes obtained with the TMA  force parameters are
compared with the available experimental data \cite{batty}, shown
by solid circles and solid squares, respectively.\mbox{} }}

In order to provide a more quantitative comparison of RMF+BCS and
the RCHB calculations, we have plotted in the upper panel of fig.
6a the differences $\Delta r_p$ and $\Delta r_n$ for the proton
and neutron rms radii obtained using the RMF+BCS and RCHB
approaches with NL-SH force. These results have been,
respectively, shown by opened circles and diamonds for the neutron
and by opened squares and triangles for the proton. It is seen
from the figure  that the difference between the results of
RMF+BCS and RCHB calculations for the proton and neutron radii are
rather quite small. Indeed, the maximum difference of about 0.15
Fermi is found for the extremely neutron rich isotopes (A$\geq$70)
at the top of the drip-line wherein already the halo like
formation has taken place and radii are rather quite large. Indeed
it is clearly seen that the value of $\Delta r_p$ and $\Delta r_n$
for the RMF+BCS and RCHB approaches are of the order of $1-2\%$
and $2-5\%$, respectively. Similarly it is found that this
difference for the RMF+BCS using two different forces (not shown
in the figure) is of the order of about $3-5\%$ for the proton and
$3-6\%$ for the neutron rms radii, respectively.

AS described earlier, in the case of neutron rich $\rm Ca$
isotopes the neutron $1g_{9/2}$ state happens to be one of the
main resonant states having good overlap with the bound states
near the Fermi level. This causes the pairing interaction to
scatter particles from the neighboring bound states to the
resonant state and vice versa. Thus, it is found that the resonant
$1g_{9/2}$ state starts being partially occupied even before the
lower bound single particle states are fully filled in. This
property of the resonant states is observed throughout for all the
nuclei considered in the present investigation. However, in the
case of neutron rich $\rm Ca$ isotopes it is found that the
neutron $3s_{1/2}$ state which lies close to the $1g_{9/2}$ state
also starts getting partially occupied before the $1g_{9/2}$ state
is completely filled. The neutron $3s_{1/2}$ state due to lack of
centrifugal barrier contributes more to the neutron rms radius as
compared to the $1g_{9/2}$ state, and thus one observes a rapid
increase in the neutron rms radius shown in fig. 6a beyond the
neutron number $N=42$ indicating the halos like formation. A
comparison of the rms neutron radii with the $r_n=r_0 N^{1/3}$
line shown in the figure suggests that these radii for the
drip-line isotopes do not follow the  simple $r_0 N^{1/3}$
systematics.

An important aspect of the heavy neutron rich nuclei is the
formation of the neutron skin.\cite{tanihata} For the nucleus
$^{66}\rm{Ca}$ this characteristic feature is  seen in Fig. 7,
where we plot the radial density distribution for protons by
hatched lines and that for the neutrons by solid line. The figure
also shows the neutron density distributions for the other two
isotopes $^{56,64}\rm{Ca}$. The neutron density distributions in
the neutron rich $^{62-72}\rm{Ca}$ nuclei are found to be widely
spread out in the space and give rise to the formation of neutron
halos as will be discussed later.

\vskip 0.15in \psfig{figure=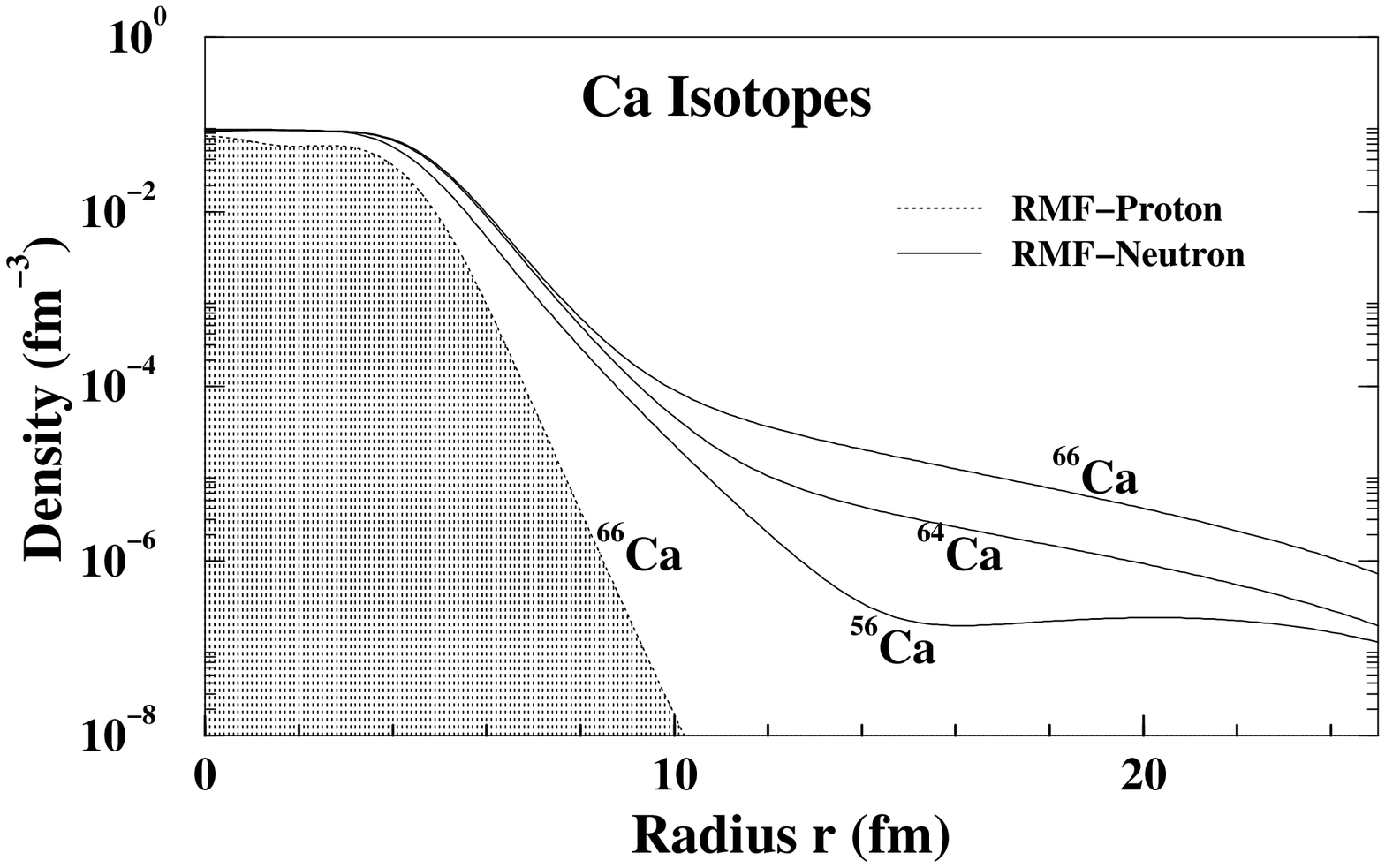,width=13cm}
\vskip 0.15in {\noindent \small {{\bf Fig. 7.} Results for the
proton and neutron density distributions obtained in the RMF+BCS
calculations employing the TMA force. The radial proton density
distribution for the isotope $^{66}\rm Ca$ has been shown by the
hatched area. The neutron radial density distribution for the
$^{56,64,66}\rm Ca$ isotopes have been shown by solid
lines.\mbox{}}} \vspace{0.5cm}

A typical example of variation in the proton radial density
distributions with increasing neutron number has been shown in
fig. 8 for the NL-SH force calculations. As is seen in the figure
the proton distributions are observed to be confined to smaller
distances. Moreover, these start to fall off rapidly already at
smaller distances ( beyond $r> 3$ fm.) as compared to those for
the neutron density distributions. In the interior as well as at
outer distances the proton density values are larger for the
proton rich $\rm{Ca}$ isotopes and decrease with increasing
neutron number N. However, in the surface region, ($r \approx 4$
fm), the proton density values reverse their trend and increase
with increasing neutron number. Due to this feature of the proton
density distributions the proton radii are found to increase,
albeit in a very small measure, with increasing neutron number
though the proton number is fixed at $Z = 20$ for the $\rm {Ca}$
isotopes. Similar features of the proton density distributions are
also exhibited by the results obtained with the TMA force
parametrization. From fig. 7 it is observed that the neutron
density distributions have wide spatial extension, and with
increasing neutron number the density values are appreciably
increased. The asymptotic behavior of the densities are influenced
by the positive energy quasi bound  states via the pairing
correlations. For the closed shell $\rm{Ca}$ isotopes ($N = 14,
20, 28$ and $40$), due to absence of this correlation, the
calculations yield sharply falling asymptotic density distribution
as is discussed below.

\vskip 0.15in \psfig{figure=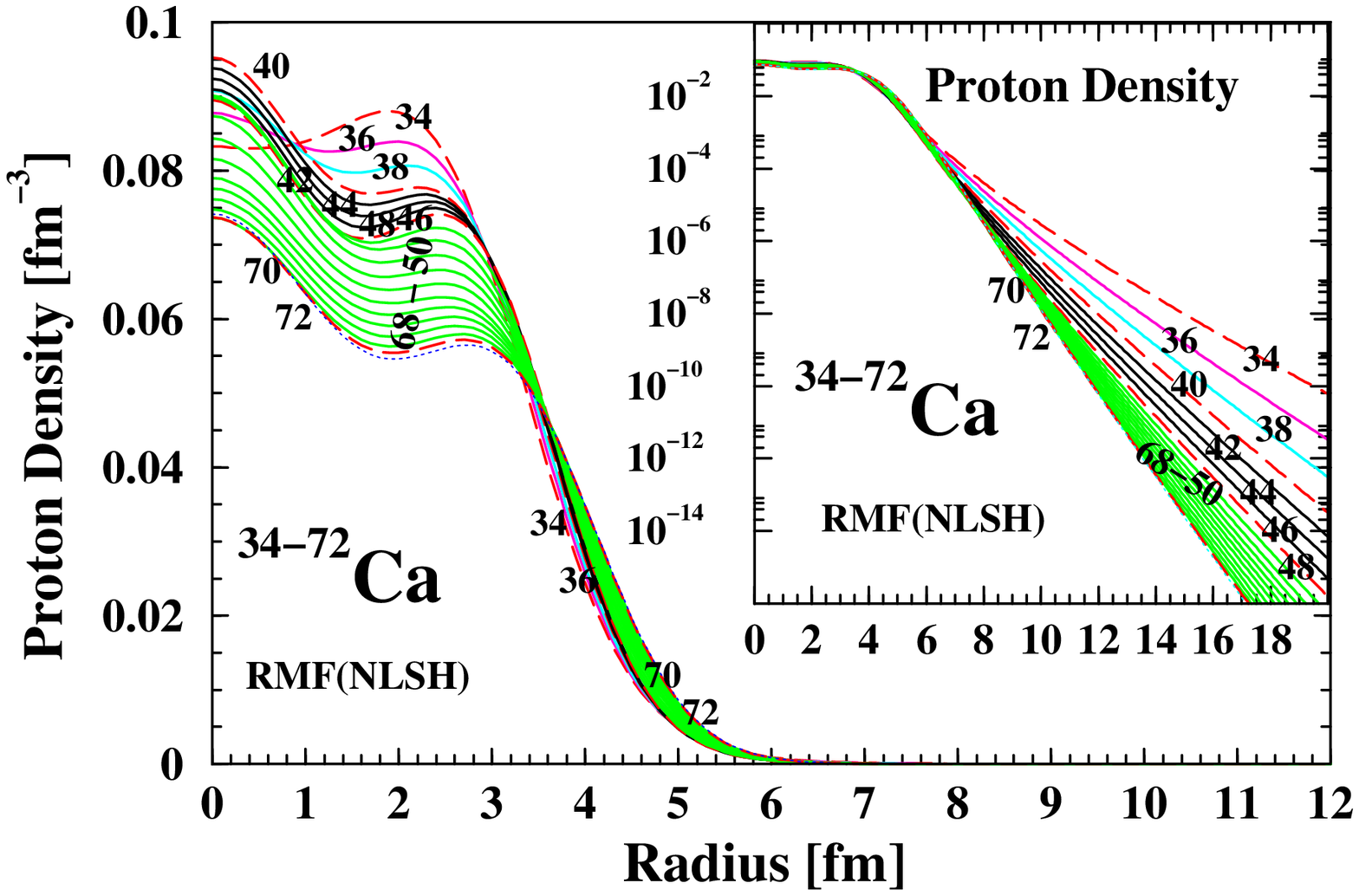,width=13cm}
\vskip 0.15in {\noindent \small {{\bf Fig. 8.} Results for the
proton density distributions for the $\rm{Ca}$ isotopes obtained
in the RMF+BCS calculations employing the NL-SH force. The numbers
on the density distribution lines indicate the mass number of the
$\rm{Ca}$ isotope. \mbox{}}} \vspace{0.5cm}

\subsubsection{Halo Formation in $\rm {Ca}$ Isotopes}

For the purpose of illustration, we have displayed in Fig. 9 the
detailed neutron density profile obtained with the TMA force for
the $\rm {Ca}$ isotopes. In our earlier discussions for the
$\rm{Ca}$ isotopes, it has been observed from the neutron
dependence properties that the neutron numbers N = 14, 20, 28 and
40 correspond to closed shells, and thus, represent the magic
numbers for these isotopes. This conclusion is found to be
consistent with the calculated densities shown in fig. 9 for the
isotopes corresponding to these neutron numbers. Evidently for N =
14, 20, 28 and 40 the neutron densities fall off rapidly and have
smaller tails as compared to the isotopes with other neutron
numbers. As remarked earlier this sharp fall in asymptotic density
values is due to the fact that for the closed shell isotopes there
are no contribution to the density from the quasi bound  states
having positive energy albeit close to zero energy near the
continuum threshold.

\vspace{0.5cm} \psfig{figure=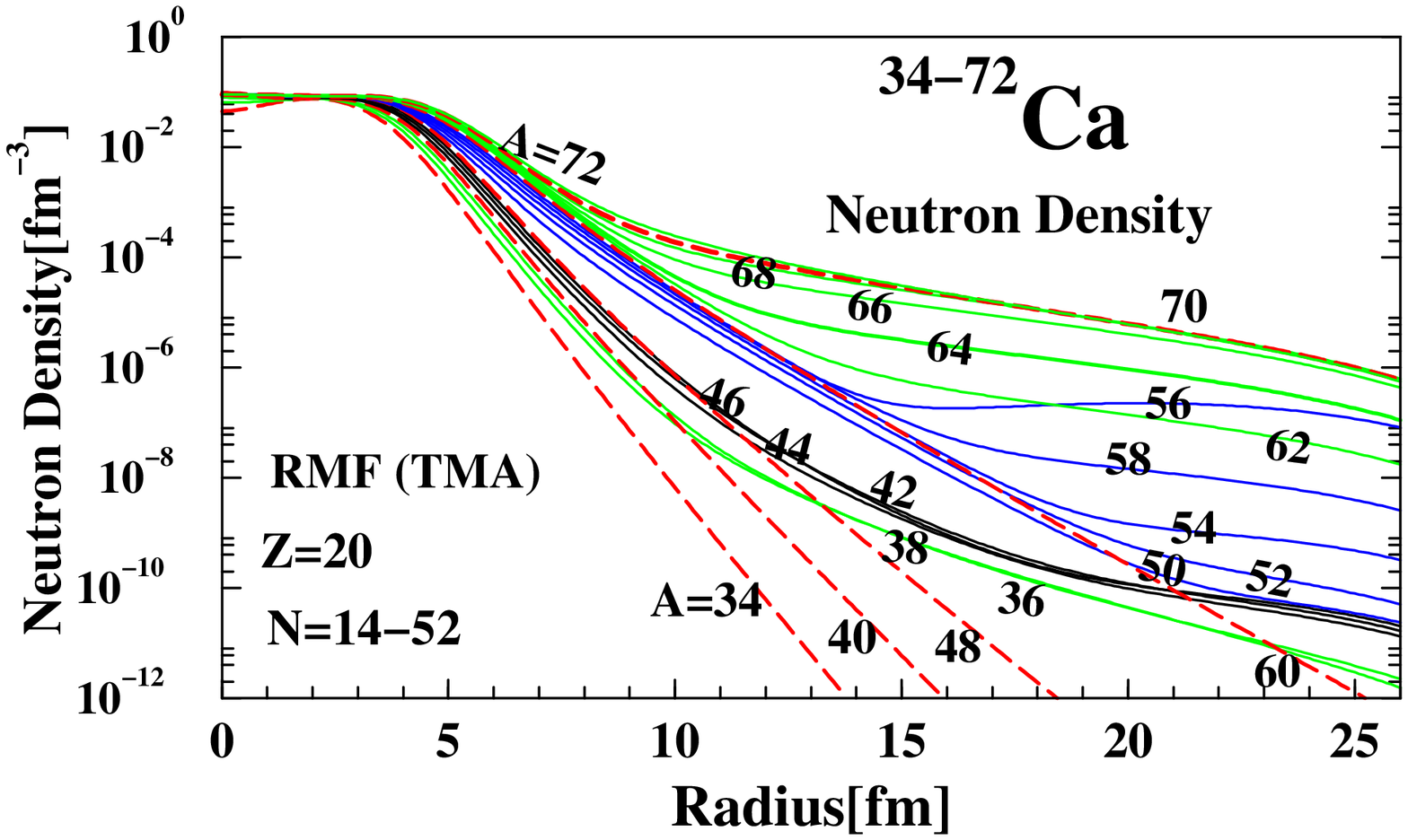,width=13cm}
\vskip 0.15in {\noindent \small {{\bf Fig. 9.} The neutron radial
density distribution for the $^{34-72}\rm{Ca}$ isotopes obtained
in the RMF+BCS calculations using the TMA force. The rapidly
falling density distributions correspond to isotopes with neutron
magic numbers at N = 14, 20, 28 and 40, and have been shown by the
dashed lines whereas for the other isotopes these are plotted as
solid lines.\mbox{} }} \vskip 0.15in

The density distribution for the $N = 50$ case (corresponding to
$^{70}\rm{Ca}$) is seen to be very different from those of the
isotopes with N = 14, 20, 28 and 40 indicating thereby that for
the neutron rich $\rm {Ca}$ isotopes the neutron number $N = 50$
does not correspond to a magic number. The neutron single particle
spectrum for the $^{70}\rm{Ca}$ isotope shows that apart from the
bound states and high lying resonant states (the lower $1g_{9/2}$
resonant state has already become bound for this isotope), the
positive energy neutron states $2d_{5/2}$, $2d_{3/2}$, $4s_{1/2}$,
$3p_{3/2}$, $3p_{1/2}$, $3d_{5/2}$, $3d_{3/2}$, all lying very
close to the neutron Fermi level ($\epsilon_f = 0.08$ MeV), are
also being occupied. The occupation probabilities for these
positive energy states are indeed very small. However, a sum of
all these accounts for about 0.6 nucleon occupying the positive
energy states. Out of this, the $2d_{5/2}$ state lying closest to
the Fermi level at $\epsilon = 0.47$ MeV has the maximum occupancy
weight which amounts to about 0.55 nucleon being in this energy
state. The contribution to the total density distribution for the
$^{70}\rm{Ca}$ isotope ($N=50$) from these positive energy states
consequently modifies the asymptotic density distribution. Thus in
contrast to the closed shell isotopes, here the density does not
fall off rapidly with increasing radial distances. This feature of
the density profile can be essentially attributed to loosely bound
states like the $2d_{5/2}$ in the $^{70}\rm{Ca}$ isotope. This
effect is seen to persist in all the isotopes irrespective of
being neutron rich or poor, exceptions being those with closed
shells wherein the pairing correlations are not able to populate
the positive energy states.

Next we consider a comparison of results obtained with the TMA and
NL-SH forces, and also of those obtained using the RCHB approach.
Results of calculations using the NL-SH force are found to be
almost similar to those obtained with the TMA Lagrangian already
shown in fig. 9. Also it is found that the RMF+BCS results for the
proton and neutron densities are almost similar to those obtained
in the RCHB approach. For the purpose of illustration we have
plotted in fig. 10 the results for the neutron densities obtained
in the RMF+BCS calculations (solid lines), and those obtained in
the RCHB calculations (dashed lines) again  using the NL-SH force
for the purpose of comparison. It is seen from the figure that the
results from both, the RMF+BCS and the RCHB, approaches for the
neutron density distributions are almost similar. In particular,
for the isotopes with neutron shell closure corresponding to $N=$
14, 20, 28 and 40 this similarity extends up to large radial
distances whereby the densities are already diminished to very
small values, rather less than $10^{-10}$ fm$^{-3}$. This has been
explicitly demonstrated in the inset of fig. 10 which shows the
results on a logarithmic scale for radial distances up to $r = 16$
fm. For the other isotopes, there are small deviations between the
RMF+BCS and the RCHB approaches beyond the radial distance $r = 8$
fm. However, beyond this distance the densities are already quite
reduced ranging between $10^{-4}$ fm$^{-3}$ to $10^{-8}$
fm$^{-3}$. These small deviations between the RMF+BCS and RCHB
results beyond $r = 8$ fm is caused due to small differences in
the single particle energies and wave functions, especially near
the Fermi surface, and also on the energy cut off being used in
the RCHB calculations. Nevertheless, similarities between the
RMF+BCS and the RCHB results are exceedingly encouraging.

\vspace{0.5cm}\psfig{figure=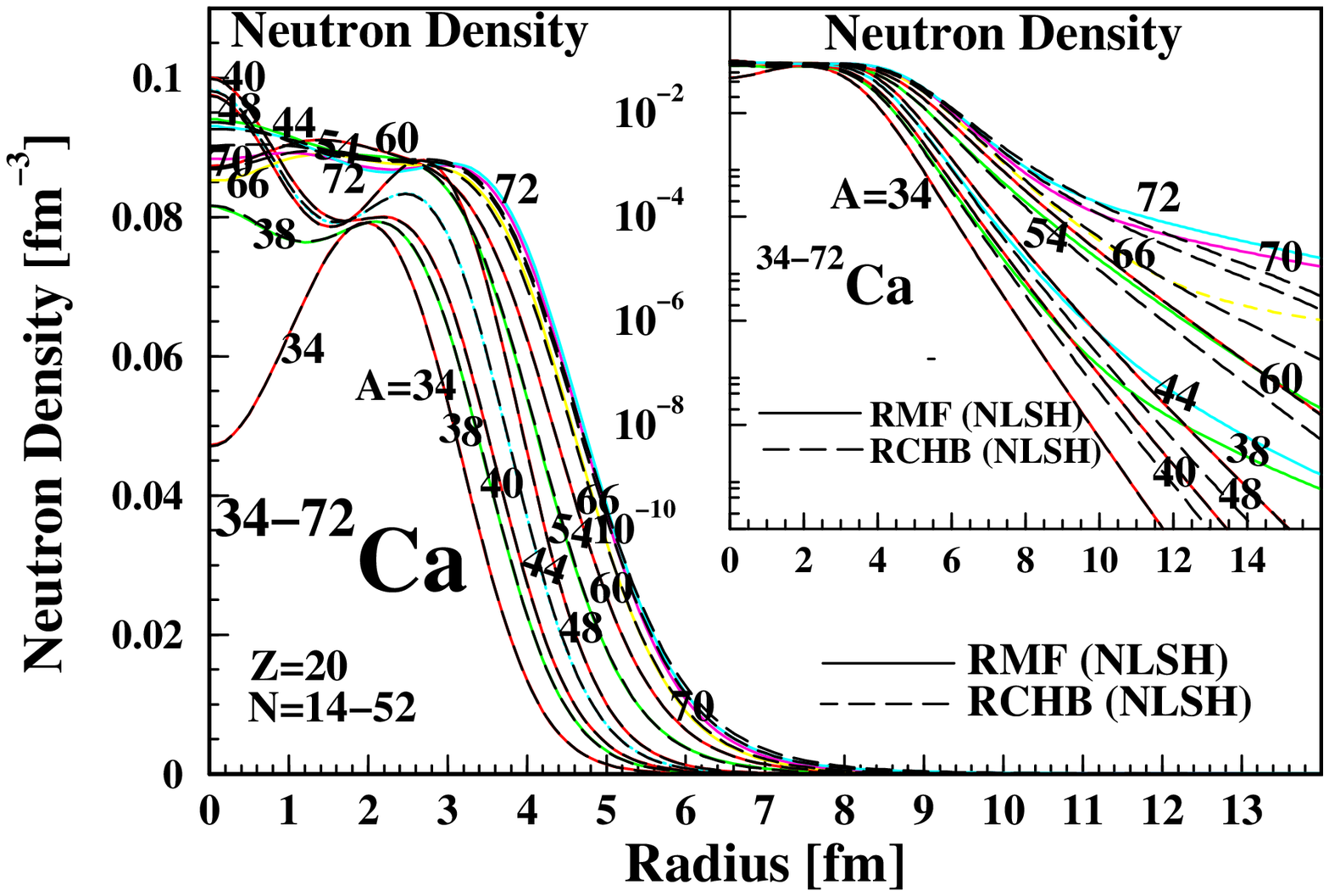,width=13cm}
\vskip 0.15in {\noindent \small {{\bf Fig. 10.} The solid line
show the neutron radial density distribution for the
$^{34-72}\rm{Ca}$ isotopes obtained in the RMF+BCS calculations
using the NL-SH force. Corresponding results of the RCHB
calculations with the NL-SH force are shown by the dashed lines
for the purpose of comparison. The inset shows the results on a
logarithmic scale up to rather large radial distances.\mbox{} }}
\vskip 0.15in

In fig. 10, it is interesting to note that the neutron density
distributions, out side the nuclear surface and at large
distances, for the neutron rich $\rm Ca$ isotopes with neutron
number $N \ge 42$ are larger by several orders of magnitude as
compared to the lighter isotopes. This behavior of the density
distribution for the neutron rich $\rm Ca$ isotopes is quite
different from the corresponding results, especially  for the
neutron rich isotopes of $\rm {Ni}$, $\rm {Sn}$ and $\rm {Pb}$
nuclei. In the latter cases, as more neutrons are added the tail
of the neutron density distributions for the neutron rich isotopes
tends to saturate.  The results described above can be easily
understood by inspecting the variation in the position of the
single particle states and that of the Fermi energy with
increasing neutron number depicted in fig. 11.

\vspace{0.5cm} \psfig{figure=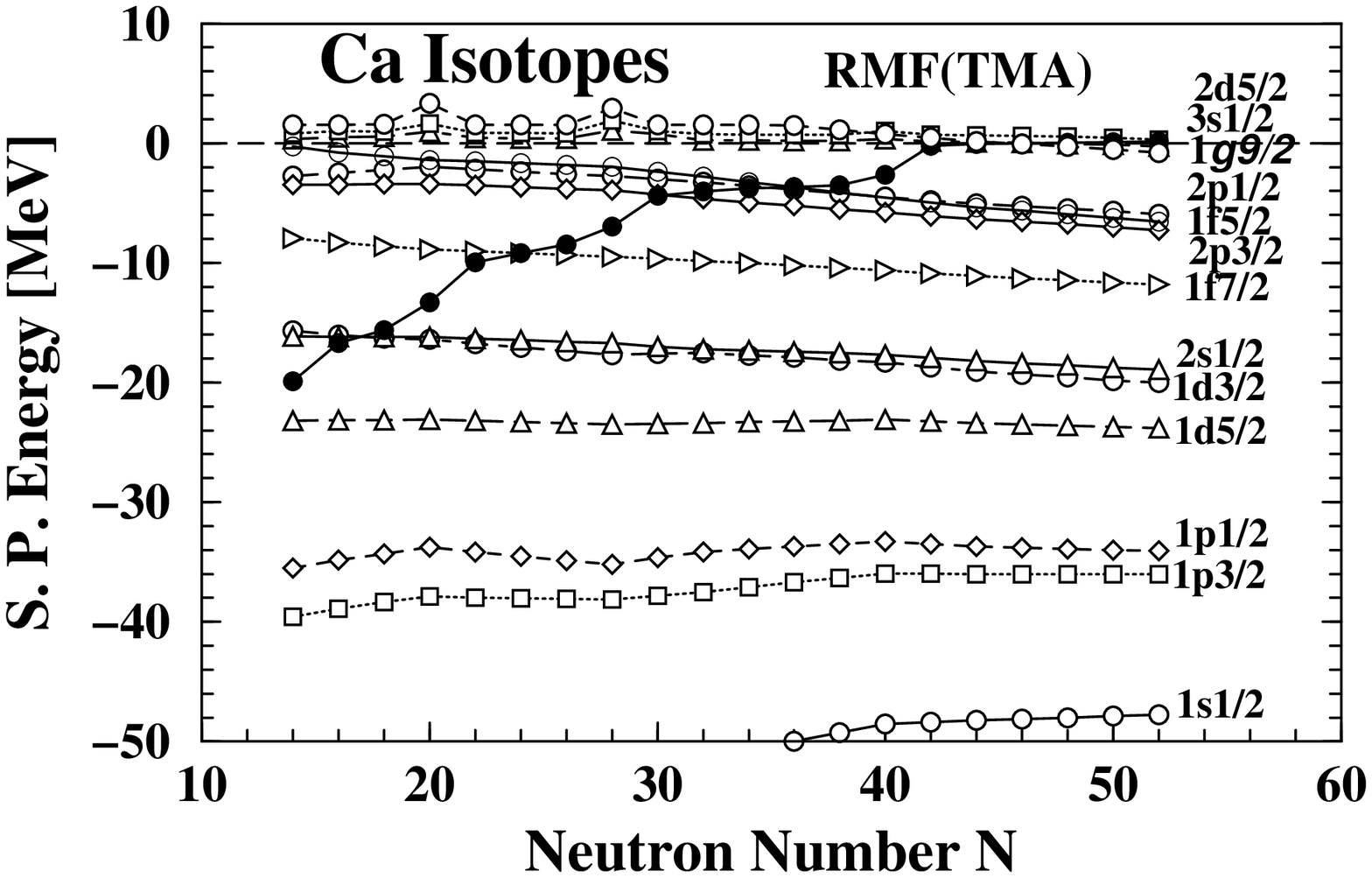,width=13cm}
\vskip 0.15in {\noindent \small {{\bf Fig. 11.} Variation of the
neutron single particle energies obtained with the TMA force for
the $\rm{Ca}$ isotopes with increasing Neutron number. The Fermi
level has been shown by filled circles connected by solid line to
guide the eyes.\mbox{} }} \vspace{0.5cm}

First of all, it is readily seen in fig.11 that the large energy
gaps between single particle levels $1d_{5/2}$ and $1d_{3/2}$, and
the levels $2p_{1/2}$ and $3s_{1/2}$ etc. are responsible for the
properties akin to shell or sub-shell closures in the
$^{34-72}\rm{Ca}$ isotopes for the neutron number N = 14 and 40
apart from the traditional magic nos. N = 20 and 28. However the N
= 50 shell closure is seen to disappear due to absence of gaps
between the $1g_{9/2}$ state and the states in the s-d shell.

For an understanding of the halo formation in the neutron rich $(N
\ge 42)$ $\rm{Ca}$ isotopes we next look  into the detailed
features of single particle spectrum and its variation as one
moves from the lighter isotope to heavier one in fig. 11. The
neutron Fermi energy which lies at $\epsilon_f = -19.90$ MeV in
the neutron deficient $^{34}\rm{Ca}$ nucleus moves to $\epsilon_f
= -0.21$ MeV in  the neutron rich $^{62}\rm{Ca}$, and to
$\epsilon_f = 0.08$ MeV (almost at the beginning of the single
particle continuum) in $^{70}\rm{Ca}$. The $1g_{9/2}$ state which
lies at higher energy in continuum for the lighter isotopes, (for
example at $\epsilon = 1.68$ MeV in $^{34}\rm{Ca}$) comes down
gradually to become slightly bound, ($\epsilon = -0.07$ MeV in
$^{66}\rm{Ca}$) for the neutron rich isotopes. Similarly, the
$3s_{1/2}$ state which lies in continuum for the lighter isotopes
(for example at $\epsilon = 0.70$ MeV in $^{34}\rm{Ca}$)also comes
down, though not so drastically, to become slightly bound
(($\epsilon = -0.05$ MeV in $^{68}\rm{Ca}$) for the neutron rich
isotopes. In the case of $^{60}\rm{Ca}$ with shell closure for
both protons and neutrons, the neutron single particle states are
filled in up to the $2p_{1/2}$ state, while the next high lying
states $3s_{1/2}$ and $1g_{9/2}$, separated by about $5$ MeV from
the $2p_{1/2}$ level are completely empty. Now on further addition
of 2 neutrons, it is observed that the $1g_{9/2}$ is filled in
first even though $3s_{1/2}$ state is slightly lower (by about
0.31 MeV) than the $1g_{9/2}$ state as has been shown in fig. 12.
Still another addition of 2 neutrons are found to fill in the
$1g_{9/2}$ state once again, though  now the $1g_{9/2}$ state is
higher to the $3s_{1/2}$ state merely by $0.08$ MeV. This
preference for the $1g_{9/2}$ state  results in the existence of
loosely bound highly neutron rich $\rm{Ca}$ isotopes. Indeed this
preference stems from the fact that in contrast to the $3s_{1/2}$
state, the positive energy $1g_{9/2}$ state being a resonant state
has its wave function entirely confined inside the potential well
akin to a bound state. This has been shown earlier in fig. 2 for
the nucleus $^{62}\rm{Ca}$. As is evident from fig. 2, the
$3s_{1/2}$ state in the continuum has its wave function spread
over mostly outside the potential well and thus the additional
neutrons do not occupy this state until it becomes a bound state.

\vspace{0.7cm} \psfig{figure=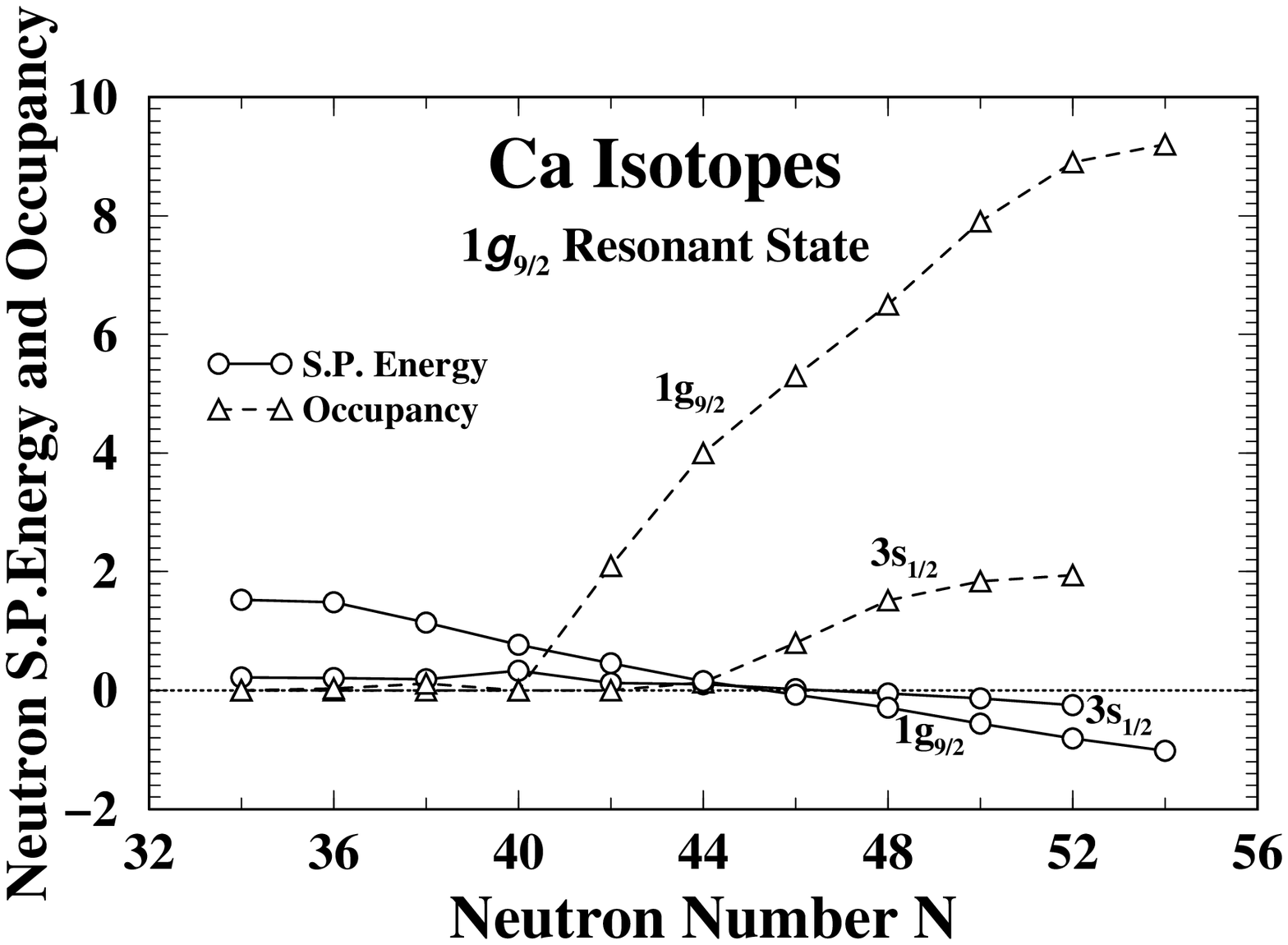,width=13cm}
\vskip 0.15in {\noindent \small {{\bf Fig. 12.} Variation in
energy shown by open circles, and occupancy (no. of neutrons
occupying the levels) depicted by open triangles for the neutron
$1g_{9/2}$ and $3s_{1/2}$ single particle states in the $\rm{Ca}$
isotopes.\mbox{} }} \vspace{0.5cm}

For the neutron number N = 48, it is found that both the
$1g_{9/2}$ and $3s_{1/2}$ states become bound and start to compete
together to get occupied on further addition of neutrons as can be
seen in fig. 12. On the other hand, the positive energy states
like $2d_{5/2}$, $2d_{3/2}$,  $3p_{3/2}$ and $3p_{1/2}$  lying
near the Fermi level at the continuum threshold are also being
occupied though with very small probability. Further it is
observed from the figure that both of these states are completely
filled in for the neutron number N = 52, and thus, the neutron
drip line is reached with a loosely bound $^{72}\rm{Ca}$ nucleus.
The next single particle state, $2d_{5/2}$, is higher in energy by
about 0.5 MeV in the continuum and further addition of neutrons
does not produce a bound system.

As mentioned above, the $1g_{9/2}$ state is mainly confined to the
potential region, and hence its contribution to the neutron radii
is similar to a bound state. In contrast, the $3s_{1/2}$ state
which has no centrifugal barrier, and thus, is spread over large
spatial extension contributes substantially to the neutron density
distribution at large distances. Similarly it turns out that the
$3p_{1/2}$ state though to a lower extent contributes to the
extended density distribution. Due to this reason, for $N
> 42$ as the $3s_{1/2}$ state starts being occupied the neutron
density distribution has large tails, and the neutron radii for
the neutron rich isotopes ($N  > 42$) grow abruptly as has been
shown in fig. 6. Thus the filling of the $3s_{1/2}$ single
particle state  and to a lesser extent that of $2d_{5/2}$,
$3p_{3/2}$ and $3p_{1/2}$ etc. with increasing neutron number in
the $^{64-72}\rm{Ca}$ isotopes causes the  neutron halos like
formation in these nuclei.

\subsection{$\rm{Ni}$ Isotopes}

Results for the $\rm{Ni}$ isotopes exhibit similar characteristics
for the pairing gap contribution due to the resonant states, and
those from other single particle states near the Fermi level.
However, the neutron rich $\rm{Ni}$ isotopes do not exhibit a
situation like the neutron rich $\rm{Ca}$ isotopes favoring for
halo formation. This is due to difference in the single particle
structure near the Fermi surface as will be discussed later. In
order to illustrate the physical situation of neutron rich
$\rm{Ni}$ isotopes, we have plotted in fig. 13 the RMF potential
(upper panel), and some of the single particle wave functions
(lower panel)for the nucleus $^{84}\rm{Ni}$.  The upper panel also
shows the spectrum for the neutron bound single particle states,
and a few of the positive energy states in the continuum. Some of
the most relevant neutron single particle states close to the
neutron Fermi level, which are found to be important for the
description of neutron rich $\rm{Ni}$ isotopes near the drip line
are: $4s_{1/2}$, $3p_{3/2}$, $3p_{1/2}$, $1g_{7/2}$, $2d_{5/2}$
and $2d_{3/2}$. Analogous to the $1g_{9/2}$ state for the case of
neutron rich $\rm{Ca}$ isotopes,  the neutron $1g_{7/2}$ state is
found to be the most important resonant state for the $\rm{Ni}$
isotopes. It has been shown in the figure along with the positive
energy $4s_{1/2}$, $3p_{3/2}$, $3p_{1/2}$ states mentioned above.
In addition, we have also depicted in fig. 13 the total mean field
potential for the resonant $1g_{7/2}$ state, obtained by adding
the centrifugal potential energy. Once again, many states high
lying in the continuum  have not been shown in the figure. Also in
fig. 13 (lower panel), we have displayed the radial wave functions
of some of the neutron single particle states close to the Fermi
surface, the neutron Fermi energy being $\lambda_n\, =\,-1.21$
MeV.  These include the bound $2d_{3/2}$ and the continuum
$3p_{3/2}$, $4s_{1/2}$ and $1g_{7/2}$ single particle states.

It is seen from fig. 13 that the resonant neutron $1g_{7/2}$ state
in the neutron rich $\rm{Ni}$ isotopes, similar to the resonant
state in the $\rm{Ca}$ isotopes, remains mainly confined to the
region of the potential well and the wave function exhibits
characteristics similar to that of a bound state. Also, it is
evident from the figure that, in contrast to the resonant state
$1g_{7/2}$, the wave function for a typical non-resonant state,
for example $3p_{3/2}$, is well spread over to large distances
outside of the potential region. Such states thus have a small
overlap with the bound states near the Fermi surface leading to
diminished value for the pairing gaps for these states. The next
important states are the loosely bound states. A representative
example of such a state is the $2d_{3/2}$ state at -0.55 MeV. From
fig. 13 it is seen that this state has a sizable part of the wave
function within the region of the potential well, and for such
states the pairing gap values, as has been shown in fig. 14, are
found to be comparable to the bound states making significant
contributions to the pairing energy.

\vskip 0.15in \psfig{figure=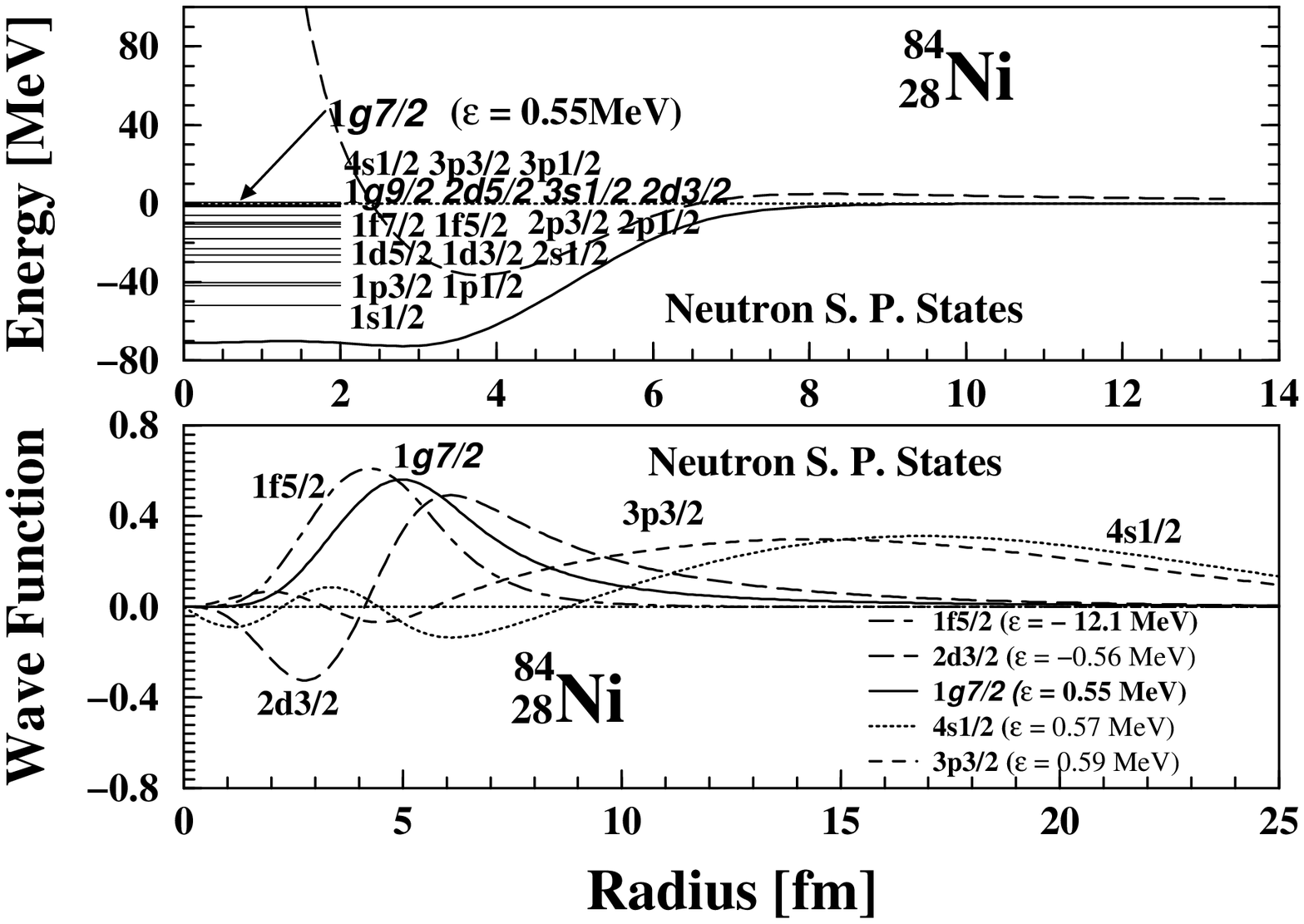,width=13
cm,height=10cm} \vskip 0.001in {\noindent \small {{\bf Fig.13.}
Upper panel: The RMF potential energy (sum of the scalar and
vector), for the nucleus $^{84}\rm{Ni}$ as a function of radius is
shown by the solid line. The long dashed line represents the sum
of RMF potential energy and the centrifugal barrier energy for the
neutron resonant state $1g_{7/2}$. The figure also shows the
energy spectrum of the bound neutron single particle states along
with the resonant $1g_{7/2}$
state at 0.55 MeV.\\
Lower panel: Radial wave functions of a few representative neutron
single particle states with energy close to the Fermi surface for
the nucleus $^{84}\rm{Ni}$. The solid line shows the resonant
$1g_{7/2}$ state at energy 0.55 MeV, while the bound $2d_{3/2}$
state at -0.56 MeV is shown by long dashed line. The $4s_{1/2}$
and $3p_{3/2}$ states with positive energies 0.57 MeV and 0.59 MeV
have been depicted by dotted and small dashed lines,
respectively\mbox{} }}

The next resonance above $1g_{7/2}$ is a $1h_{11/2}$ state.  This
resonance appears quite high in energy, at $E=4.6$ MeV, and has a
width of about 0.13 MeV.  We estimated the position and the width
of this resonance by calculating the phase shift corresponding to
the RMF single particle potential.  It is worth mentioning that in
a box calculation one could find always a $1h_{11/2}$ state at
approximately 4.5 MeV, but the order of its appearance depends on
the box radius. However, after inclusion of the $1h_{11/2}$ state,
it is found the results remain essentially unchanged. Indeed, its
contribution is much smaller than that of the $1g_{7/2}$ state due
to the large energy difference of the $1h_{11/2}$ state with
respect to the Fermi level.

\vskip 0.15in \psfig{figure=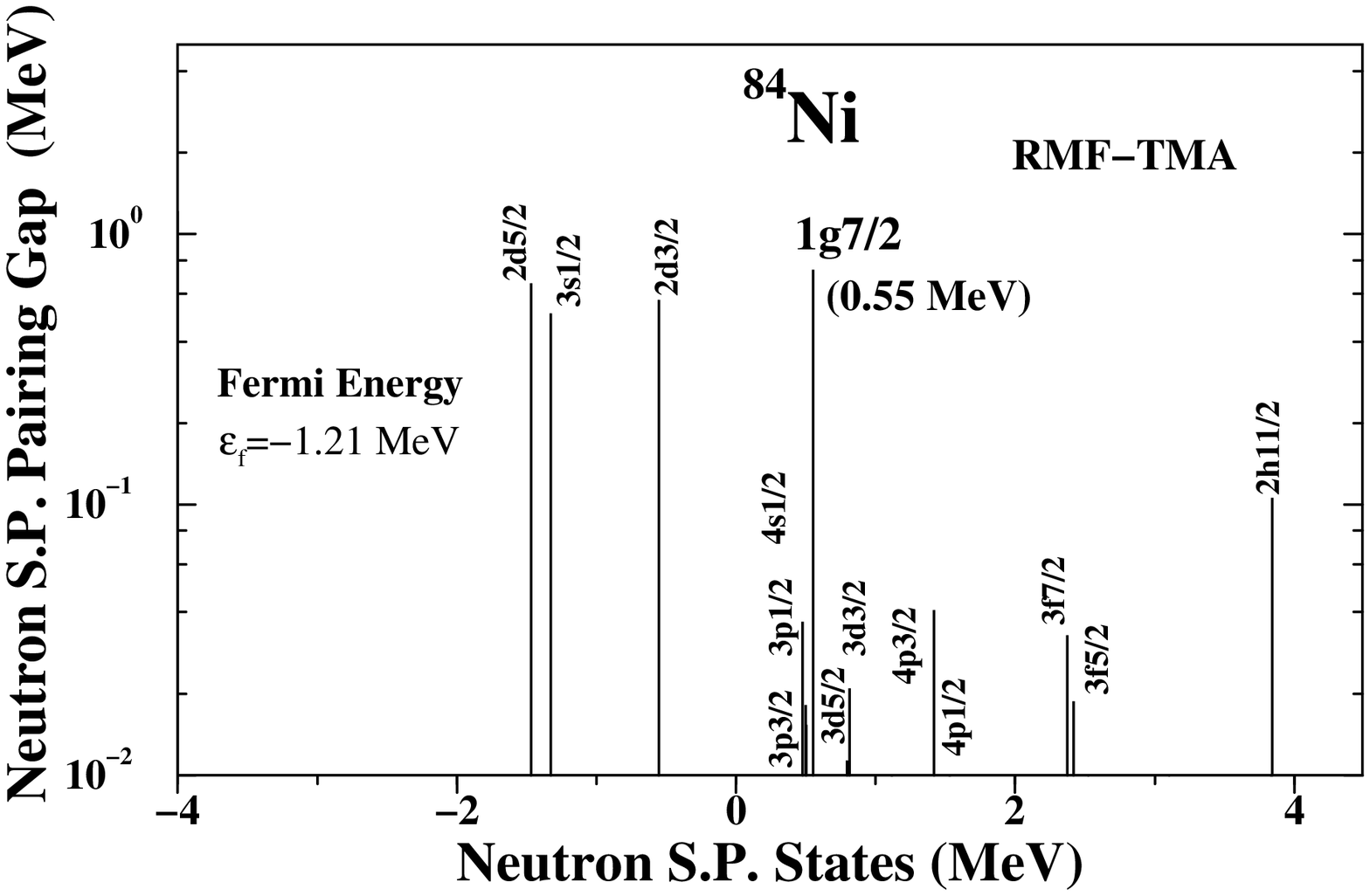,width=13cm} \vskip
0.15in {\noindent \small {{\bf Fig. 14.} Pairing gap energy
$\Delta_{j}$ of neutron single particle states with energy close
to the Fermi surface for the nucleus $^{84}\rm{Ni}$. The resonant
$1g_{7/2}$ state at energy 0.55 MeV has the gap energy of about
$1$ MeV which is close to that of bound states like $2d_{5/2}$
etc. \mbox{} }}

 The pairing gap values obtained in our RMF+BCS calculations as
shown in fig 14 turn out to be of the order of 1 MeV and are
consistent with the results of other mean field calculations and
those deduced from experimental data on the binding
energies.\cite{dobac96,lala,meng4,dobac95}. The RHB calculations
of Lalazissis et al.\cite{lala} have been carried out using the
finite range Gogny interaction D1S\cite{gogny} for the treatment
of pairing correlations. These calculations provide the average
values of canonical pairing gaps as function of canonical single
particle energies. These RHB averaged canonical gaps correspond to
the pairing gaps in  BCS theory, although the two are not
completely equivalent.  In the publication of Lalazissis et
al.\cite{lala} detailed results for the gaps are available only
for the $^{70}\rm{Ni}$ isotope. The averaged gaps decrease
approximately monotonically with increasing single particle
energies and have values ranging from about 2.2 MeV for the deep
hole states, to about 0.5 MeV for the high lying states in the
continuum. In comparison to these values our RMF+BCS calculations
yield a little smaller gaps ranging from about 1.2 MeV for the
well bound states, to about 0.8 MeV for the continuum states near
the Fermi surface. Similarly, the HFB calculations of Dobaczewski
et al.\cite{dobac96} have been carried out using three different
interactions for the pairing correlations. These are the finite
range density dependent Gogny interaction D1S\cite{gogny}, the
effective Skyrm interaction\cite{dobac,dobac95} SKP, and the
interaction SKP$^{\delta}$ which uses SKP parametrization in the
p-h channel and a $\delta$-interaction for the pairing component.
These extensive calculations carried out for the $\rm{Sn}$
isotopes show that the results for the HFB with D1S interaction
overestimate the gap values as compared to those obtained with
either the SKP or the SKP$^{\delta}$ interaction. Indeed our
calculated gaps for the $\rm{Sn}$ isotopes are found to be very
close to those of HFB using the SKP$^{\delta}$ interaction as will
be elaborated later while discussing the results of $\rm{Sn}$
isotopes. With this in view, our RMF+BCS results seem quite
consistent with RHB and HFB calculations using similar pairing
interaction.

\vskip 0.15in \psfig{figure=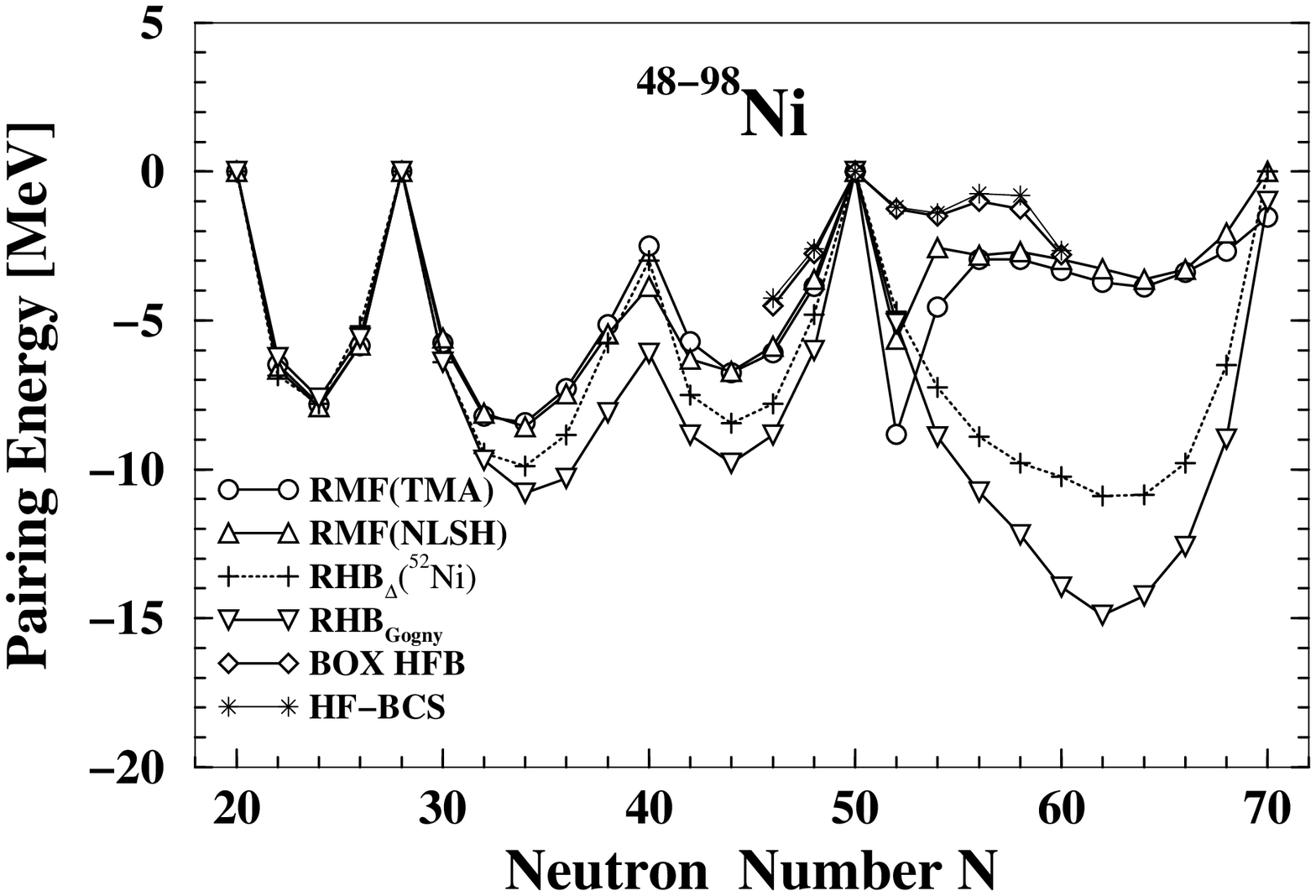,width=13 cm} \vskip
0.15in {\noindent \small {{\bf Fig. 15.}Present RMF results for
the pairing energy  for the $^{48-98}\rm{Ni}$ isotopes obtained
with the TMA (open squares) are compared with those obtained using
the NL-SH (open triangles) force parameters. The results of
different RHB calculations (see text for details) are also shown
for the purpose of comparison. \mbox{} }}

 In fig. 15 we compare our results for the contribution to the
pairing energy with those obtained using other mean-field
approaches for the $^{48-98}\rm{Ni}$. The choice of comparison
with the $^{48-98}\rm{Ni}$ isotopes is guided by the fact that for
these nuclei several theoretical results are available. As before,
our RMF+BCS calculations using the TMA and NL-SH forces yield
similar values for the pairing energy. These two results have been
shown in fig. 15 by open circles and triangles, respectively.
Further, we note that the pairing energy vanishes for the nuclei
with neutron numbers N = 20, 28, 50 and 70 indicating shell
closure for these neutron numbers. This indicates the appearance
of new shell corresponding to neutron number N = 70. The variation
of the neutron pairing energy with the neutron number N exhibits
features well known from the case of stable nuclei. Thus, as
expected, between the two neutron magic numbers the pairing energy
increases in magnitude from zero to reach a maximum for the half
filled shell nuclei. In the present case, the pairing energy for
the neutron half filled shell nuclei is found to vary between -10
MeV for  N= 24 and -15 MeV for the neutron rich case with N= 62 as
seen in fig. 15. Also, it is observed from this figure that N=40
represents a sub-shell with a small value of total pairing energy
which is approximately 3 MeV.

We have also plotted in fig. 15, for the purpose of comparison,
the results of RHB calculations carried out with a zero range
delta force, shown by plus symbols, and that with the finite range
Gogny interaction \cite{meng4} depicted by inverted triangles.
Also, are shown the results obtained in the continuum HF-BCS,
depicted  by stars, and the Box HFB calculations, depicted by
diamonds, of Grasso et al.\cite{grasso}. The RMF+BCS results for
the N dependence of the pairing energy exhibit similar trend as
those obtained from the RHB calculations, However, the values of
the pairing energy for the neutron rich nuclei are rather large in
the case of RHB calculations as is seen in fig. 15. The results
for the continuum HF-BCS and HFB calculations, though available
only for limited isotopes \cite{grasso}, are closer to our
calculations.

In fig. 16 we plot the results of two neutron separation energy
for the entire chain of $\rm{Ni}$ isotopes up to the neutron
drip-line. The figure also shows the HFB results of Terasaki {\it
et al}.\cite{terasaki} and that of the RHB calculations of
Ref.~19, along with the experimental data available\cite{audi} for
the $^{50-78}\rm{Ni}$ isotopes. Since the RHB calculations of
Ref.~20 are performed only for 24 $\leq\,\rm N\,\leq$ 50 and
because these results are similar to that of the RHB calculations
of Ref.~19, we have shown in Fig. 16 only the results of Ref.~19.
Further, we have chosen the calculations in Ref.~3 using the HFB
theory to be a representative one, therefore the results of other
HFB calculations\cite{dobac96,mizu} are not depicted in the
figure. It is gratifying to note that our RMF results are in
excellent agreement with the data. The strong variations in the
experimental separation energy near the neutron magic numbers N =
28, 50 are well accounted for by our calculations. The figure
shows that the overall trends of results obtained using different
theories are similar, though there are minor differences in
details. However, it is clearly seen that the RHB calculations
compare very well with our results for the two neutron separation
energy and the slope of the curve all the way up to A=100 in the
two cases are very similar. This good agreement with the RHB
results\cite{meng,lala} provides another strong support, in
addition to that of the HF+BCS calculations with continuum of
Ref.~6, to the validity of a RMF+BCS approach for the
investigation of the drip-line nuclei.

\vskip 0.15in \psfig{figure=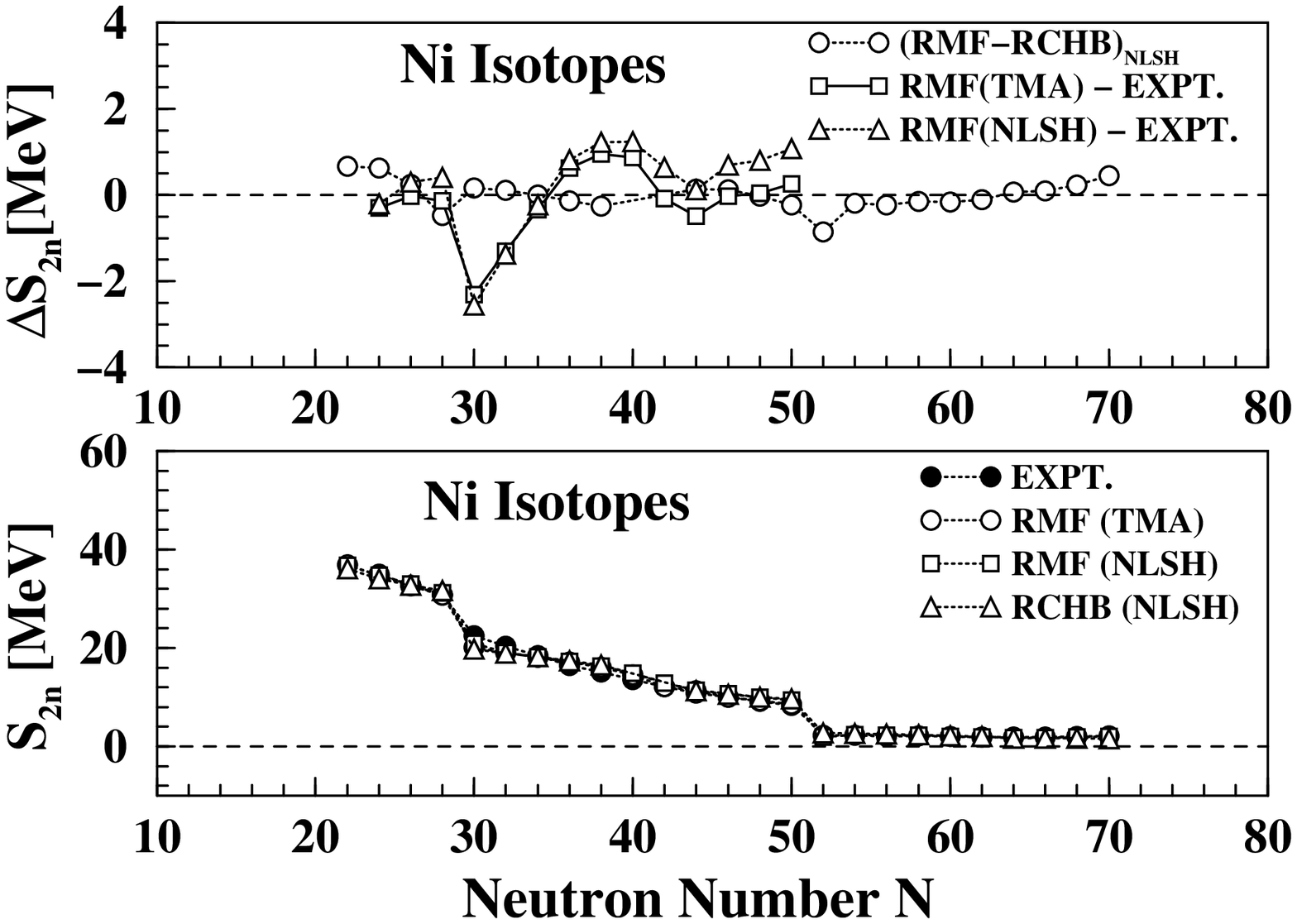,width=13 cm} \vskip
0.15in {\noindent \small {{\bf Fig. 16.} In the lower panel the
present RMF results for the two neutron separation energy  for the
$^{50-98}\rm{Ni}$ isotopes obtained with the TMA (open circles)
and the NL-SH (open squares) force parameters are compared with
the continuum relativistic Hartree-Bogoliubov (RCHB) calculations
of Ref.~26 carried out with the NL-SH force (open triangles). The
lower panel also depicts the available experimental
data\cite{audi} ($^{50-78}\rm{Ni})$ for the two neutron separation
energy (solid circles) for the purpose of comparison. The upper
panel depicts the difference in the RMF+BCS results  and the RCHB
results of Ref.~26  obtained for the NL-SH force. This part of the
figure also shows the difference of the calculated results with
respect to the available experimental data\cite{audi}.\mbox{} }}
\vspace{0.1cm}

In both relativistic calculations, i.e. RMF+BCS and RCHB, the
two-neutron drip line is predicted at A=98 (N=70). This is quite
different from the result of non-relativistic HFB calculations of
Ref.\cite{terasaki}, which predicts the two-neutron drip line
around A=90 (N=62). This difference between the two predictions
can be traced back to the spin orbit splitting of the neutron
$1g_{9/2}$ and $1g_{7/2}$ states, which is smaller in the
relativistic calculations. Thus in RMF+BCS the positive energy
state corresponding to the $1g_{7/2}$ resonance is much closer to
the continuum threshold than in the HFB calculations and is
becoming a bound states beyond N=58 as has been shown in fig. 17.
Consequently, in variance with HFB, in RMF+BCS calculations one
can put 8 neutrons more on the bound state $1g_{7/2}$ before we
reach the two neutron drip line.

\vspace{0.5cm} \psfig{figure=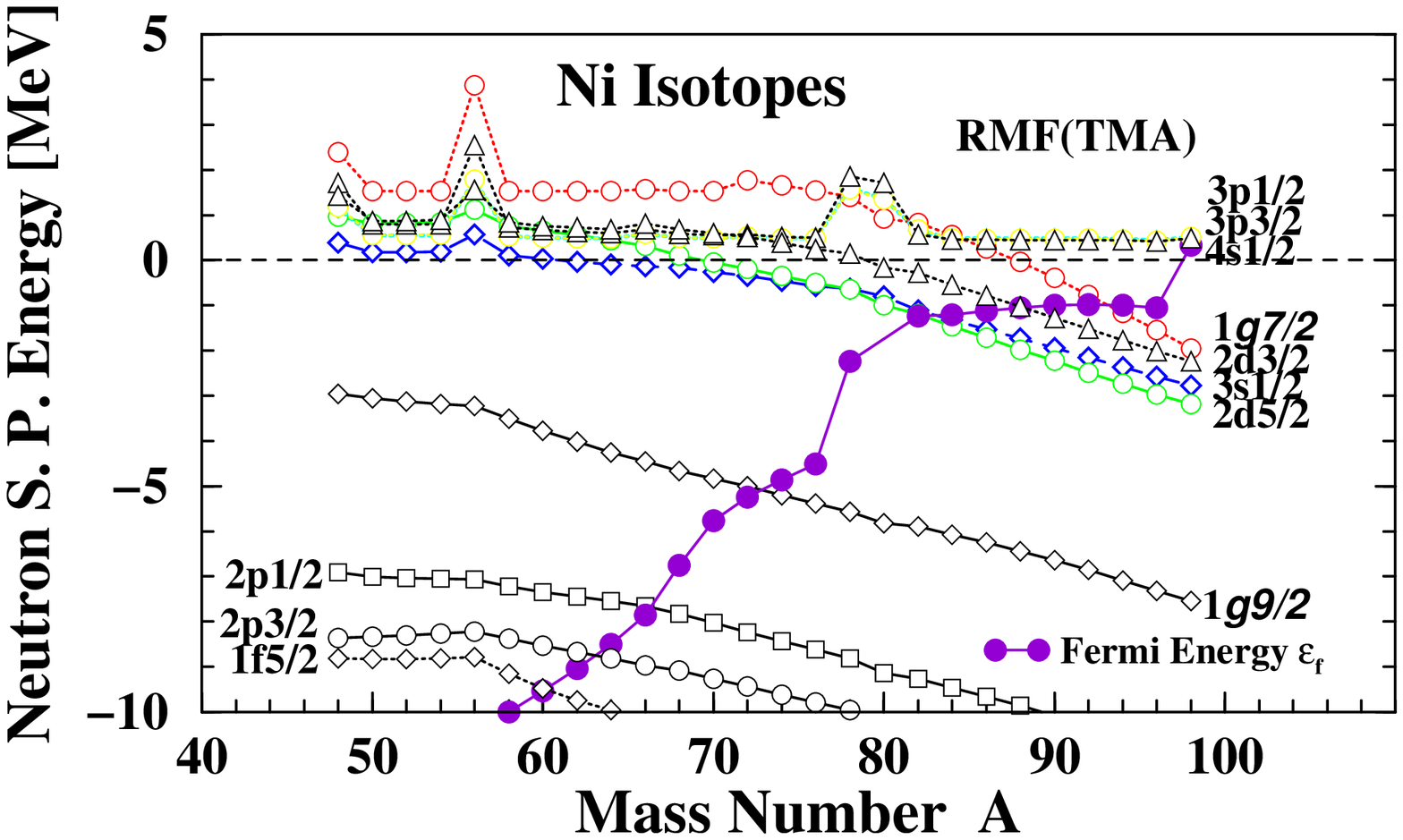,height=10
cm,width=13 cm} \vskip 0.15in {\noindent \small {{\bf Fig. 17.}
Variation of the neutron single particle energies for the
$\rm{Ni}$ isotopes with increasing mass number A. The Fermi level
has been indicated by solid circles.\mbox{} }} \vspace{0.1cm}

A detailed study of the proton, neutron and matter rms radii has
also been carried out. Our RMF+BCS results for the proton and
neutron rms radii and their comparison with the RCHB calculations
have been displayed in fig. 18. The results obtained with the TMA
and NL-SH force parameters are clearly seen to be very similar to
each other. The RMF+BCS results obtained by using the NL-SH force
and those of the RCHB approach using the same force are also quite
close to each other as can be seen in the figure.

\vskip 0.15in \psfig{figure=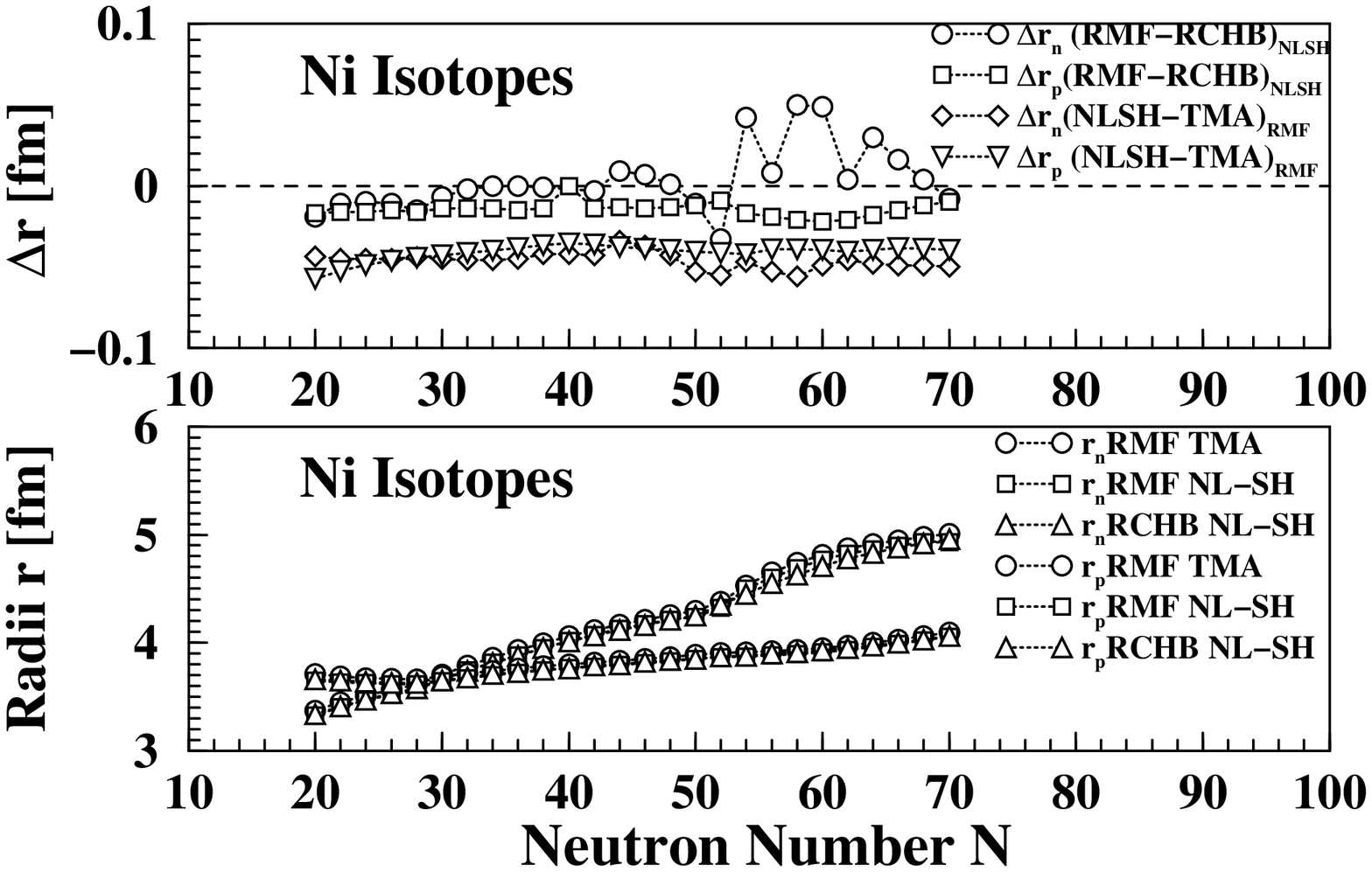,height=9
cm,width=13 cm} \vskip 0.15in {\noindent \small {{\bf Fig. 18.}
The lower panel presents the RMF results for the proton and
neutron rms radii ($r_p$ and $r_n$) for the $^{48-98}\rm{Ni}$
isotopes obtained with the TMA (open circles) and the NL-SH (open
squares) force parameters. These are compared with the continuum
relativistic Hartree-Bogoliubov (RCHB) calculations of Ref.~26
carried out with the NL-SH force (open triangles). The upper panel
displays the difference between the our RMF, and the RCHB results
of Ref.~26 for the proton and the neutron rms radii obtained for
the NL-SH force. This panel also shows the difference between the
calculated results for the TMA and NL-SH forces within the RMF+BCS
approach.\mbox{} }}

Further, the RMF+BCS results for the rms proton radii and the
neutron radii are found to be in close agreement with the
available experimental data.  The experimental data on the neutron
and proton rms radii are available only for a few of the $\rm{Ni}$
isotopes. Our RMF+BCS results obtained with the TMA force have
been plotted in fig. 19 with the the data on the proton rms radii
of $^{58-64}\rm{Ni}$, and that for the neutron rms radii of
$^{58,64}\rm{Ni}$ explicitly. It is gratifying to note that the
theoretical results are in good agreement with the measured radii.

\vskip 0.15in \psfig{figure=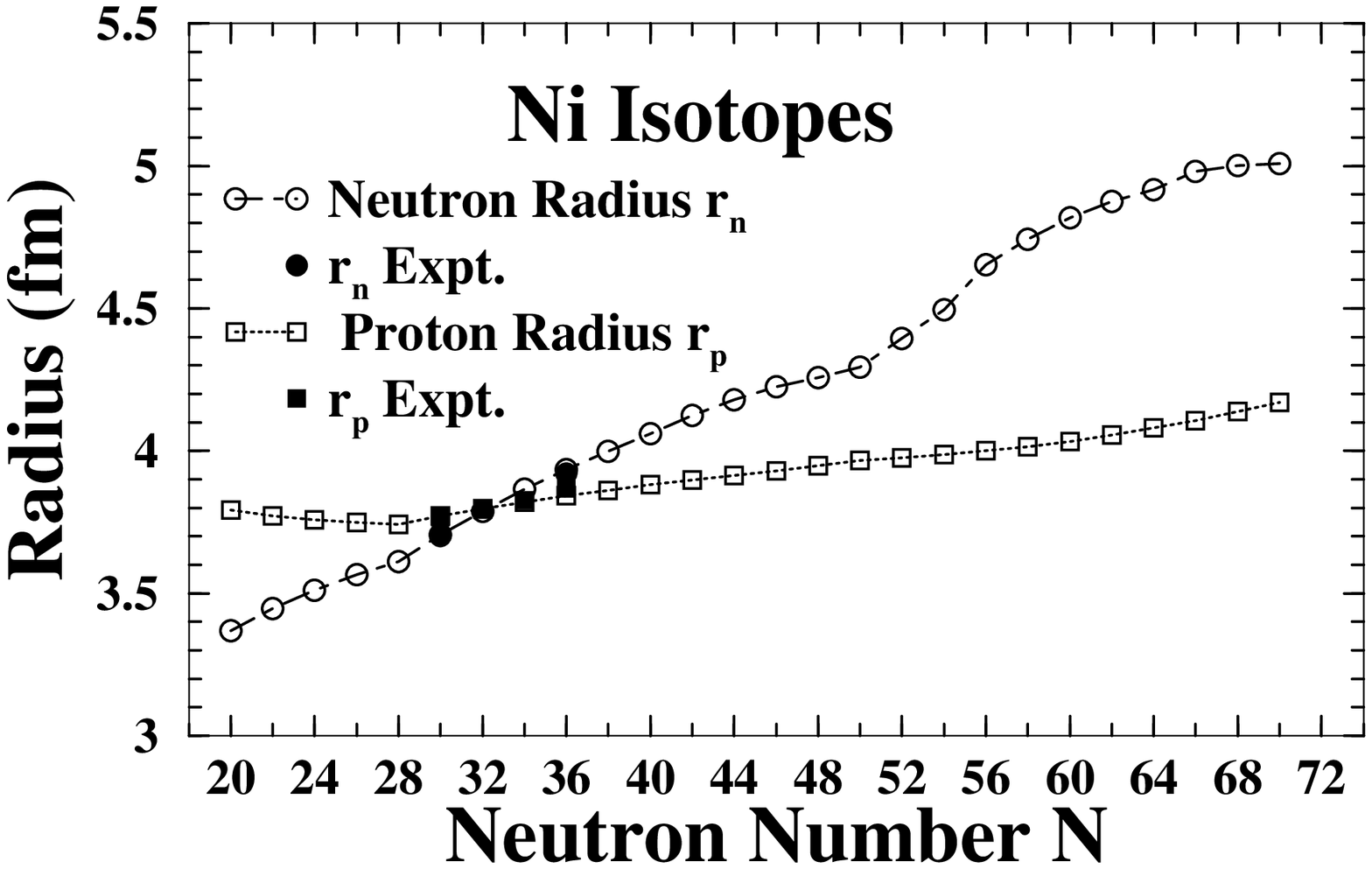,width=13 cm} \vskip
0.15in {\noindent \small {{\bf Fig. 19.} The  RMF results for the
neutron, $r_n$ (open circles) and proton, $r_p$ (open squares) rms
radii  for the $^{48-98}\rm{Ni}$ isotopes obtained with the TMA
force parameters. These are compared with the available
experimental data shown by solid circles and solid squares for the
neutron and proton radii, respectively. \mbox{} }}

The neutron skin formation is easily seen in the case of neutron
rich $\rm{Ni}$ isotopes showing large spatial extension of the
neutron density distributions. As a typical example, for
$^{84}\rm{Ni}$ this characteristic feature is seen in Fig. 20,
where we plot the density distribution for protons by hatched
lines and that for the neutrons by solid line.

\vskip 0.15in \psfig{figure=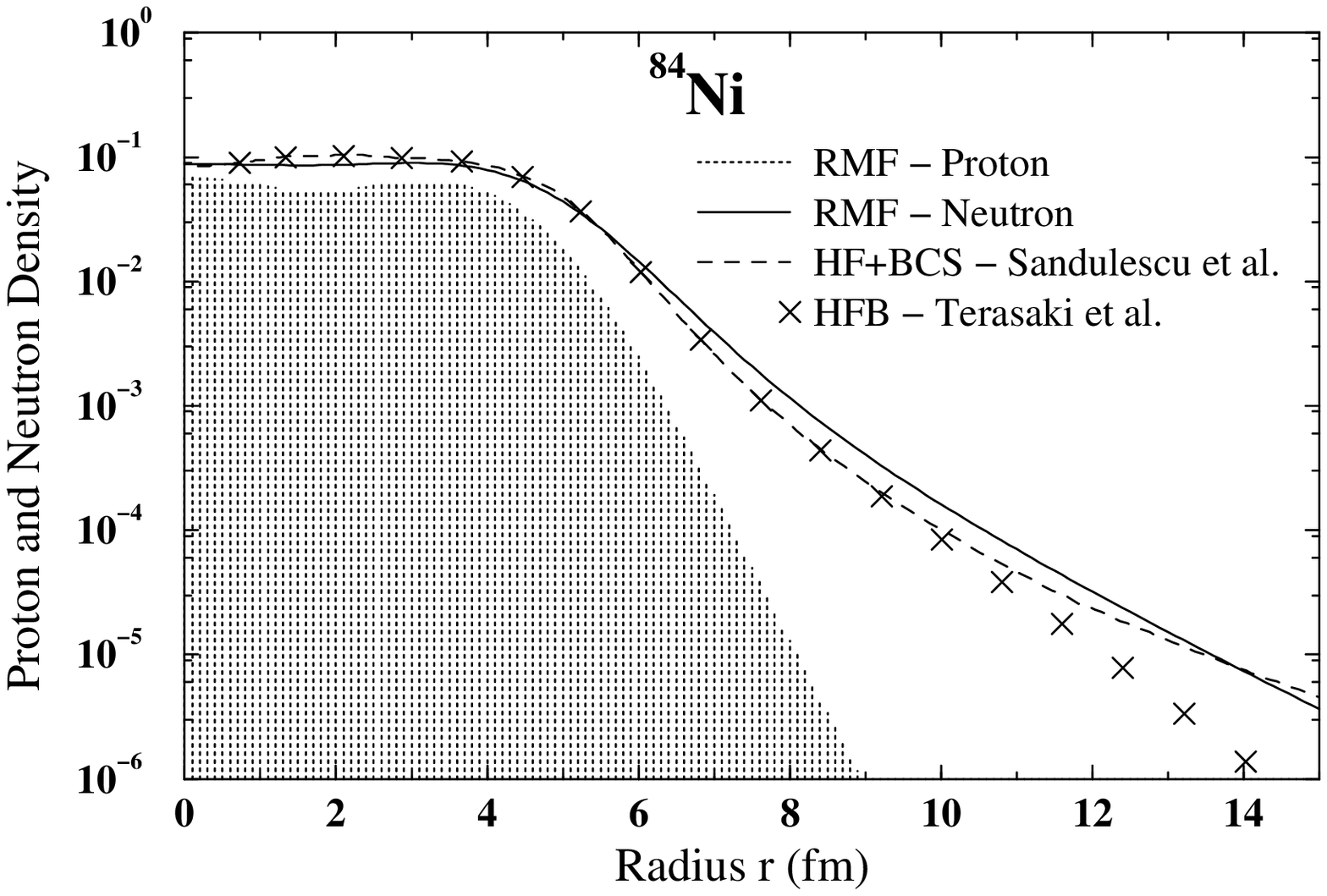,width=13cm}
\vskip 0.15in {\noindent \small {{\bf Fig. 20.} A comparison of
the neutron radial density distribution for $^{84}\rm Ni$ obtained
in different models. The results of the present calculations shown
by solid line show good agreement especially with those obtained
within the resonant continuum HF+BCS approach of Ref.~6 (dashed
line). The crosses correspond to the HFB
results\protect\cite{terasaki}. The radial density distribution
for the protons has been shown by the hatched area. \mbox{} }}
\vspace{0.5cm}

The tail of the neutron density extends far beyond the mean field
potential and has appreciable value up to about 14 fm. In order to
compare our calculations with other theoretical approaches we show
in fig. 20 the results of neutron density obtained using
HF+BCS\cite{sand2} and HFB\cite{terasaki} approximations. The tail
of the density given by our RMF+BCS calculations is much closer to
the resonant continuum HF+BCS results of Ref. 6. The main
contribution to the tail of the density comes from the loosely
bound states $3s_{1/2}$ and $2d_{3/2}$. The fact that in the HFB
calculations the tail is smaller can be due to the contribution of
the pairs scattered in states which are not time reversed
partners.\cite{grasso} This contribution is not included in a BCS
approach.

The neutron density distributions for the neutron rich $\rm{Ni}$
isotopes are distinct from those for the neutron rich $\rm {Ca}$
isotopes. This is clearly seen from fig. 21 which shows the
neutron densities for some selected $\rm{Ni}$ isotopes. The inset
in the figure displays the distribution on a logarithmic scale up
to radial distance $r=16$ fm. A comparison of this neutron density
distribution with that for the $\rm{Ca}$ isotopes displayed
earlier in fig. 9  shows  that in the case of neutron rich
$\rm{Ca}$ isotopes we have densities with wide spatial extensions
up to large distances, whereas in the neutron rich $\rm{Ni}$
isotopes it saturates. This is due to the fact that in the case of
$\rm{Ni}$ isotopes the $3s_{1/2}$ state is a bound state lying
below the $1g_{7/2}$ state which gradually itself becomes bound
while reaching towards the neutron drip line (see fig. 17). In
contrast the $3s_{1/2}$ state in  the $\rm{Ca}$ isotopes remains
unbound until just before the neutron drip line is reached. A
partial filling in of this unbound $3s_{1/2}$ state in the neutron
rich  $\rm{Ca}$ isotopes causes the formation of halos as has been
discussed earlier. The single particle structures in the case of
neutron rich $\rm{Sn}$ and $\rm{Pb}$ isotopes also exhibit
situation similar to that in the $\rm{Ni}$ isotopes and,
therefore, in these nuclei a large growth in neutron radii or halo
formation is not expected to occur as one moves towards the
neutron drip line.

\vspace{0.5cm}\psfig{figure=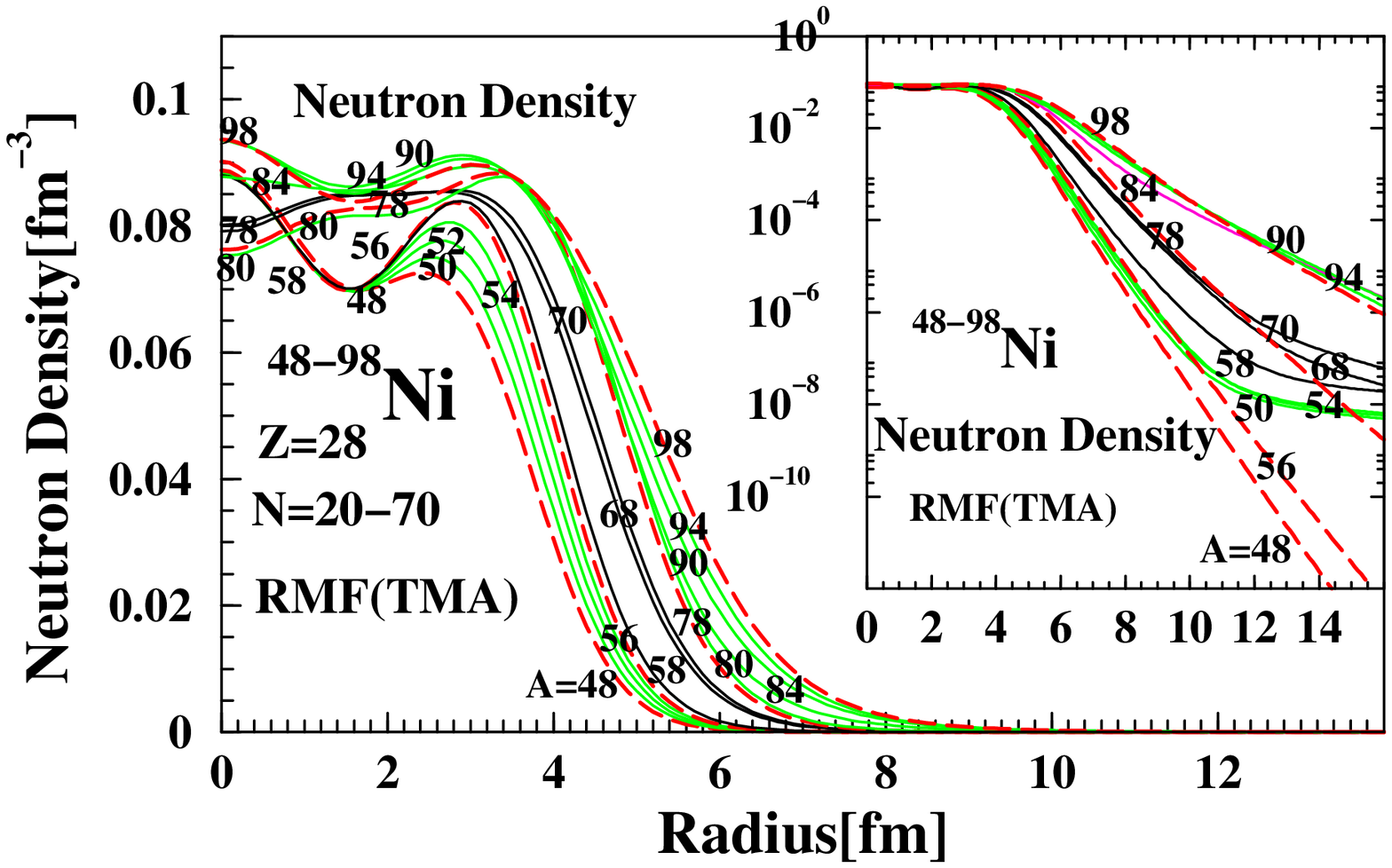,width=13cm}
\vskip 0.15in {\noindent \small {{\bf Fig. 21.} Neutron radial
density distribution for the $^{48-98}\rm{Ni}$ isotopes obtained
in the RMF+BCS calculations using the TMA force. The dashed lines
have been used to distinguish the closed neutron shell isotopes
with others. The inset shows the results on a logarithmic scale up
to rather large radial distances.\mbox{} }} \vskip 0.5cm

\subsection{Isotopes of O, Zr, Sn and Pb Nuclei}
Results for the isotopes of proton magic nuclei $\rm{Sn}$ and
$\rm{Pb}$ are similar to those of  proton magic $\rm{Ni}$ nuclei
discussed above. The $\rm{Zr}$ nuclei are not proton magic and,
therefore, there is always a contribution to the pairing energy
from the protons. Also, it is found that the neutron rich
$\rm{Zr}$ isotopes near the drip line have very small two neutron
separation energy and constitute an example of very loosely bound
system similar to the neutron rich $\rm{Ca}$ isotopes. In contrast
to these nuclei, the isotopes of $\rm{O}$ show slightly different
behavior as will be elaborated later. With the above mentioned
similarities in view, in the following we discuss briefly  the
results of $\rm{O, Zr, Sn}$ and $\rm{Pb}$ nuclei taken together in
order to save space. Also, as stated earlier we restrict our
treatment to spherical shapes for simplicity, even though it is
likely that some of the isotopes of $\rm O$ and that of $\rm {Zr}$
could be actually deformed.

In fig. 22 we have displayed for some of the neutron rich
representative isotopes of $\rm{O, Zr, Sn}$ and $\rm{Pb}$ nuclei
the pairing gap energies of single particle states located close
to the Fermi level. For our purpose we have chosen $^{20}\rm{O}$,
$^{138}\rm{Zr}$, $^{150}\rm{Sn}$ and $^{250}\rm{Pb}$ to elucidate
the general results for these nuclei. It is seen from fig. 22 that
the pairing gaps for the single particle neutron states  in the
$^{20}\rm{O}$ isotope near the Fermi level range between $0.9$ MeV
to $1.8$ MeV, where the Fermi energy  lies at $-5.22$ MeV. In
contrast to $\rm{Ca}$ and $\rm{Ni}$ and other nuclei discussed
here, it is seen that only the lowest neutron single particle
states are occupied according to the number of neutrons, and the
pairing interaction does not couple the bound states near the
Fermi level with the positive energy states . Thus in the case of
neutron rich $\rm{O}$ isotopes the positive energy states are not
populated. It is found from our calculations that the isotopes
$^{14}\rm{O}$, $^{16}\rm{O}$, $^{22}\rm{O}$, $^{24}\rm{O}$ and
$^{28}\rm{O}$ are doubly magic. And as expected the total pairing
energy for these isotopes is zero as can be seen from fig. 23.

In the case of neutron rich  $\rm {zr}$ isotopes, the single
particle state $1h_{9/2}$ is found to be a resonant state. This
state  has large pairing gap energy akin to that of bound sates.
As a representative case of the neutron rich $\rm {zr}$ isotopes,
we have plotted in fig. 22 the pairing gaps for the nucleus
$^{138}\rm {zr}$. In this plot only the neutron states lying near
the Fermi surface have been considered. The low lying resonant
state at $\epsilon$ = 0.6 MeV shown in the plot is found to play
an important role. The other important neutron single particle
states near the Fermi level ($\epsilon_f$ = -0.09 MeV) shown in
the plot are $2f_{7/2}$, $3p_{3/2}$, $3p_{1/2}$, $2f_{5/2}$ and
$4s_{1/2}$. For all these states the pairing gap energy $\Delta_j$
varies between about $0.5$ MeV and $1$ MeV. These gap energies are
consistent with other mean field calculations and the experiments.
It is further found that the high lying state $2i_{13/2}$ at
$\epsilon$ = 4.13 MeV is also a resonant state, though its
contribution to the pairing energy is insignificant as it lies far
above the Fermi level.

Similarly the important resonant state in the case of neutron rich
$\rm {Sn}$ nuclei is found to be the neutron $1i_{13/2}$ state. As
a representative example of $\rm {Sn}$ isotopes we have plotted in
fig. 22 the pairing gaps for the neutron rich $^{150}\rm{Sn}$
isotope. For this nucleus the $1i_{13/2}$ state lies at $\epsilon$
= -1.59 MeV, whereas  the Fermi energy is seen to be at
$\epsilon_f$ = 0.35 MeV. Some of the important neutron single
particle states near the Fermi surface for the neutron rich
nucleus $^{150}\rm{Sn}$, include the positive energy states
$4s_{1/2}$ at $\epsilon$ = 0.59 MeV, the resonant state
$1i_{13/2}$ at $\epsilon$ = 0.35 MeV, and the negative energy
bound states $2f_{5/2}$ at $\epsilon$ = -0.90 MeV. As is observed
from Fig. 22, the pairing gap energy of the resonant $1i_{13/2}$
state has a value $\Delta_{13/2}\approx 1.5$ MeV, which is close
to that of bound states like $1h_{9/2}$, $1h_{11/2}$ and
$2d_{3/2}$ etc. The other state in the continuum which has
somewhat appreciable gap energy is the $2g_{9/2}$ state at around
2 MeV. However, this state being higher in energy as compared to
the Fermi energy, plays only a minor role in the contribution to
the total pairing energy of the nucleus.

The gap energies for the single particle states shown in fig. 22
are found to be consistent with the results obtained from the HFB
and RHB calculations\cite{dobac96,lala,meng4}. To elaborate, the
HFB+D1S results of Lalazissis et al.\cite{dobac96} for the pairing
gaps $\Delta_{nlj}$ in the nucleus $^{150}\rm{Sn}$ for the deep
bound states are about $2$ MeV, while those for states near the
Fermi surface are about $1.75$ MeV, and those for the high lying
continuum states it decreases from about 1.5 MeV to about 0.5 MeV.
In contrast, the HFB+SKP results yield gaps ranging from very
small values of about $0.5$ MeV for deep hole states, to about 1.5
MeV for the states near the Fermi level, to about 0.75 MeV for the
high lying particle states in the continuum. Further, the
HFB+SKP$^{\delta}$ calculations provide results wherein the gaps
for the states throughout have almost a constant value of about
1.2 MeV with a small variation of about $\pm 0.2$ MeV. A
comparison of these results with our RMF+BCS calculations shows
the closest agreement with the HFB+SKP$^{\delta}$ calculations. As
has been pointed out by Dobaczewski et al.\cite{dobac96}, the D1S
interaction somewhat overestimates the gaps due to large magnitude
of the strength of D1S interaction.  Similarly, the detailed
results published by Dobaczewski et al.\cite{dobac96} for the
nucleus $^{120}\rm{Sn}$ are found to be in close agreement for the
various single particle states. Indeed, the results for the
average neutron gaps quoted in the calculations of Dobaczewski et
al.\cite{dobac96} using the SKP$^{\delta}$ interaction  for the
entire chain of $\rm{Sn}$ nuclei up to neutron number N= 126,
which represents the neutron drip line, are found to be consistent
with our RMF+BCS calculations.

\vskip 0.15in \psfig{figure=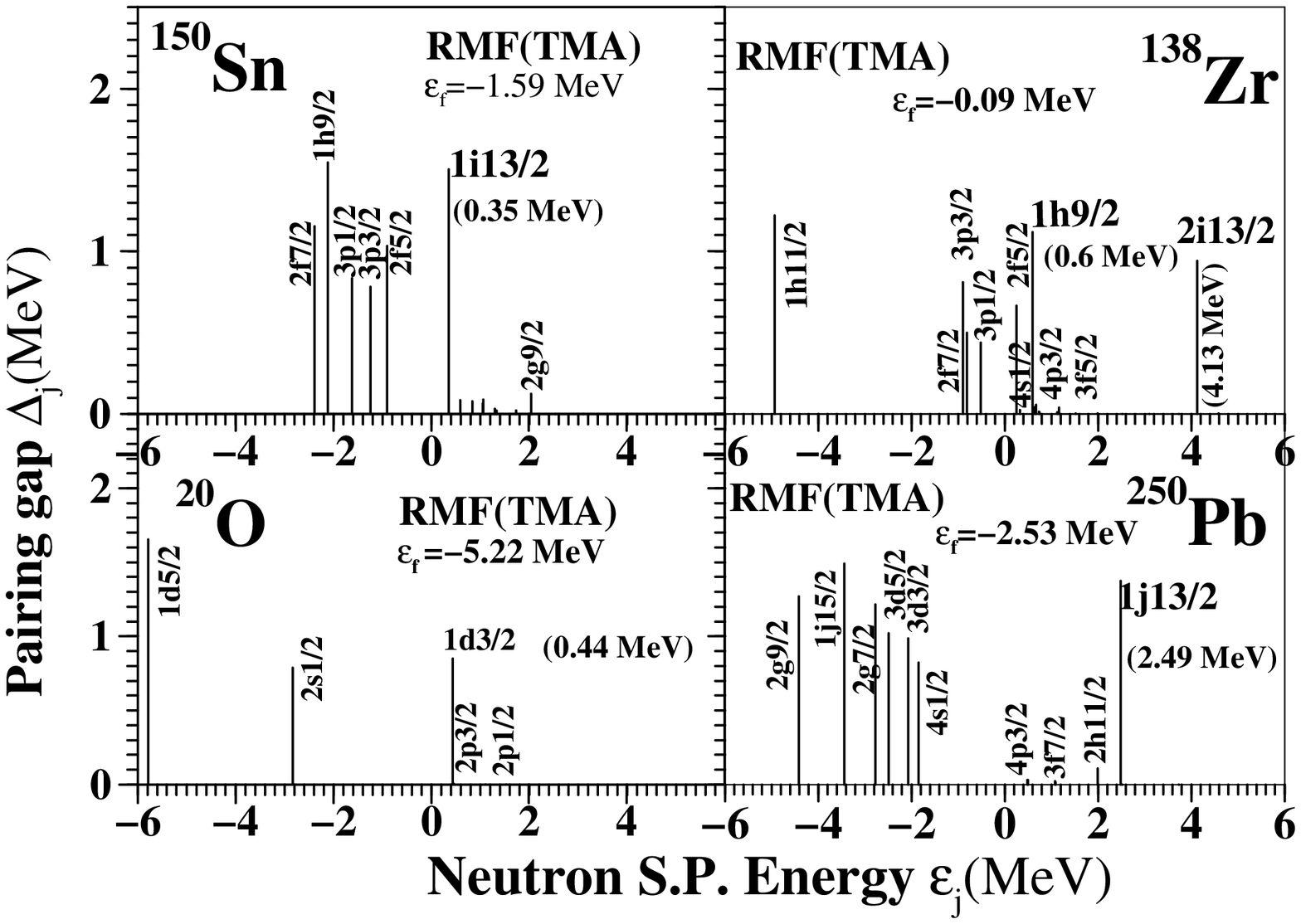,width=13cm} \vskip
0.15in {\noindent \small {{\bf Fig. 22.} Pairing gap energy
$\Delta_{j}$ of neutron single particle states with energy close
to the Fermi surface for the neutron rich nuclei $^{20}\rm{O}$,
$^{138}\rm{Zr}$, $^{150}\rm{Sn}$ and $^{250}\rm{Pb}$.\mbox{} }}

Finally, fig. 22 also shows a similar plot for the nucleus
$^{250}\rm {pb}$ representing the neutron rich case of the
$\rm{Pb}$ isotopes. The neutron state $1j_{13/2}$ shown in the
plot at energy $\epsilon$ = 2.48 MeV is a resonant state. As is
seen from the plot this state has large pairing gap, $\Delta_j
\approx 1.5$ MeV comparable to the bound single particle states
$3d_{5/2}$, $3d_{3/2}$ etc. near the Fermi level. The pairing gaps
values for the $\rm{Pb}$ isotopes are found to be consistent with
those obtained in other mean field
calculations\cite{lala,meng4,dobac96,dobac95}.

\subsubsection{Pairing Energy}

Apart from the case of $\rm{Zr}$ isotopes,  $\rm{O, Ca, Ni, Sn}$
and $\rm{Pb}$ nuclei considered here are all proton magic and,
therefore, for these nuclei the contribution to total pairing
energy is mostly from the neutron single particle states. Thus for
the proton magic nuclei, the pairing energy vanishes for $N$
values corresponding to the neutron shell closures. This can be
seen in fig. 23 which displays the total pairing energy
contribution for the nuclei $\rm{O, Zr, Sn}$ and $\rm{Pb}$ as a
function of neutron number $N$.

\vskip 0.11in \psfig{figure=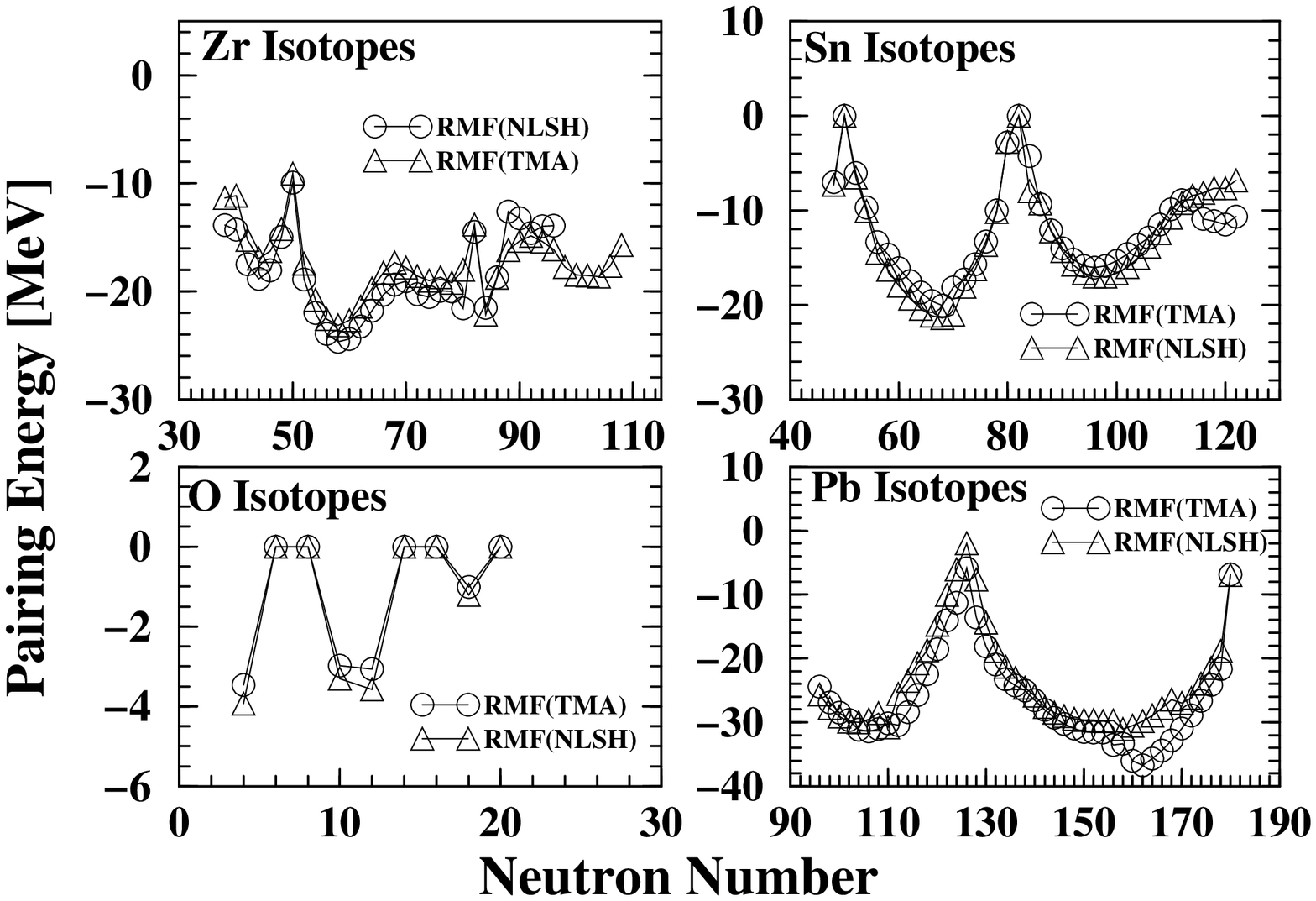,width=13 cm}
\vskip 0.15in {\noindent \small {{\bf Fig. 23.}Present RMF results
for the pairing energy  for the $\rm{O}$, $\rm{Zr}$, $\rm{Sn}$ and
$\rm{Pb}$ isotopes obtained with the TMA (open squares) are
compared with those obtained using the NL-SH (open triangles)
force parameters.\mbox{} }}

It is seen from the plot for the $\rm{O}$ isotopes in fig. 23 that
the neutron shell closures for $\rm{O}$ isotopes occur for N= 6,
8, 14, 16 and 20. Similarly, for the $\rm{Sn}$ isotope it is
observed to occur for N=50 and 82, whereas in the case of
$\rm{Pb}$ isotopes N=126 is found to correspond to the neutron
shell closure. However, some neutron rich isotopes of $\rm{Sn}$
and $\rm{Pb}$, due to reorganization of single particle states, do
not continue to be proton magic and, hence, there are
contributions  to the total pairing energy from proton single
particle states as well. In contrast, $\rm{Zr}$ is not a proton
magic nucleus. Thus, as such both the proton and neutron single
particle states contribute to the total pairing energy. Therefore
one observes slightly different feature of the N dependence of the
pairing energy as compared to that of the proton magic nuclei.
This can be seen from the plot for the $\rm{Zr}$ isotopes in fig.
23. The figure shows that for $N=50$ and $N=82$ the total pairing
energy is close to 10 MeV and 15 MeV, respectively, and it is
solely due to the protons. Similarly, for $N=40$ and $N=70$ due to
sub-shell behavior of the isotopes, the pairing energy is reduced.
Midway between $N=50$ and $N=70$, it is found to be maximum as
expected.

Further, from the plot of the  $\rm{Sn}$ isotopes in fig. 23, it
is seen that the pairing energy vanishes for $N=50$ and $N=82$ due
to shell closure. The maximum pairing energy values are found to
occur for isotopes in the midway between the neutron magic numbers
$50$ and $82$. Similar behaviour for the $N$ dependence of the
pairing energy is approximately seen for isotopes lying between
the magic numbers $N=82$ and $N=126$. For large neutron number
approaching $N=126$ it is found that the proton single particle
states are reorganized such that the $\rm{Sn}$ isotopes are no
more proton magic and, therefore, there is finite contribution
from the protons to the total pairing energy. This makes the shape
of the curve for the pairing energy for $N > 112$ flatter than
expected while approaching the neutron magic number $N=126$. The
pairing energy values obtained in our RMF+BCS calculations are
found to be of similar magnitude as those obtained in other mean
field calculations\cite{lala,dobac96,dobac95}. Further from fig 23
it is seen that the results obtained using the TMA and NL-SH
forces are very close to each other. However, slight differences
occur in some of the $\rm{Sn}$ and $\rm{Pb}$ isotopes with large
neutron numbers.

\subsubsection {Two Neutron Separation Energy}

In the lower panel of Fig. 24 we have shown the two neutron
separation energy $S_{2n}$ for the $\rm{O}$ isotopes. Further, the
results of RMF+BCS calculations employing the TMA\cite{suga}  and
the NL-SH\cite{sharma} forces have also been shown along with
those obtained in the RCHB approach for the purpose of comparison.

\vskip 0.5cm \psfig{figure=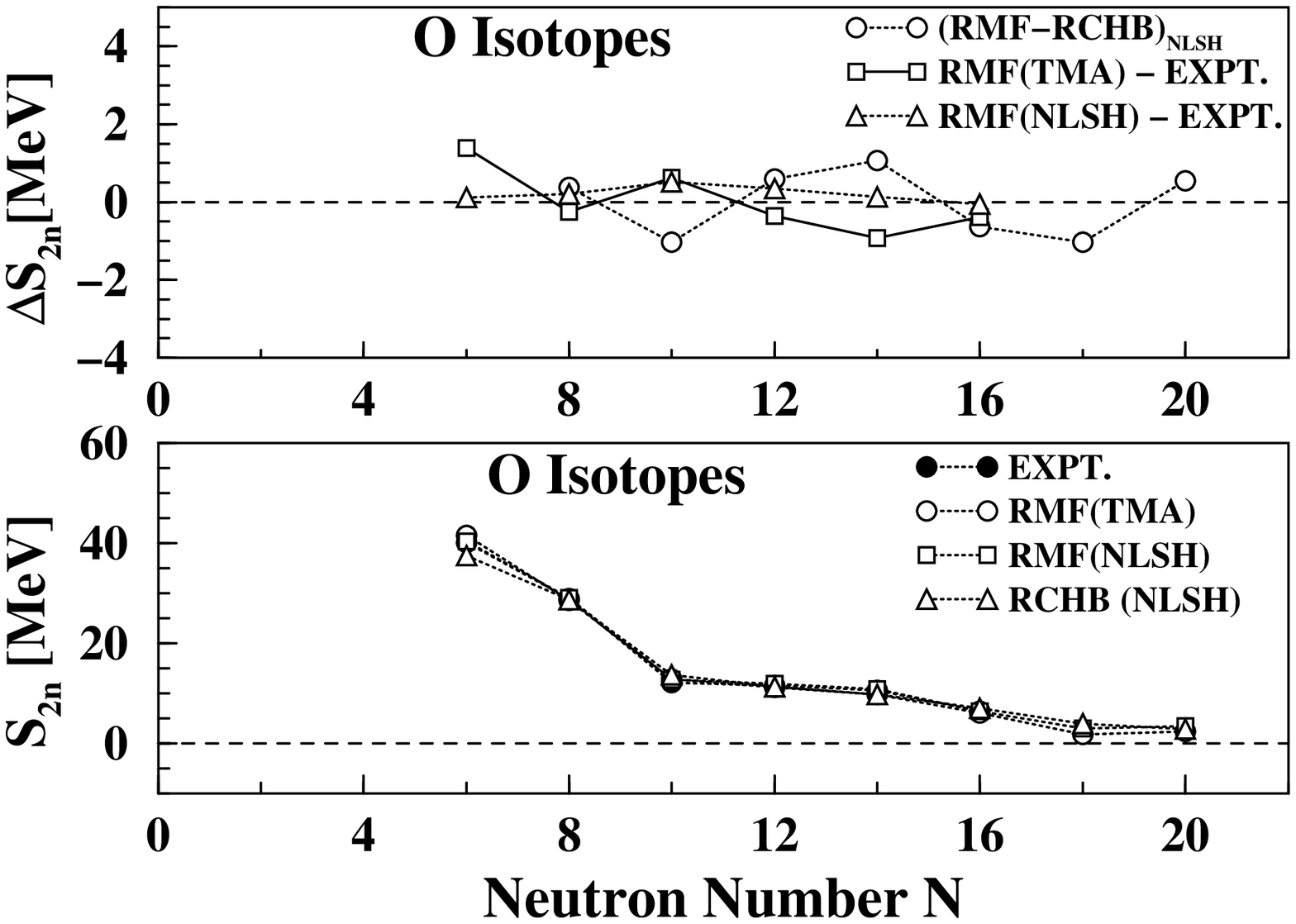,width=13 cm} \vskip
0.15in {\noindent \small {{\bf Fig. 24.} In the lower panel the
present RMF results for the two neutron separation energy  for the
$^{12-28}\rm{O}$ isotopes obtained with the TMA (open circles) and
the NL-SH (open squares) force parameters are compared with the
continuum relativistic Hartree-Bogoliubov (RCHB) calculations of
Ref.~26 carried out with the NL-SH force (open triangles). The
lower panel also depicts the available experimental
data\cite{audi} (solid circles) for the purpose of comparison. The
upper panel depicts the difference in the RMF+BCS results  and the
RCHB results of Ref.~26 for the two neutron separation energy
obtained for the NL-SH force. This plot also shows the difference
of the calculated results with respect to the available
experimental data\cite{audi}.\mbox{} }} \vspace{0.5cm}

It is seen from the figure that the calculations using two
different force parameters, the TMA and the NL-SH Lagrangian,
yield almost similar results as in the case of $\rm{Ca}$ and
$\rm{Ni}$ isotopes described earlier. Also, it is observed that
the results for the NL-SH force using the RCHB  approach and the
present RMF+BCS calculations are quite close to each other. The
upper panel depicts the difference between the results obtained
using the RMF+BCS and the RCHB approaches. The difference is
indeed small, a maximum difference is seen for $^{18}\rm{O}$ and
it is less than 1 MeV. It is further seen that the calculated
results for $S_{2n}$ are in good agreement with the experimental
data. The upper panel depicts the difference between the
experimental and calculated values. The maximum difference is
about 1.5 MeV. The calculated two neutron drip line is found to
occur at $N=20$ in both the RMF+BCS and the RCHB calculations.
However, in contrast it is  experimentally\cite{fauer} observed to
occur at $N=16$. The discrepancy between the theoretical
prediction and the experimental data is found almost in all the
mean field calculations employing both relativistic as well as
non-relativistic approaches.

 A better insight into the position of the neutron drip
line can be gained by looking at the dependence of the neutron
single particle states around the Fermi level as a function of
increasing neutron number $N$. This has been shown in fig. 25 for
the bound neutron $1d_{5/2}$, $2s_{1/2}$ and $1d_{3/2}$ states, as
well as for the unoccupied $2p_{3/2}$, $2p_{1/2}$ and $1f_{7/2}$
states lying in the continuum.

\vspace{0.5cm} \psfig{figure=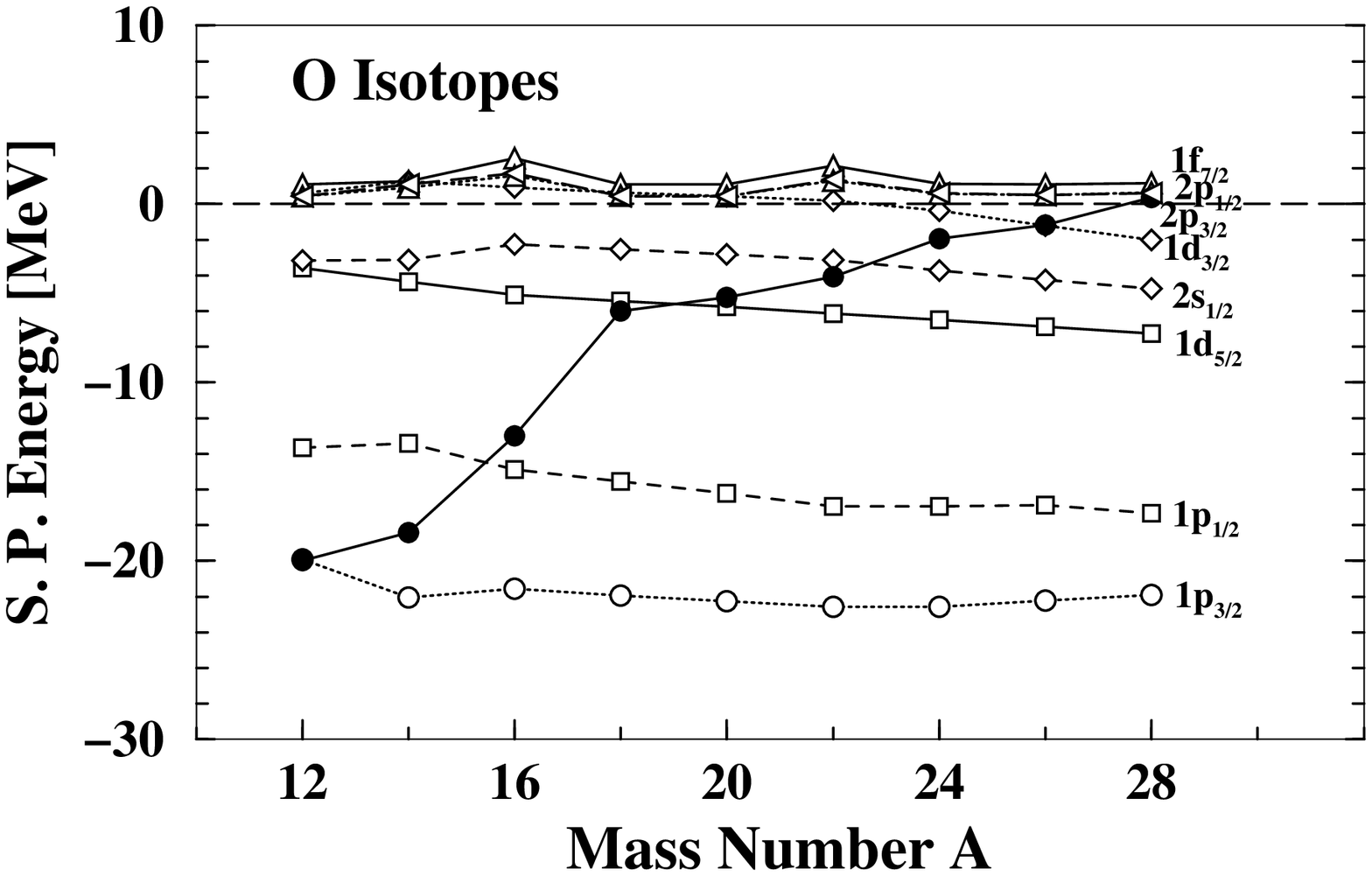,width=13cm}
\vskip 0.15in {\noindent \small {{\bf Fig. 25.} Variation of the
neutron single particle energies for the $\rm{O}$ isotopes with
increasing mass number A.\mbox{} }} \vspace{0.5cm}

It is seen from fig. 25 that with increasing neutron number $N$,
the bound states  near the Fermi level, for example, $1d_{3/2}$
and $2s_{1/2}$ etc. have a tendency to come down in energy. For
$N\le 14$ the crucial $1d_{3/2}$ state remains unbound and lies
close to other unbound states $2p_{3/2}$, $2p_{1/2}$, and
$1f_{7/2}$, while the neutron Fermi energy increases with
increasing neutron number $N$. Around $N=16$, the $1d_{3/2}$ state
comes down  to become bound, whereas the other neighboring states
$2p_{3/2}$, $2p_{1/2}$, $1f_{7/2}$ and $1g_{9/2}$ continue to
remain unbound. With further addition of two neutrons, at $N=18$,
the $1d_{3/2}$ state becomes even more bound while the neutron
Fermi energy approaches to zero. For $N=20$, the Fermi energy
becomes positive and the $1d_{3/2}$ state is fully occupied. This
results in a stable $^{28}\rm O$ isotope, the heaviest one. Beyond
this the neutron Fermi energy becomes positive and any further
addition of neutrons makes the isotope unstable. Thus the
essential point for the position of the neutron drip line is that
beyond $A=24$, the $1d_{3/2}$orbital is slightly too bound and
that the continuum states $2p_{3/2}$, $2p_{1/2}$, $1f_{7/2}$ and
$1g_{9/2}$, though close to zero energy, remain entirely
unoccupied.  As shown in fig. 26, a study of $N$ dependence of the
spin-orbit splitting energy $E_{ls}$ = $(E_{nlj=l-1/2}
-E_{nlj=l+1/2})/(2l+1)$ of the spin-orbit doublet $1d_{3/2}$ and
$1d_{5/2}$ indicates that the neutron doublet splitting energy is
slightly reduced beyond $A=24$ and, therefore, the $1d_{3/2}$
orbital is pushed down and accommodates further addition of $4$
neutrons.

\vspace{0.5cm} \psfig{figure=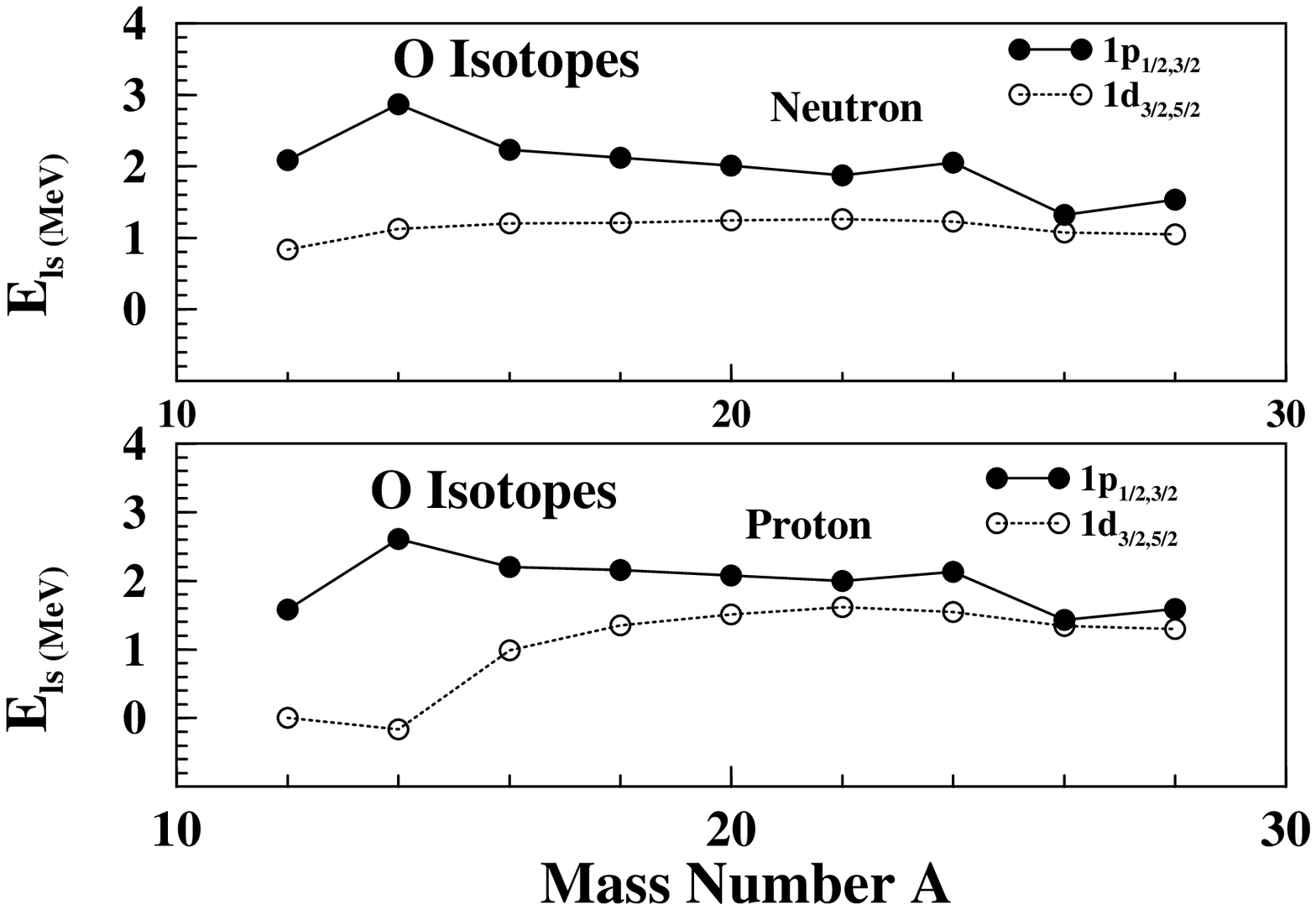,width=13cm} \vskip
0.15in {\noindent \small {{\bf Fig. 26.} Variation of the
spin-orbit splitting energy $E_{ls}$ for the neutron (upper panel)
and proton (lower panel) doublets $(1p_{1/2}, 1p_{3/2})$ and
$(1d_{3/2}, 1d_{5/2})$ in the $\rm{O}$ isotopes with increasing
mass number A.\mbox{} }} \vspace{0.5cm}

From the above discussion it appears that for the neutron rich
$\rm O$, the pairing interaction  beyond $A=24$ is not strong
enough to populate the positive energy states near the Fermi
level, for example, $2p_{3/2}$, $2p_{1/2}$ and $1f_{7/2}$ etc., to
make the isotopes heavier than $A=24$ unstable. This discrepancy
between the theoretical results and the measurements for the two
neutron drip line in the $\rm O$ isotopes is found in most of the
mean field calculations \cite{dobac96,lala,meng4,dobac95,leja} and
needs further investigations.

Next we consider the $\rm{Zr}$ isotopes for which the two neutron
separation energies have been plotted in the lower panel of fig.
27, whereas the upper panel shows the differences between the
$S_{2n}$ values obtained from theoretical predictions and the
measurements. It is seen from fig. 27 that the degree of agreement
between the available experimental data and the theoretical
results using TMA and NLSH force parameterizations are of similar
quality as for the isotopes of other nuclei discussed earlier. In
this case the maximum difference between the measured $S_{2n}$ and
the predicted value for it is found to occur for a few isotopes
around $N=54$, and is less than $3$ MeV. Again, as for the results
of other nuclei and their isotopes,  the RCHB and the RMF+BCS
calculations for the $\rm{Zr}$ isotopes are also found to be in
excellent agreement with each other as is evident from fig. 27.
With regard to the discrepancy around $N=54$ between the data and
the results obtained in the RMF+BCS as well as in the RCHB
calculations, it may be pertinent to remark that the present
calculations treat these nuclei as spherical whereas actually some
of the $\rm{Zr}$ isotopes are believed to be well deformed.
Whether an inclusion of deformation in the present RMF+BCS
approach would markedly improve the calculated results needs
further investigations.

\vskip 0.15in \psfig{figure=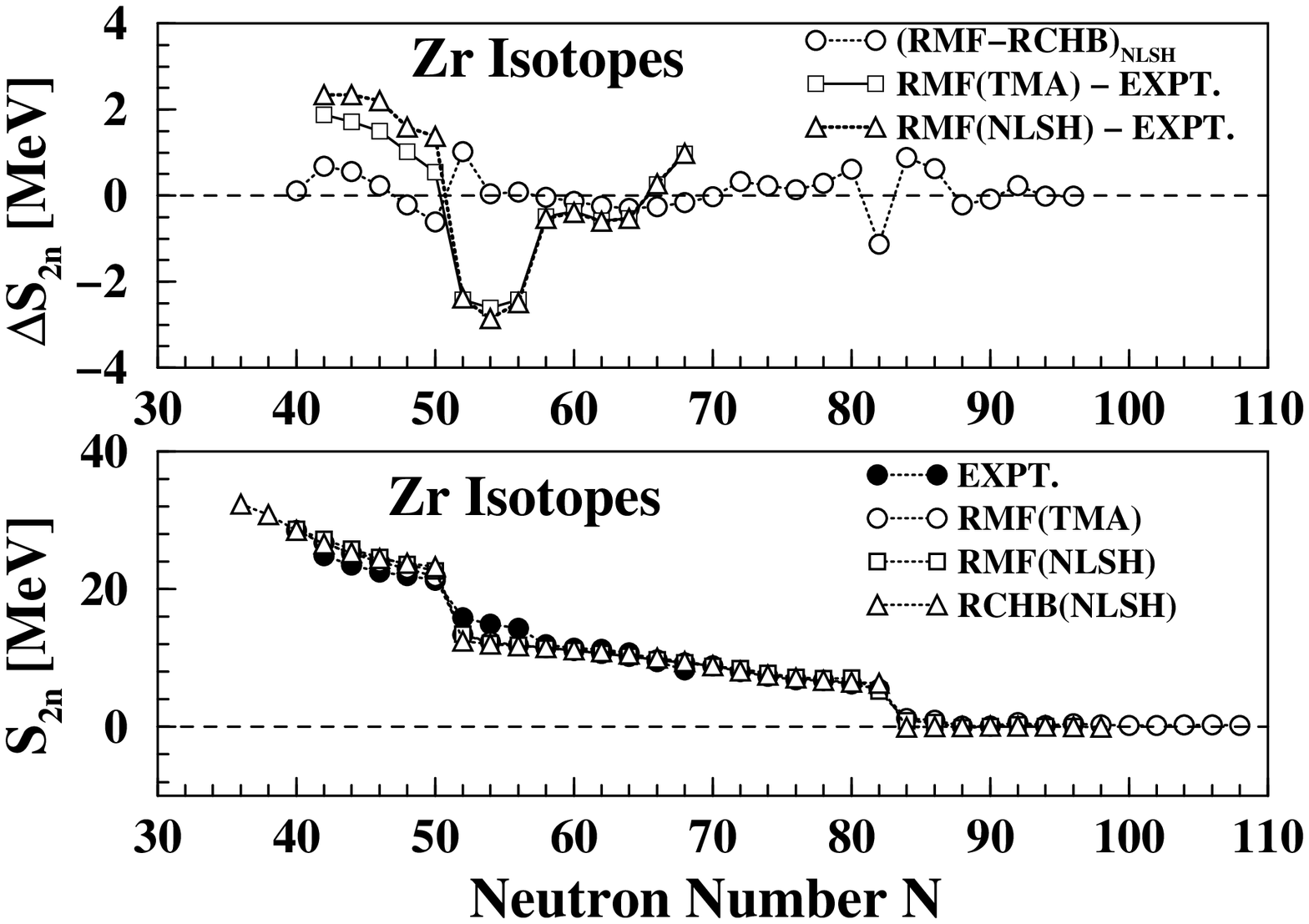,width=13 cm} \vskip
0.15in {\noindent \small {{\bf Fig. 27.} The present RMF results
for the two neutron separation energy  for the $\rm{Zr}$ isotopes
obtained with the TMA (open circles) and the NL-SH (open squares)
force parameters are compared with the continuum relativistic
Hartree-Bogoliubov (RCHB) calculations of Ref.~26 carried out with
the NL-SH force (open triangles), and with the available
experimental data\cite{audi}(solid circles). The upper panel
depicts the difference in the RMF+BCS results  and the RCHB
results of Ref.~26  obtained for the NL-SH force along with the
difference of the calculated results with respect to the available
experimental data\cite{audi}.\mbox{} }} \vspace{0.5cm}

The calculations employing the NL-SH force for the mean field
description predict the heaviest stable isotope to be
$^{136}\rm{Zr}$. This is true for both the RMF+BCS as well as the
RCHB calculations as can be seen from fig. 27. The results
obtained using the TMA force are at variance with this prediction
and the two neutron drip line is found to occur at $N=112$
corresponding to $^{152}\rm{Zr}$ isotope. This difference in the
predictions of the two neutron drip line is due to the
dissimilarities in the spectrum of single particle states near the
neutron Fermi level. Further, it is seen from fig. 27 that the
$S_{2n}$ values for $N \ge 84$ is rather small, and is reduced
further more with increasing number of neutrons. For neutron
numbers around $N \ge 84 $ the $S_{2n}$ values are close to a few
hundred keV. Such a physical situation for very loosely bound
systems has already been seen to occur for highly neutron rich
$\rm {Ca}$ isotopes.

In order to have a better understanding of  such loosely bound
extremely neutron rich $^{124-152}\rm {Zr}$ isotopes, we have
shown in fig. 28 the variation of neutron single particle energy
for the states near the Fermi level obtained with the TMA force.

\vspace{0.5cm} \psfig{figure=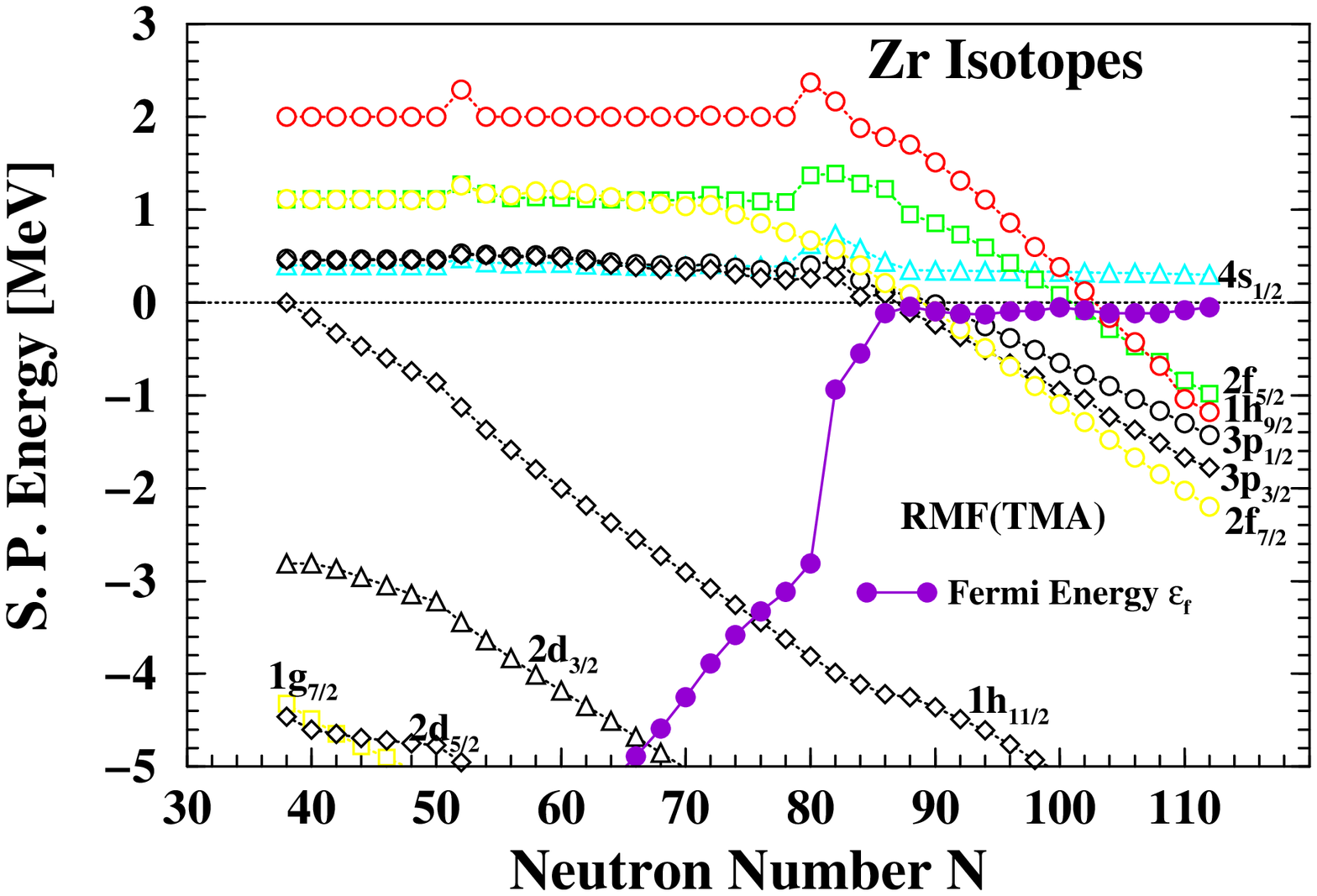,width=13cm}
\vskip 0.15in {\noindent \small {{\bf Fig. 28.} Variation of the
neutron single particle energies for the $\rm{Zr}$ isotopes with
increasing neutron number N for the TMA force.\mbox{} }}
\vspace{0.5cm}

It is seen from  fig. 28 that for these neutron rich $\rm {Zr}$
isotopes the neutron Fermi level lies almost at zero energy. The
important neutron single particle states with finite occupancy
near the Fermi level are $1h_{9/2}$, $2f_{5/2}$, $2f_{7/2}$,
$3p_{1/2}$, $3p_{3/2}$ and $4s_{1/2}$. For $N=82$ corresponding to
$^{122}\rm {Zr}$, we have shell closure for the neutron states.
For this nucleus the neutron states up to $1h_{11/2}$ are
completely filled  and the higher states are completely empty as
is expected of a closed shell system. On further addition of two
neutrons, the occupation pattern of states gets changed as
compared to that of $^{122}\rm {Zr}$ nucleus in the sense that the
added neutron simply do not occupy the next higher state. Instead
the neutrons are redistributed amongst the $1h_{11/2}$ and the
states $3p_{3/2}$, $2f_{7/2}$, $3p_{1/2}$, $2f_{5/2}$, $1h_{9/2}$
and the high lying resonant state $1i_{13/2}$. Also as we keep
adding more and more neutrons the positive energy states
$3p_{3/2}$, $2f_{7/2}$, $3p_{1/2}$, $2f_{5/2}$ and $1h_{9/2}$
gradually come down in energy and subsequently become bound for
larger neutron number $N$ as is seen in fig. 28. To illustrate
this we consider the detailed single particle structure of, for
example, $^{126}\rm {Zr}$ and $^{138}\rm {Zr}$ nuclei. For the
nucleus $^{126}\rm {Zr}$ the occupation weight of the state
$3p_{3/2}$ at energy $\epsilon =0.05$ MeV is $0.33$, of $2f_{7/2}$
at energy $\epsilon =0.22$ MeV is $0.27$, of $3p_{1/2}$ at energy
$\epsilon =0.28$ MeV is $0.10$, of $2f_{5/2}$ at energy $\epsilon
=1.20$ MeV is $0.03$, of $1h_{9/2}$ at energy $\epsilon =1.82$ MeV
is $0.03$, and that of the high lying resonant state $1i_{13/2}$
at energy $\epsilon =5.14$ MeV is $0.03$. All these positive
energy states accommodate together about 4.5 neutrons. However,
the first three of these states lying very close to the Fermi
level ($\epsilon_f = - 0.06$ MeV) have the major share of about
$3.6$ neutrons. In comparison consider the nucleus $^{138}\rm
{Zr}$ which is reached after an addition of 12 neutrons to the
nucleus $^{126}\rm {Zr}$. Due to the additional neutrons, the
$2f_{7/2}$, $3p_{3/2}$ and $3p_{1/2}$ states become bound whereby
$2f_{7/2}$ state becomes lower in energy as compared to that of
$3p_{3/2}$. Also, the neutron Fermi energy is slightly changed to
$\epsilon_f = -0.09$ MeV. Now the occupation weight of the state
$2f_{7/2}$ at energy $\epsilon = -0.90$ MeV is $0.85$, of
$3p_{3/2}$ at energy $\epsilon = -0.81$ MeV is $0.90$, of
$3p_{1/2}$ at energy $\epsilon = -0.51$ MeV is $0.84$, of
$2f_{5/2}$ at energy $\epsilon =0.25$ MeV is $0.27$, of $1h_{9/2}$
at energy $\epsilon =0.60$ MeV is $0.24$, and that of another high
lying resonant state $2i_{13/2}$ at energy $\epsilon =4.13$ MeV is
$0.01$. In this case the positive energy states accommodate
together about 4 neutrons. On further addition of neutrons, the
positive energy states $2f_{5/2}$ and $1h_{9/2}$  also gradually
become bound and thus can accommodate neutrons up to $N=112$ which
correspond to a shell closure. Thus one is able to reach a bound
$^{152}\rm {Zr}$ nucleus. It is remarkable that the energy of the
neutron $4s_{1/2}$ single particle state remains positive and
almost constant with increasing neutron number beyond $N=90$ as is
seen from fig. 28. And, therefore, further increase of neutrons
which fill in the next available $4s_{1/2}$ state make the nucleus
$^{154}\rm {Zr}$ unbound. Further, due to the fact that for the
neutron rich ($N = 84$ to $N = 90$)  $^{124-130}\rm {Zr}$
isotopes, one of the last  partially occupied states is the
positive energy $3p_{1/2}$ state having low orbital angular
momentum and a reduced centrifugal barrier, the density
distribution of these exotic nuclei assumes a wider extension. It
is in turn reflected in neutron radii becoming rather large as can
be seen in fig. 33. This may give rise to halo formation as in the
case of $\rm {Ca}$ isotopes discussed earlier. However, it should
be emphasized that the case of neutron rich $\rm {Zr}$ isotopes is
slightly different. In the case of $\rm {Ca}$ isotopes, one of the
outermost positive energy  single particle states which is
partially occupied is the $3s_{1/2}$ state. This state with
orbital angular momentum $l=0$ and having no centrifugal barrier
provides a more favorable condition for a hallo formation, as
compared to the $3p_{1/2}$ state in the neutron rich $\rm {Zr}$
isotopes. Beyond $^{130}\rm {Zr}$ the neutron  $3p_{1/2}$ state
becomes bound and the positive energy resonant states $2f_{5/2}$
and $1h_{9/2}$ continue together to accommodate more neutrons. As
the nucleus $^{142}\rm {Zr}$ is reached, the $2f_{5/2}$ becomes
bound. On further addition of two neutrons the $1h_{9/2}$ state
also becomes bound and continues to accommodate more neutrons
until completely filled. For the $^{144-152}\rm {Zr}$ isotopes the
occupation weights of the positive energy states like $4p_{3/2}$
and $4p_{1/2}$ located close to the continuum threshold remains
almost negligible.

The single particle structure obtained with increasing neutron
number in the case of RMF+BCS calculations carried out with the
NL-SH force is slightly at variance with that seen above for the
TMA force. In this case one observes that the neutron states
$3p_{3/2}$, $2f_{7/2}$, and $3p_{1/2}$ as for the TMA results
gradually become bound with increasing neutron number. However,
the states $2f_{5/2}$ and $1h_{9/2}$ do not come down sufficiently
in energy and, thus, remain positive energy states as has been
shown in the lower panel of fig. 29. Consequently, with the NL-SH
force one has the nucleus $^{136}\rm {Zr}$ as the heaviest bound
system. In order to gain more insight into the role of resonant
and other states near the Fermi level in accommodating more and
more neutrons to have extremely neutron rich isotopes, we have
shown in the upper panel of fig. 29 the occupancy in terms of
number of neutrons occupying the above mentioned crucial states
$1h_{9/2}$, $2f_{5/2}$, $3p_{3/2}$, $2f_{7/2}$, and $3p_{1/2}$. As
is seen in this part of the figure, all these states at the
threshold of the continuum are partially occupied for $N=84$
($^{124}\rm {Zr}$). The total number of nucleons occupying these
positive energy states is about two, approximately one each in the
$2f_{5/2}$ and $1h_{9/2}$ states. The occupancy of other states is
almost negligible. We note that the neutron states $3p_{3/2}$,
$2f_{7/2}$, and $3p_{1/2}$ are hardly bound. With addition of two
neutrons at $N=86$ ($^{126}\rm {Zr}$) all the states come down in
energy. Now there are more than three neutrons in the positive
energy states, mainly the states $3p_{3/2}$ and $2f_{7/2}$ are
occupied with one and two neutrons, respectively. With further
addition of two neutrons at $N=88$ ($^{128}\rm {Zr}$), the
$3p_{3/2}$ state becomes bound, and the $3p_{1/2}$ and $2f_{7/2}$
states are now hardly bound. For this case the number of neutrons
in the continuum is now less than three occupying the $3p_{1/2}$
and $2f_{7/2}$ states. In a similar manner for $N=90$ ($^{130}\rm
{Zr}$) it is seen from the figure that the state $2f_{7/2}$ also
becomes just a bound one and there are about a little more than
one neutron in the continuum, mostly in the $3p_{1/2}$ state.
Finally for $N=92$ corresponding to ($^{132}\rm {Zr}$) the
$3p_{3/2}$, $2f_{7/2}$, and $3p_{1/2}$ become bound states and the
states $2f_{5/2}$ and $1h_{9/2}$ continue to remain unbound. It is
seen from the figure that for the isotopes ($^{132-136}\rm {Zr}$)
the number of nucleons occupying the continuum states is less than
one. From these considerations one clearly sees the important role
of resonant states and that of loosely bound states like
$1h_{9/2}$, $2f_{5/2}$, $3p_{3/2}$, $2f_{7/2}$, and $3p_{1/2}$ in
accommodating more and more nucleons to produce exotic neutron
rich loosely bound isotopes of ($^{124-136}\rm {Zr}$).

Further, on the other hand, the single particle state $4s_{1/2}$
is found to exhibit similar behavior of variation for both the TMA
and NL-SH forces and with addition of neutrons it remains
throughout an unbound state without significant occupancy. This
prohibits the very heavy neutron rich isotopes of $\rm {Zr}$ to
exhibit as prominent halo formation as predicted for the heavy
neutron rich $\rm {Ca}$ isotopes.

Also, it should be emphasized that the difference in the binding
energy of the loosely bound $^{132-152}\rm {Zr}$ isotopes obtained
in the case of TMA force is in fact very small, of the order of
100 keV for the neighboring two isotopes. With this in view, the
difference in the positions of the two-neutron drip line as
predicted for the TMA and NL-SH forces should not be treated too
seriously.

\vspace{0.5cm} \psfig{figure=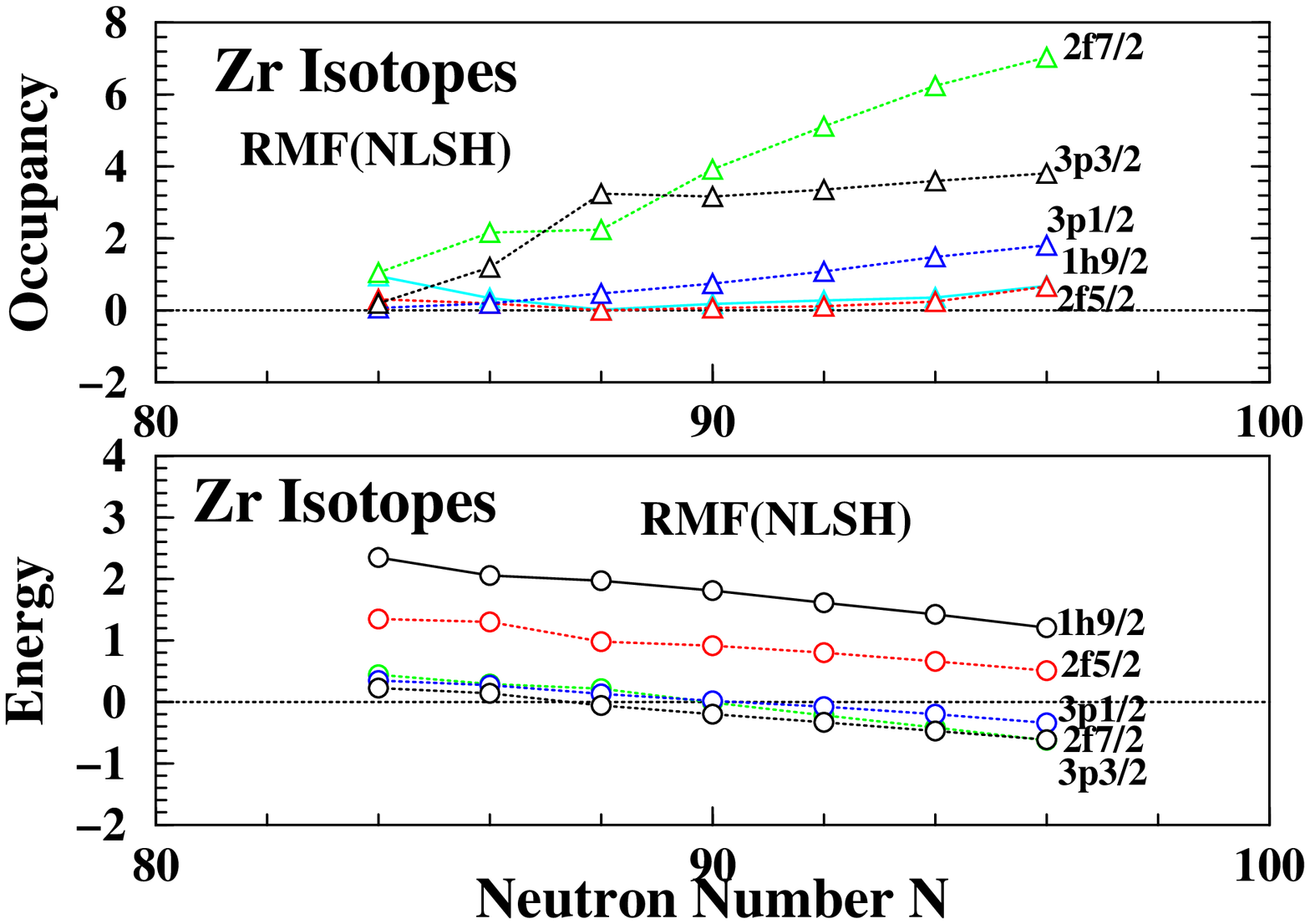,width=13cm}
\vskip 0.15in {\noindent \small {{\bf Fig. 29.} Lower Panel:
Variation in energy of the neutron single particle states for the
$^{124-136}\rm{Zr}$ isotopes with increasing neutron number N for
the NL-SH force.\ Upper Panel: Occupancy of the single particle
states in terms of number of neutrons with increasing neutron
number.\mbox{} }} \vspace{0.5cm}

In Fig. 30 we have displayed results of the two neutron separation
energy $S_{2n}$ for the $\rm{Sn}$ isotopes. The difference between
the RCHB and the RMF+BCS results have been shown in the upper
panel of the  figure. It is indeed small for most of the isotopes.
The maximum difference between the two approaches, which occurs in
a few case is approximately $1$ MeV. Similarly a comparison with
the available experimental data shows that both the TMA and NL-SH
forces provide a good description of the two neutron separation
energies. However, the maximum difference of approximately  $2$
MeV is found for a few $\rm{Sn}$ isotopes as can be seen in the
upper panel of fig. 30. It is found that the RMF+BCS calculations
using either TMA or NL-SH forces  predict $^{176}\rm{Sn}$ to be
the heaviest stable isotope.  In complete agreement, the RCHB
calculations also predict the two neutron drip line to occur at
$N=126$ corresponding to $^{176}\rm{Sn}$.

\vskip 0.15in \psfig{figure=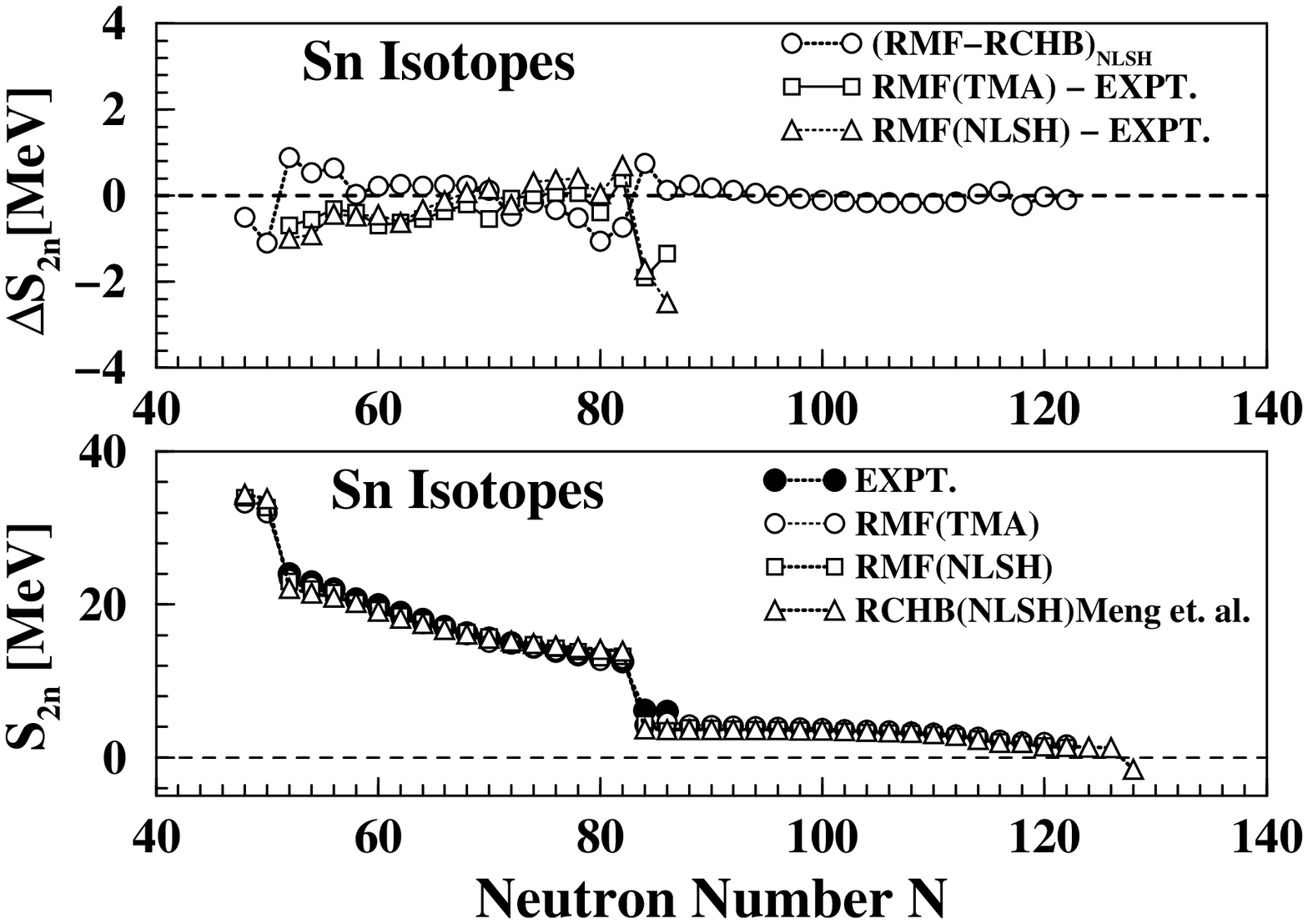,width=13 cm} \vskip
0.15in {\noindent \small {{\bf Fig. 30.} RMF+BCS results  and
their comparisons with that of RCHB calculations of ref.~26, and
available experimental data\cite{audi} $^{104-136}\rm{Sn}$ for the
two neutron separation energy for the $^{98-172}\rm{Sn}$ isotopes.
For details see the caption for fig. 27.\mbox{} }} \vspace{0.5cm}

Similar results of the two neutron separation energy $S_{2n}$ for
the $\rm{Pb}$ isotopes are displayed in fig. 31. As before the
upper panel shows the difference between different results plotted
in the lower panel. It is to be noted that the RCHB calculations
for the $\rm{Pb}$ isotopes by Meng et al.\cite{meng} have been
reported only up to $N=150$ which correspond to the
$^{232}\rm{Pb}$ isotope. Due to this we have shown only the
limited results for the RCHB calculations in fig. 31. It is seen
from the figure that the RMF+BCS results are very close to those
obtained using the RCHB approach. The difference between the RCHB
and the RMF+BCS results have been shown in the upper panel of the
figure. It is indeed small for most of the isotopes. The maximum
difference between the two approaches, which occurs in a few case
is approximately $1$ MeV. Similarly a comparison with the
available experimental data shows that both the TMA and NL-SH
forces provide an equally good description of the two neutron
separation energies for the $\rm{Pb}$ isotopes. The two neutron
drip line for the $\rm{Pb}$ isotopes is found to occur at $N=184$
corresponding to $^{266}\rm{Pb}$ isotope.

\vskip 0.15in \psfig{figure=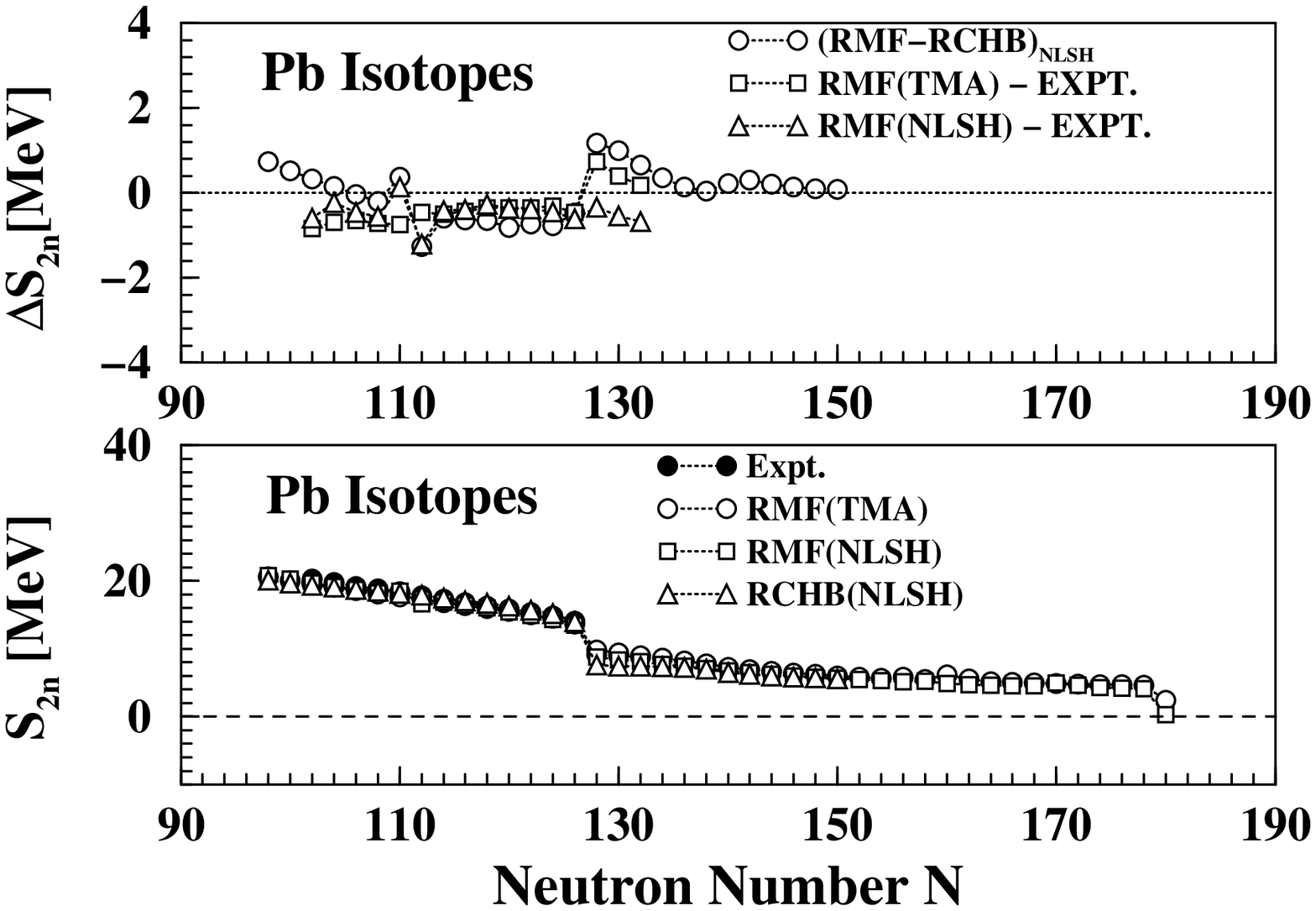,width=13 cm} \vskip
0.15in {\noindent \small {{\bf Fig. 31.} RMF+BCS results and their
comparisons with that of RCHB calculations of ref.~26, and the
available experimental data\cite{audi} ($^{184-214}\rm{Pb})$ for
the two neutron separation energy  for the $^{180-262}\rm{Pb}$
isotopes. For details see the caption for fig. 27.\mbox{} }}

\subsubsection{RMS Radii for Proton and Neutron Distributions}

Earlier we have made a detailed comparison of RMF+BCS results with
those obtained in the RCHB calculations\cite{meng2}, and with the
available experimental data for the rms radii of proton and
neutron distributions for the $\rm Ca$ and $\rm Ni$ isotopes as
displayed in figs. 6 and 18. Similar comparisons for the  results
of these rms radii for $\rm O, Zr, Sn$ and $\rm Pb$ isotopes lead
us to conclude that the rms proton, neutron and matter radii as
obtained in the RMF+BCS description are quantitatively in good
agreement with those obtained in a more complete RCHB approach.
Also, the results for these radii using the TMA and the NL-SH
forces are observed to be very close to each other. Furthermore,
it is found that the calculated proton and neutron rms radii are
in good agreement with the available experimental data. As an
illustration we have shown in fig. 32 the calculated RMF+BCS
results (open triangle) obtained with the TMA force for the proton
rms radii in comparison with the available measured data (open
circles). The agreement with the data is indeed very satisfactory.

\vskip 0.15in \psfig{figure=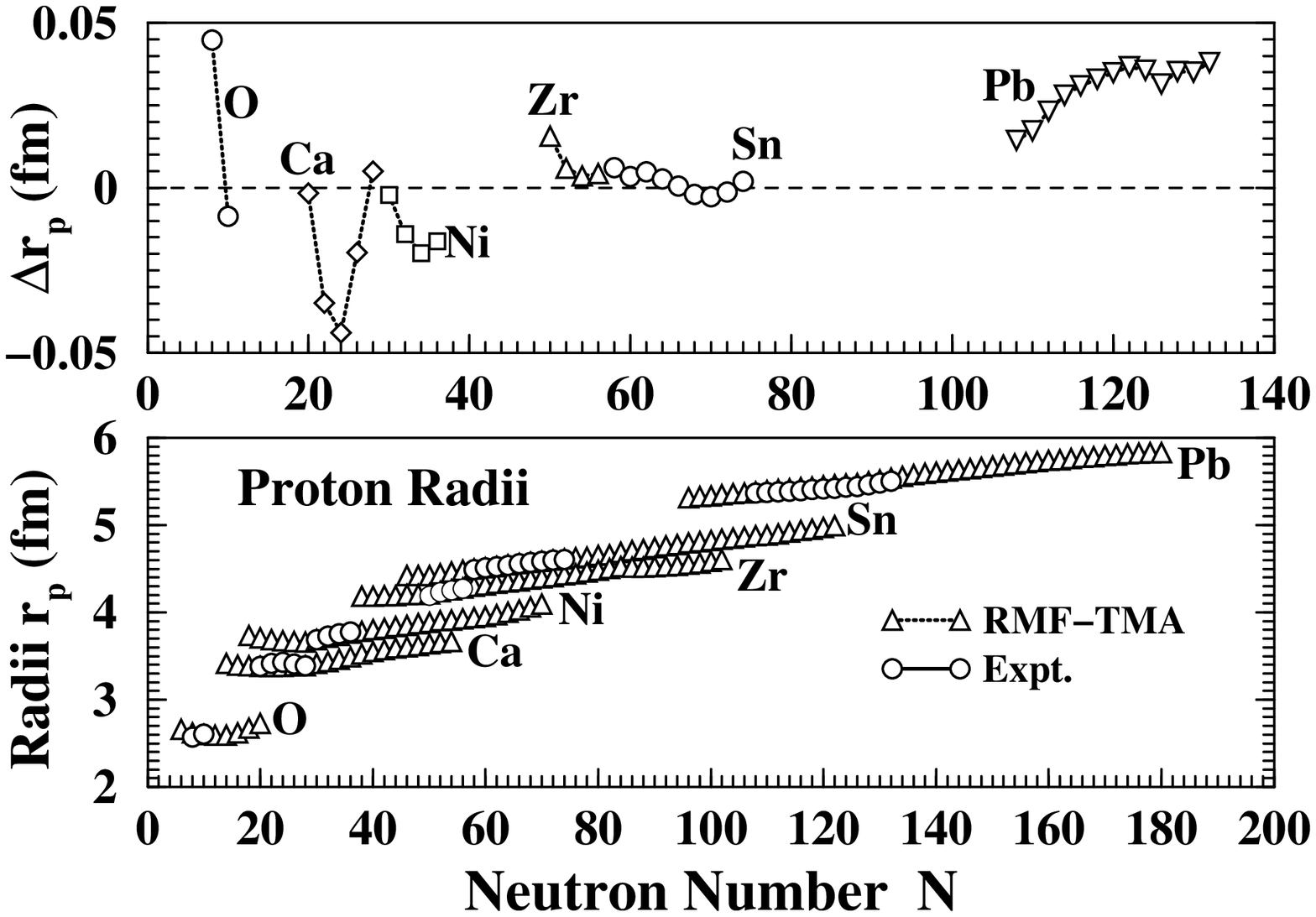,width=13 cm}
\vskip 0.15in {\noindent \small {{\bf Fig. 32.} In the lower panel
present RMF results for the proton rms radii $r_p$  for the
isotopes of various nuclei obtained with the TMA  force parameters
are compared with the available experimental data\cite{audi} (open
circles). The upper panel shows the difference between the
measured and calculated radii for the isotopes for which data are
available. \mbox{} }}

Further, in fig. 33 we have shown in various columns the
theoretical predictions for the rms neutron radii. It is found
that the variation of the neutron radii for the chains of isotopes
of nuclei studied here only approximately follows the $N^{1/3}$
law ($r_n = r_0 N^{1/3}$) as has been shown in fig. 33. One
observes deviation from the law for the case of rich proton as
well as rich neutron isotopes. The figure also shows the value of
radius constant $r_0$ providing the best fit to the radii. As is
seen from fig. 33, this best fit is obtained with  gradually
reduced value of the constant $r_0$ as we move from the lightest
$\rm O$ nuclei to the heaviest $\rm {Pb}$ isotopes. However, it is
possible to provide a better description by choosing the radius
constant to have isospin dependence. As noted above these
predictions are found to be in good agreement with the available
measurements for the neutron rms radii. We have, however, not
shown the measured data to keep the figure uncluttered.   Also,
from fig. 33 it is seen that the best candidate for experimental
observation of halo formation in the neutron rich nuclei is
offered by the $\rm{Ca}$ isotopes. For the neutron rich heavy
$\rm{Zr}$ isotopes this effect is less dramatic as is seen from
fig. 33.

 \vskip 0.15in \psfig{figure=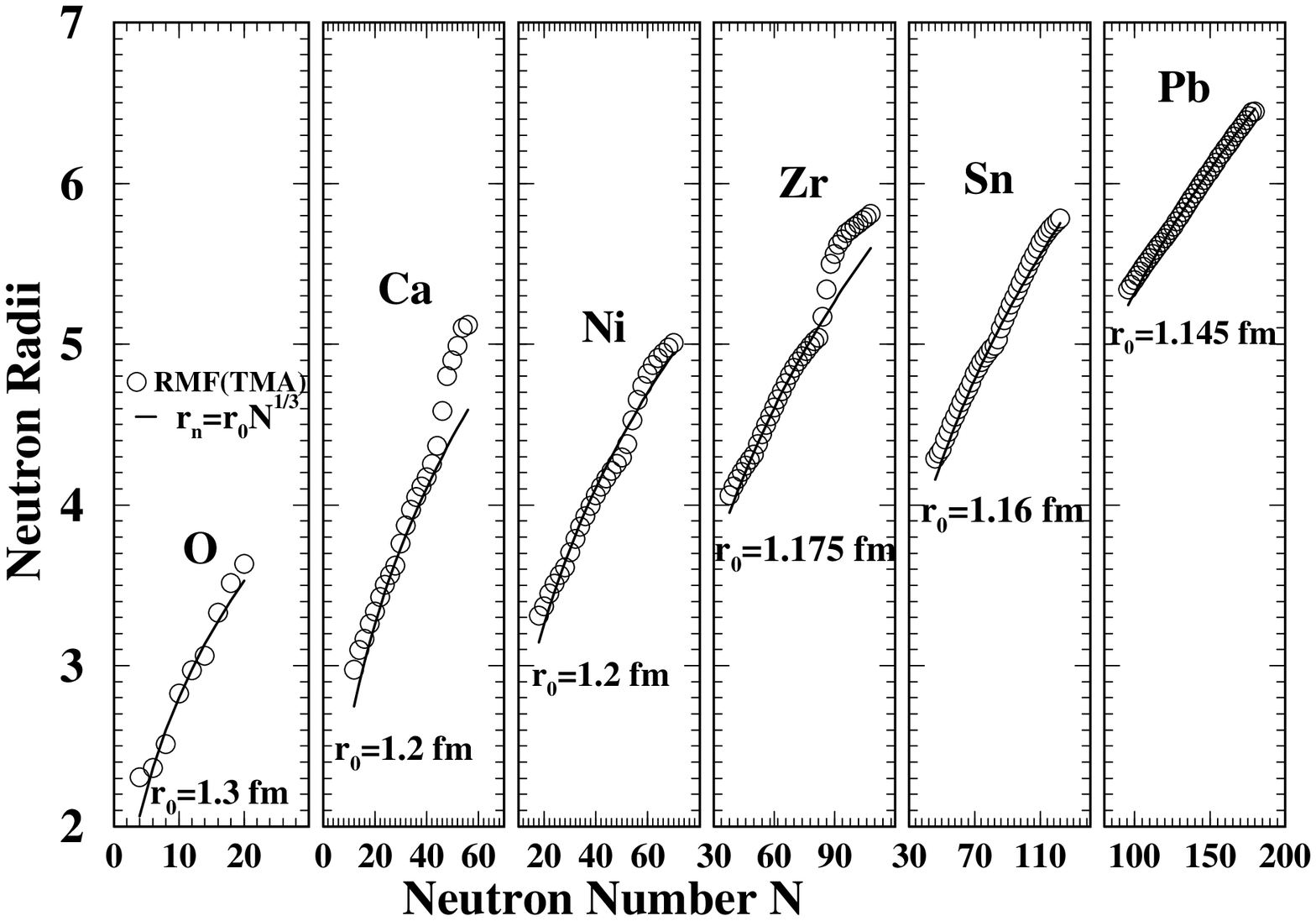,width=13
cm} \vskip 0.15in {\noindent \small {{\bf Fig. 33.} The present
RMF results for the neutron rms radii $r_p$  for the isotopes of
various nuclei obtained with the TMA  force parameters. These
results are compared with a rough estimate of neutron distribution
radius given by $r_n$ = $r_0 N^{1/3}$ wherein the radius constant
$r_0$ is chosen to provide the best fit to the theoretical
results. Halo formation in the case of neutron rich $\rm {Ca}$ and
also to some extent in $\rm{Zr}$ isotopes is clearly seen.\mbox{}
}}

\subsubsection{Proton and Neutron Densities}

We have already discussed the radial proton and neutron densities
for $^{66}\rm{Ca}$ and $^{84}\rm{Ni}$ isotopes as  representative
examples of neutron rich $\rm{Ca}$ and  $\rm{Ni}$ nuclei. Similar
plots for the proton radial densities shown by hatched areas and
neutron densities depicted by solid lines have been displayed in
fig. 34 for selected neutron rich isotopes of $^{20}\rm{O}$,
$^{138}\rm{Zr}$, $^{170}\rm Sn$ and $^{250}\rm Pb$ nuclei.  In
addition, the plots also depict the neutron densities of several
other neighboring isotopes of these nuclei for the purpose of
comparison. For completeness we have included the plots for the
neutron rich $^{62}\rm{Ca}$ and $^{96}\rm{Ni}$ isotopes as well.

\vspace{0.5cm} \psfig{figure=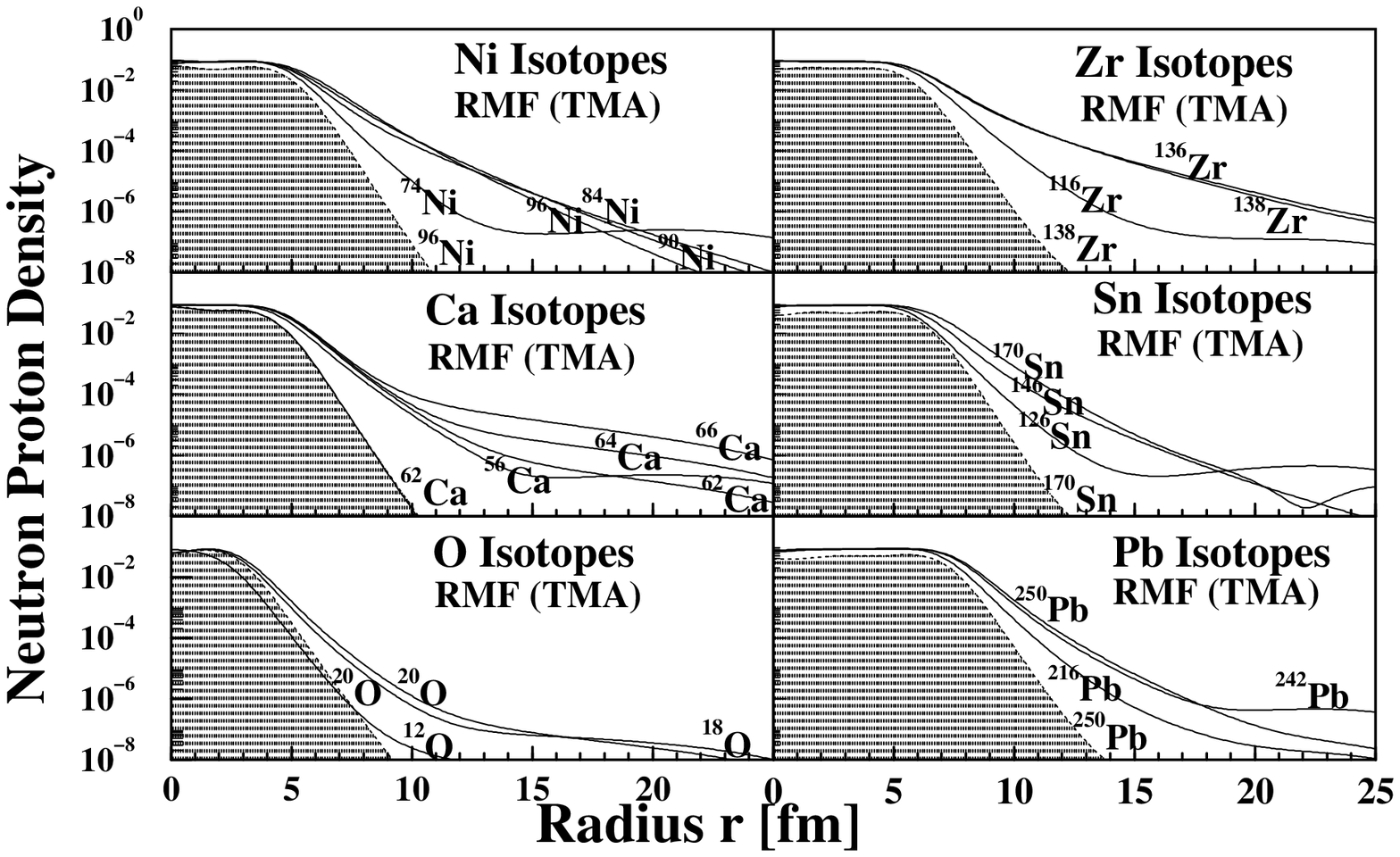,width=13cm}
\vskip 0.15in {\noindent \small {{\bf Fig. 34.} The neutron radial
density distributions for some selected isotopes of $\rm{O}$,
$\rm{Ca}$, $\rm{Ni}$, $\rm{Zr}$, $\rm{Sn}$  and $\rm{Pb}$ nuclei
have been shown by solid lines. The hatched area represents the
proton radial density distribution for the nuclei $^{20}\rm{O}$,
$^{62}\rm{Ca}$, $^{96}\rm{Ni}$, $^{138}\rm{Zr}$, $^{172}\rm{Sn}$
and $^{250}\rm{Pb}$.\mbox{} }}

It is seen from fig. 34 that the proton densities for these nuclei
have gross features similar to each other, albeit with increasing
atomic number from $\rm{O}$ to $\rm{Pb}$ the distribution assumes
wider spatial extension as expected. However, in the case of
$\rm{Zr}$ isotopes, the distribution is relatively broader due to
the fact that it is not a proton magic nucleus as compared to
other nuclei shown in the figure. This can be easily seen from the
difference in the slope of the density curves. As remarked
earlier, for the open shell isotopes the asymptotic behavior of
the density profile is affected by the contributions from the
partially occupied positive energy resonant states and those
loosely bound ones just above the continuum threshold near the
Fermi level. The positive energy states  have wide spread wave
functions and yield a density profile with wide spatial extension
and  long tail asymptotically. This has been described earlier for
the $\rm{Ca}$ and $\rm{Ni}$ isotopes and is found valid for other
nuclei as well. Accordingly for further demonstration we have
shown explicitly in figs. 35 through 40 the neutron and proton
density profiles for the $\rm{Zr}$, $\rm{Sn}$ and $\rm{Pb}$
isotopes. The inset in these plots shows the density distributions
on a logarithmic scale for larger radial distances and exhibits
its asymptotic behavior. In order to keep the plots uncluttered we
have drawn the results for limited number of isotopes for a given
chain.

\vspace{0.5cm}\psfig{figure=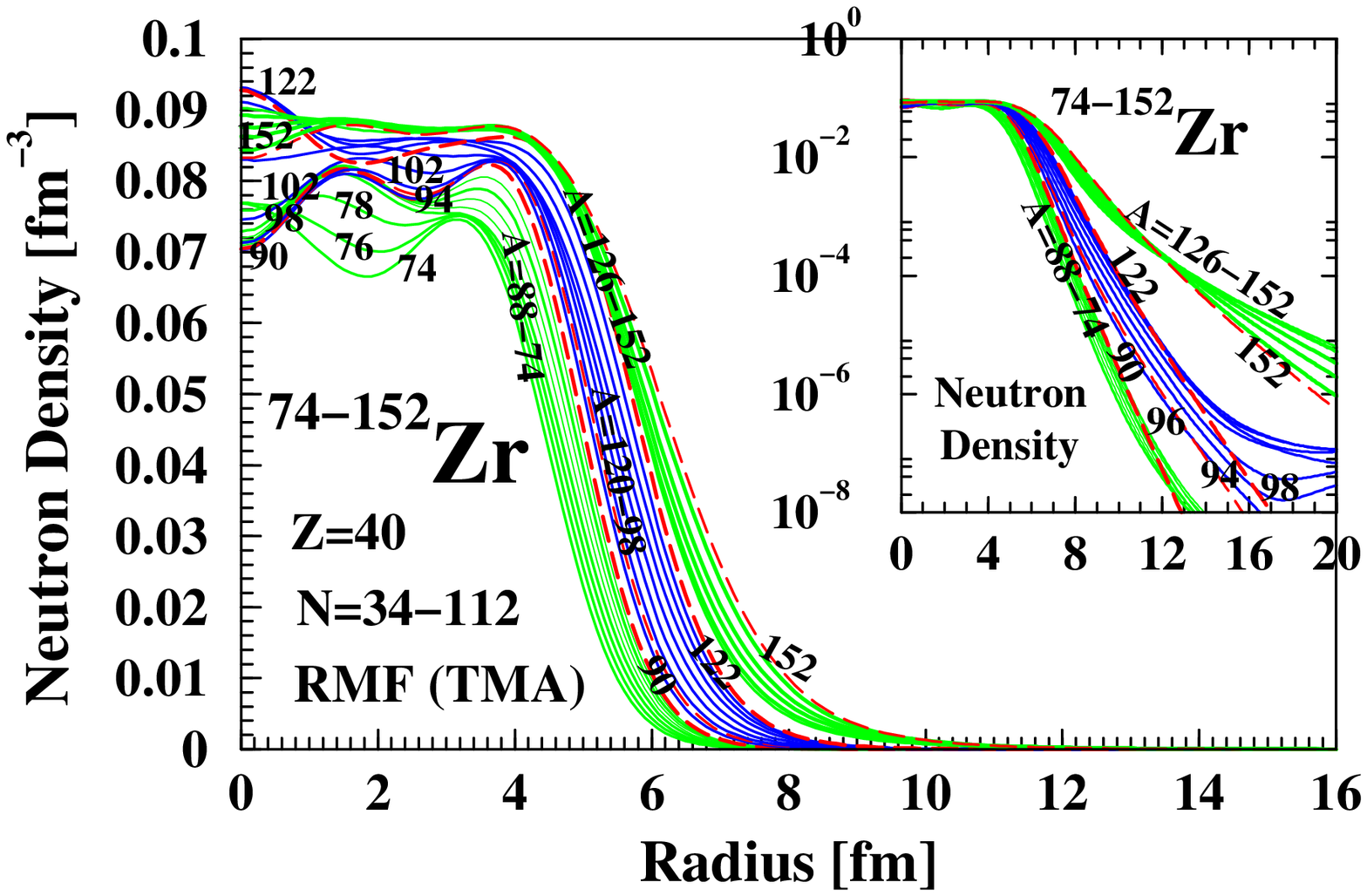,width=13cm}
\vskip 0.15in {\noindent \small {{\bf Fig. 35.} The solid line
show the neutron radial density distribution for the
$^{74-152}\rm{Zr}$ isotopes obtained in the RMF+BCS calculations
using the TMA force. To keep the figure uncluttered the
distributions for only some selected isotopes covering the entire
range have been plotted.\mbox{} }} \vskip 0.15in

It is seen from the neutron densities of $\rm{Zr}$ isotopes in
fig. 35 that for the neutron number $N=50, 82$ and $112$ which
correspond to shell closures in the $^{90,122,152}\rm{Zr}$ neutron
rich nuclei, the density distribution is rather confined to
smaller radial distances and diminishes quickly as is indicated by
the slope of these curves shown in the inset. For other isotopes
the distribution exhibits tendency to fall off sharply up to a
radial distance of about $10$ to $15$ fm and then becomes flat. In
the case of highly neutron rich isotopes $N\ge 84$ corresponding
to $^{124-150}\rm{Zr}$ the density profile has slightly altered
feature in that it is characterized by a much slower fall as a
function of radial distance and  has wider spatial extension. This
makes the curves for these isotopes to be well separated from the
others as is seen from the inset in fig. 35. As explained earlier
it is due to contributions from the positive energy states which
are partially occupied, though with very small occupancy weights,
and are either resonant or loosely bound states lying near the
Fermi level. However, the density distributions tend to be similar
to those found for the $\rm{Ca}$ isotopes.

Results for the proton density distributions for the chain of
$\rm{Zr}$ isotopes have been depicted in fig. 36. The density
distributions tend to show deviation from sharp fall off  after
about $r=10$ fm and is caused due to the fact that $\rm{Zr}$ not
being a  proton magic has contributions to the pairing
correlations from the low lying resonant states. However, these
deviations are unimportant as the magnitude of density for large
distances has already diminished to the order of $10^{-8}$
fm$^{-3}$. Further, fig. 36 shows that the central proton density
decreases with increasing neutron number and may be attributed to
the neutron-proton interaction. The maximum to minimum ratio of
the central proton density value turns out to be higher than that
for the corresponding central neutron densities. It is also seen
from fig. 36 that with increasing neutron number the proton radius
increases.

\vspace{0.5cm}\psfig{figure=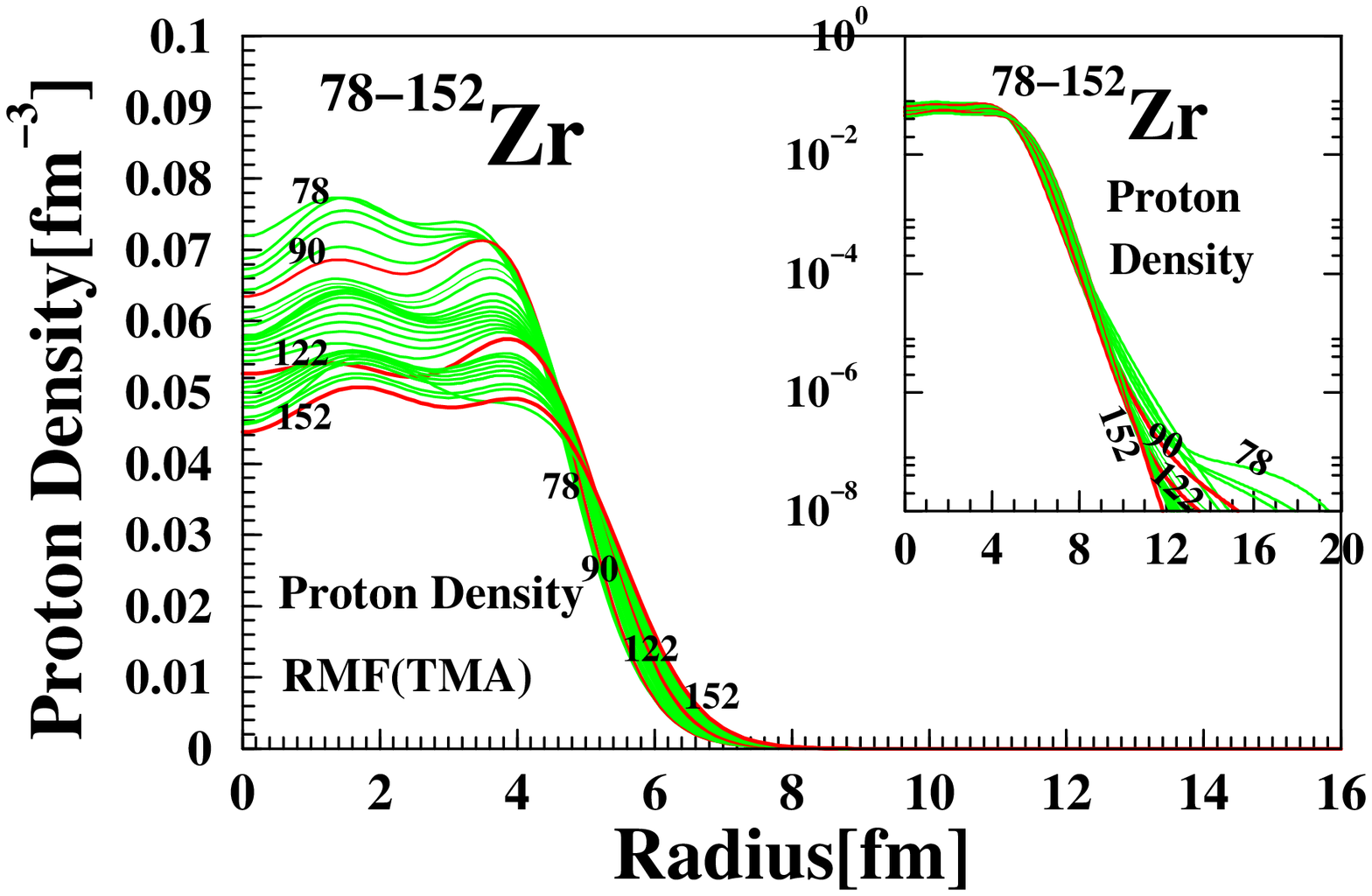,width=13cm}
\vskip 0.15in {\noindent \small {{\bf Fig. 36.} The solid line
show the proton radial density distribution for the
$^{74-152}\rm{Zr}$ isotopes obtained in the RMF+BCS calculations
using the TMA force. To keep the figure uncluttered the
distributions for only some selected isotopes covering the entire
range have been plotted.\mbox{} }} \vskip 0.15in

In contrast to the $\rm{Ca}$ and $\rm{Zr}$ nuclei, the density
profile of $\rm {Sn}$ isotopes is less extended and for heavier
$\rm{Sn}$ isotopes it falls off rather rapidly as is seen from the
inset in fig. 37. for smaller distances the density profile
exhibits peak around $r=4-5$ fm. Between $r=5$ and $10$ fm the
density curves show identical radial dependence with a rapid
smooth fall off. The distributions for isotopes with neutron
number $N = 100, 132$ and $N = 176$ corresponding to neutron shell
closures in the $\rm {Sn}$ isotopes are distinctly characterized
by sharp fall off as can be seen in the inset of the figure. For
these isotopes there are no contributions to the wave functions
coming from resonant states in the continuum.  For other isotopes
one observes the flattening of the distribution at large distances
indicating the contribution from the resonant and other positive
energy states near the Fermi level.

\vspace{0.5cm}\psfig{figure=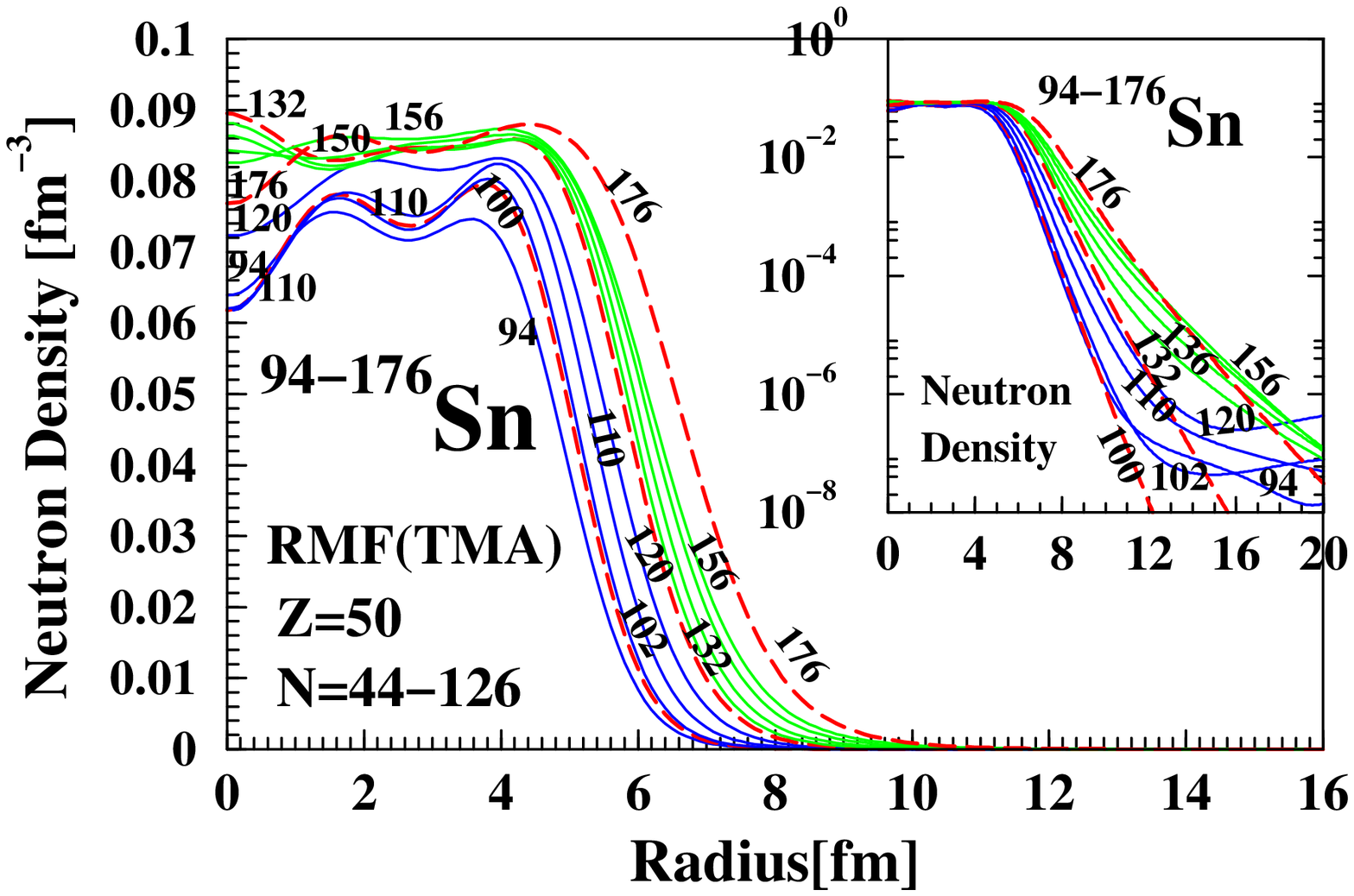,width=13cm}
\vskip 0.15in {\noindent \small {{\bf Fig. 37.} The solid line
show the neutron radial density distribution for some selected
$\rm{Sn}$ isotopes obtained in the RMF+BCS calculations using the
TMA force.\mbox{} }} \vskip 0.15in

The proton density profile for the  $\rm {Sn}$ isotopes depicted
in fig. 38 shows spatially confined distributions. Again, with
increasing neutron number the proton distribution radii are seen
to grow. It is found that for heavy neutron rich $\rm {Sn}$
isotopes the proton single particle potential is changed and the
$\rm {Sn}$ isotopes no longer have proton shell closure for
$Z=50$. For such isotopes then  there are contributions from
resonant states to the proton pairing correlations. Consequently
for neutron rich isotopes the asymptotic proton density
distributions becomes flat at the tail end as can be seen in the
inset of fig. 38.  The ratio of central maximum and minimum proton
densities in this case has almost similar value ($\approx 1.6$) as
for the proton densities of $\rm {Zr}$ isotopes. The lightest
isotope $^{78}\rm {Zr}$ is seen to have maximum proton density at
the center whereas the minimum is that of heaviest isotope
$^{176}\rm {Zr}$. In contrast, the neutron density for the
heaviest isotope is not maximum at the center as is evident from
fig. 37.

\vspace{0.5cm}\psfig{figure=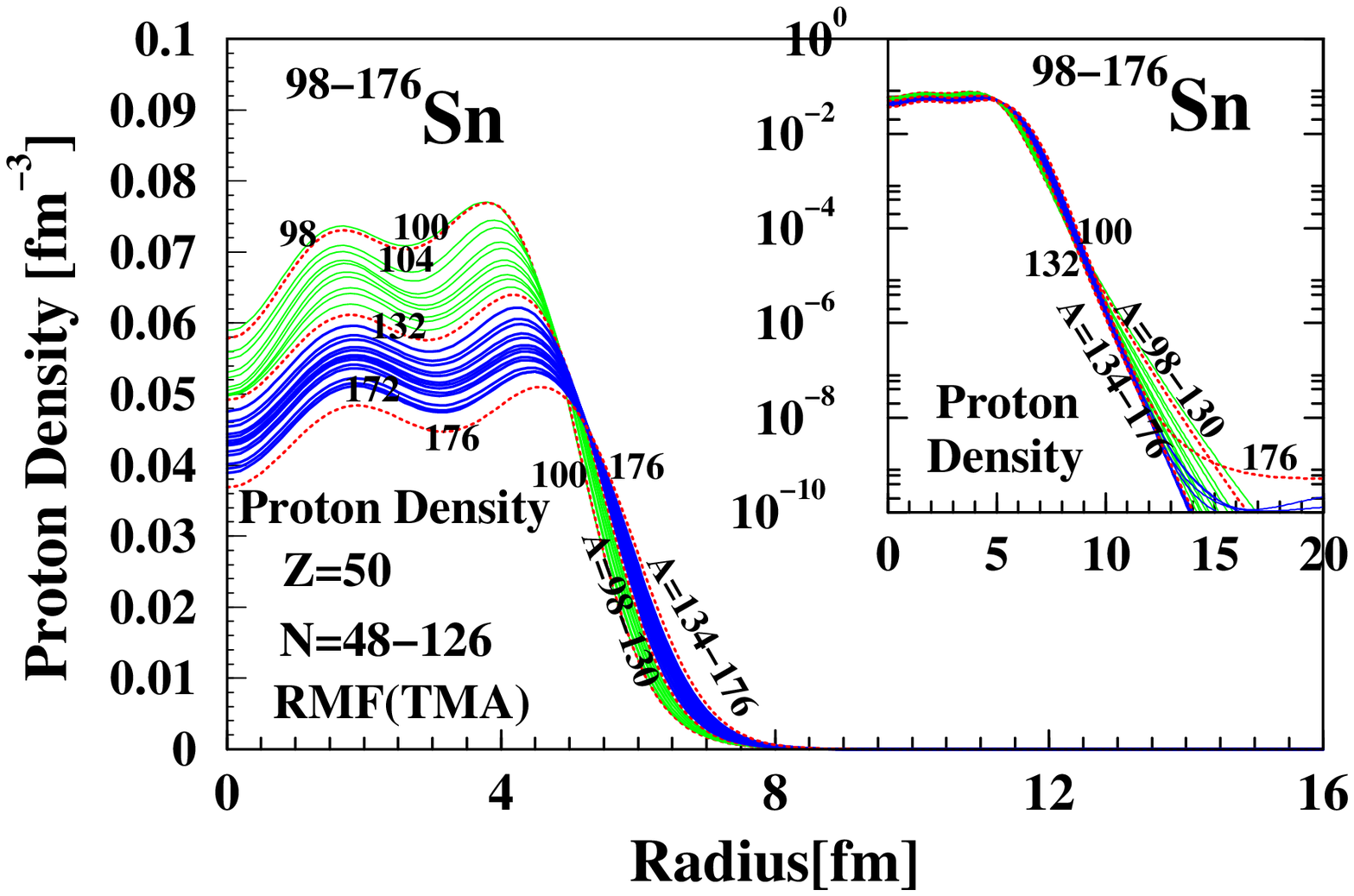,width=13cm}
\vskip 0.15in {\noindent \small {{\bf Fig. 38.} The solid line
show the proton radial density distribution for some selected
$\rm{Sn}$ isotopes obtained in the RMF+BCS calculations using the
TMA force.\mbox{} }} \vskip 0.15in

The neutron density profile of $\rm {Pb}$ isotopes plotted in fig.
39 shows much less spatial spread of the distributions. However,
one clearly sees the slight flattening of the tails as depicted in
the inset in fig. 39. In order to have some feeling of single
particle states being filled in for  we consider for example the
isotope $^{242}\rm{Pb}$. For this isotope the low lying neutron
resonant state $1j_{15/2}$ becomes bound and the positive energy
$1j_{13/2}$ state at $\epsilon = 3.07$ MeV acts as resonant state.
The occupancy weight of this state indicates that there are about
0.2 nucleons occupying this state in the continuum. This state
contributes to the density at larger distances and, thus, the
asymptotic density does not show a rapid fall off for such
isotopes. Similar results are also obtained for other  $\rm {Pb}$
isotopes. However, the isotopes $^{208}\rm{Pb}$ and
$^{266}\rm{Pb}$ corresponding to $N=126$ and $N=184$ are neutron
closed shell nuclei and for these two the spatial spread in the
neutron density is seen to be minimum as is evident from the
asymptotic density distribution shown in the inset of fig. 39. For
smaller distances the density profile exhibits peak around $r=5.5$
fm. As in other cases the neutron density falls off quickly
between $r=6$ and $10$ fm.

\vspace{0.5cm}\psfig{figure=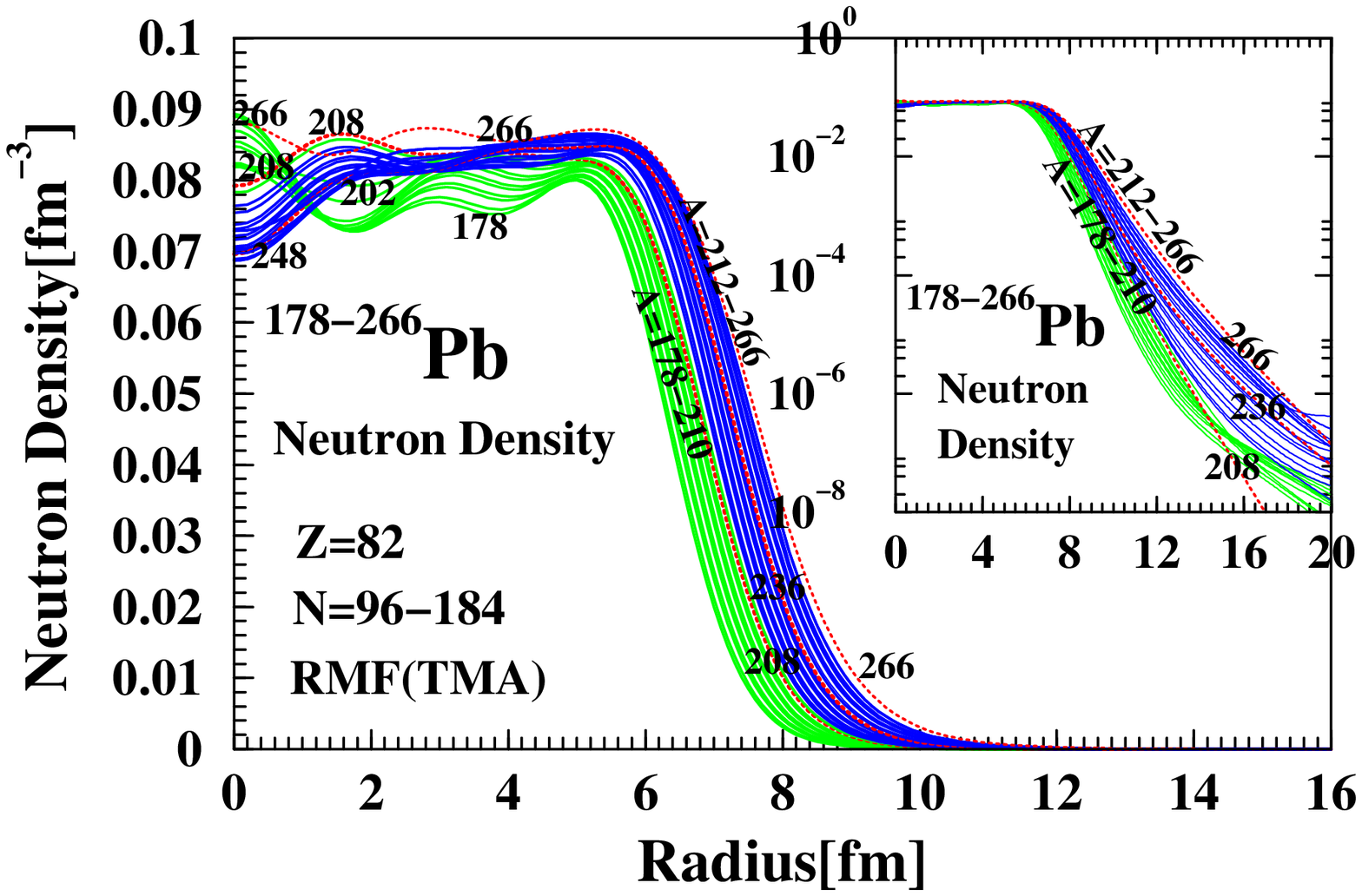,width=13cm}
\vskip 0.15in {\noindent \small {{\bf Fig. 39.} The solid line
show the neutron radial density distribution for some selected
$\rm{Pb}$ isotopes obtained in the RMF+BCS calculations using the
TMA force.\mbox{} }} \vskip 0.15in

Similar to the neutron densities, and also as the $\rm{Pb}$
isotopes are proton magic, the proton densities of $\rm {Pb}$
isotopes have smaller spatial widening at the tail end as is seen
in the inset of fig. 40. However, it is found,  as  for the other
heavier isotopes of $\rm{Sn}$, that in the case of heavy neutron
rich $\rm{Pb}$ isotopes the single particle potential for the
protons become much deeper and proton single particle energies are
slightly changed and $Z=82$ no more corresponds to a proton shell
closure. Due to contributions from proton positive energy states
this is found to change the asymptotic proton density for the
neutron rich $\rm{Pb}$ isotopes as can be seen in the inset of
fig. 40. For small radial distances ($r < 1$ fm), in contrast to
$\rm {Zr}$ and $\rm {Sn}$ proton densities, here one observes an
upward growth of the proton density near the center for the
lighter isotopes ($A=178-208)$. However, the neutron densities are
in general larger than the proton densities. At the surface the
proton densities have values between $0.05$ and $0.07$ fm$^{-3}$
whereas the neutron densities for the $\rm {Pb}$ isotopes at the
surface range between $0.08$ and $0.088$ fm$^{-3}$.

\vspace{0.5cm}\psfig{figure=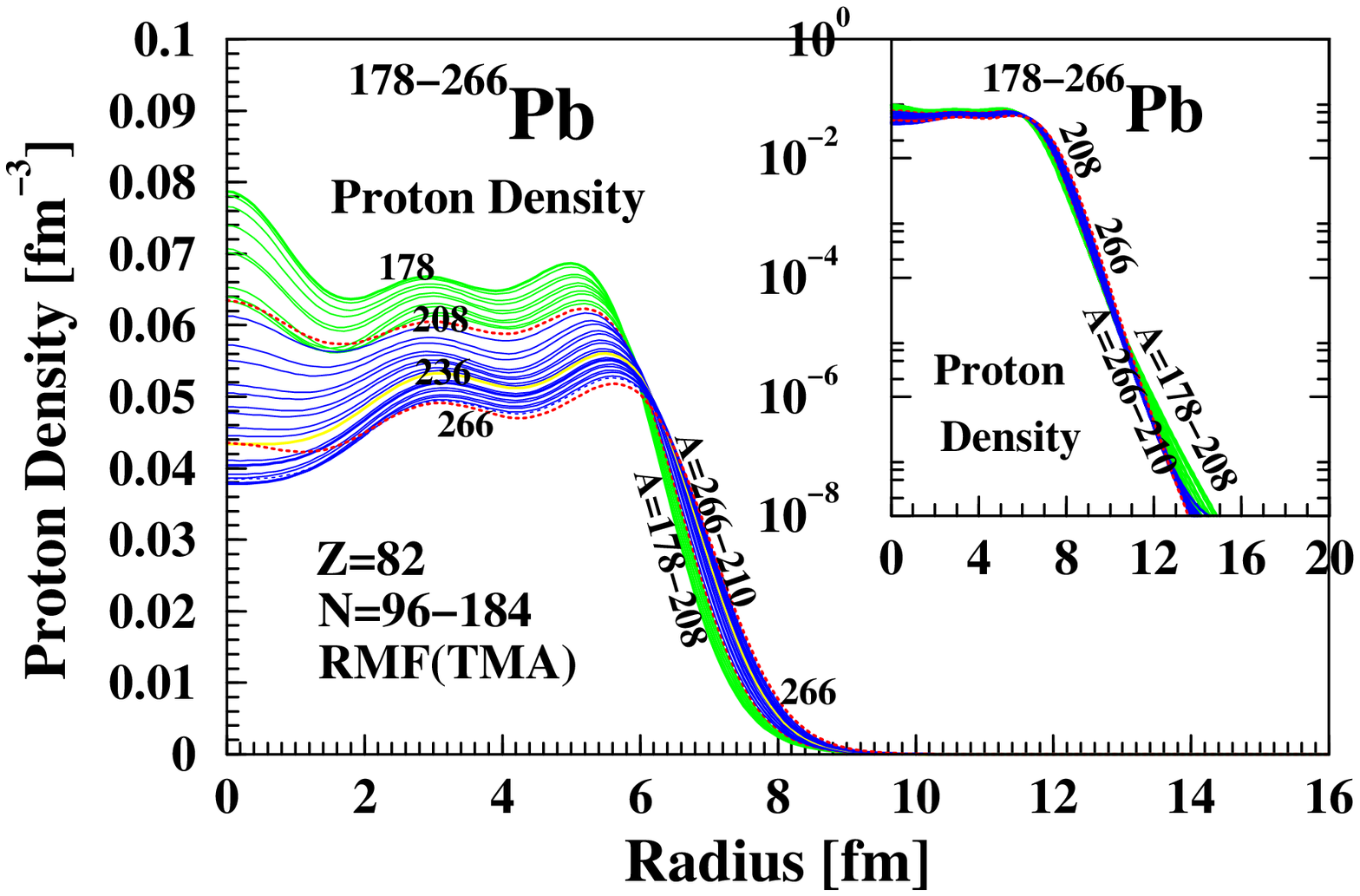,width=13cm}
\vskip 0.15in {\noindent \small {{\bf Fig. 40.} The solid line
show the proton radial density distribution for some selected
$\rm{Pb}$ isotopes obtained in the RMF+BCS calculations using the
TMA force.\mbox{} }} \vskip 0.15in

\section{Summary}
In the present investigation we have carried out a detailed study
of the chains of isotopes of proton magic nuclei $\rm O, Ca, Ni,
Sn$ and $\rm Pb$, as well as that of $\rm Zr$ considered to be
proton sub-magic within the framework of RMF+BCS. Our present
calculations have been restricted to spherical shapes for the sake
of simplicity. However, guided by our experience, we believe that
calculations with the inclusion of deformation would not change
the main conclusions reached for the proton magic nuclei
investigated here.

In view of the fact that the pairing correlations play an
important role for the description of neutron rich isotopes, we
have also studied in detail the pairing properties of these nuclei
close to the neutron drip line within the RMF+BCS framework. In
the BCS calculations we have replaced the continuum by a set of
positive energy states determined by enclosing the nucleus in a
spherical box. It is found that from amongst the positive energy
states, apart from the single particle states adjacent to the
Fermi level, the dominant contribution to the pairing correlations
is provided by a few states which correspond to low-lying
resonances. As the number of neutrons increases the low lying
resonant states become bound and the higher ones become more
crucial for the description of pairing correlations. The loosely
bound single particle states lying at the continuum threshold near
the Fermi level are found to be important especially for the
neutron rich isotopes. For the case of $\rm O, Ca, Ni, Zr, Sn$ and
$\rm Pb$ isotopes we have shown that the RMF+BCS and the RHB
calculations\cite{meng} as well as other mean-field
descriptions\cite{dobac}$^-$\cite{grasso}, give similar results
for the two-neutron separation energies, and for the proton,
neutron and matter rms radii  up to the neutron drip line.
Further, it is found that these results are in good agreement with
the available experimental data.

For the the case of $\rm Ca$ and $\rm Ni$ isotopes, and to some
extent for the $\rm Zr$ nuclei, we have discussed the results in
greater details. These RMF+BCS calculations have been explicitly
shown to agree very well with the recent continuum relativistic
Hartree-Bogoliubov (RCHB) calculations\cite{meng2}. Moreover, the
results obtained from two different popular RMF parameterizations,
the TMA and the NL-SH forces, are also shown to be very similar to
each other and with those of the RCHB. Small differences between
the results for the two forces are traced back to the differences
in the single particle spectrum near the Fermi surface. The
difference is seen to be more pronounced for the neutron rich $\rm
Zr$ isotopes which are not exactly proton magic. Further, this
difference in the single particle structure leads to difference in
the prediction of the two-neutron drip line for the TMA and NL-SH
forces for the $\rm Zr$ isotopes. However since the binding energy
difference amongst the neutron rich two isotopes of $\rm Zr$ is
found to be less than 100 keV this difference should not be taken
too seriously. The results for the $\rm Ca$ isotopes showing a
sudden increase in the radii of neutron distribution for the
neutron rich isotopes with $N\ge42$ provide evidence for a halo
formation in the $^{62-72}\rm Ca$ isotopes. For halo formation in
the neutron rich $^{62-72}\rm Ca$ isotopes the neutron $3s_{1/2}$
state turns out to be the most crucial one. Calculations for the
$\rm {Zr}$ isotopes yield similar results with indications, though
less pronounced, for the halo formation in the neutron rich ($\rm
Zr$) isotopes. In this case the neutron $3p_{1/2}$ state in the
neutron rich $\rm {Zr}$ isotopes is found to play the key role.
The results for the neutron rich isotopes of other nuclei, $\rm O,
Ni, Sn$ and $\rm Pb$ do not show such a tendency of abrupt growth
in the radii of the neutron distribution.

Similar indications with regard to halo formations and other
properties are also provided by the calculated neutron and proton
density profiles for these nuclei. The neutron density
distributions for the isotopes with halo formation are seen to
have a wide spatial extension. Further, it is found that the
resonant states and the loosely bound states near the continuum
threshold tend to accommodate a very small number of neutrons
which in turn affect the radial dependence of the density profile
at large distances. This contribution, though extremely small,
makes the asymptotic density fall off  less rapidly. However, for
the neutron magic nuclei the distribution falls off sharply as it
has no contributions from the resonant states. Similarly, the
proton density distributions for the proton magic nuclei are found
to be confined to smaller distances and fall off rapidly. However,
for neutron rich  proton magic nuclei the proton single particle
potential becomes deeper and the shell closure property is
destroyed. For such heavy neutron rich isotopes the asymptotic
proton density is slightly affected by the contributions from the
positive energy  and resonant states. The $\rm Zr$ isotopes are
not proton magic and thus  in this case the proton density are
found to have wider spatial extension and rather a slow fall off
due to pairing correlations from positive energy states near the
continuum threshold.

The detailed comparison of different results obtained from the
RMF+BCS and the Hartree-Bogoliubov frameworks on the one hand, and
a reasonably good agreement of these with the available
experimental data together confirm our earlier
findings\cite{yadav}, and provide ample indications that the
RMF+BCS approach can be considered as a good approximation, in
addition to being neat and transparent, to the full Relativistic
Hartree-Bogoliubov (RHB) treatment for the drip-line neutron rich
nuclei. This in fact is not surprising in view of similar
conclusion reached very recently by Grasso {\it et al.}\cite
{grasso} for the non-relativistic mean field descriptions.

\nonumsection{Acknowledgments} \noindent We thank  N. Sandulescu
and G. Hillhouse for fruitful discussions. We are indebted to J.
Meng for communicating some of his RCHB results before
publication. One of the authors (HLY) gratefully acknowledges the
support provided to him under the COE Professorship Program of
Monbusho of the Osaka University, Japan, and the support through a
grant by the Department of Science and Technology(DST), India. HLY
would also like to thank Prof. Faessler for his kind hospitality
while visiting Institut f\"{u}r Theoretische Physik der
Universit\"{a}t T\"{u}bingen, Germany where part of this work was
carried out.

\nonumsection{References}
\noindent
%References in the text are to be numbered consecutively in
%Arabic numerals, in the order of first appearance. They are to
%be typed in superscripts after punctuation marks,
%e.g.~``$\ldots$ in the statement.$^5$''.
%
%References are to be listed in the order cited in the text. Use
%the style shown in the following examples. For journal names,
%use the standard abbreviations. Typeset references in 9-pt Times
%Roman.

\newpage
\begin{table}[htb]
\centering \caption{Parameters of the Lagrangian TMA and NL-SH
together with the nuclear matter properties obtained with these
effective forces.}
\bigskip
\bigskip
%\begin{tabular}{|c|c|c|c|c|c|c|c|c|c|}
\begin{tabular}{cccccccccc}
\hline
%\multicolumn{1}{c}{Nucleus}&
\multicolumn{1}{c}{$Param.$}&\multicolumn{1}{c}{}&
\multicolumn{1}{c}{$TMA$}&
\multicolumn{1}{c}{$NL-SH$}\\
\hline
M &(MeV)&938.9&939.0  \\
m$_{\sigma}$ &(MeV) &519.151&526.059\\
m$_{\omega}$ &(MeV) &781.950&783.0\\
m$_{\rho}$ &(MeV) &768.100&763.0\\
g$_{\sigma}$& & 10.055 + 3.050/A$^{0.4}$&10.444\\
g$_{\omega}$& & 12.842 + 3.191/A$^{0.4}$&12.945\\
g$_{\rho}$ & &  3.800 + 4.644/A$^{0.4}$&4.383\\
g$_{2}$ &(fm)$^{-1}$ & -0.328 - 27.879/A$^{0.4}$&-6.9099\\
g$_{3}$& & 38.862 - 184.191/A$^{0.4}$&-15.8337 \\
c$_{3}$& &151.590 - 378.004/A$^{0.4}$& \\
\\
\hline
\\

Nuclear Matter Properties\\
\\
saturation density $\rho_{0}$&(fm)$^{-3}$ &0.147&0.146\\
Bulk binding energy/nucleon (E/A)$_{\infty}$ &(MeV) &16.0&16.346\\
Incompressibility K  &(MeV) &318.0& 355.36 \\
Bulk symmetry energy/nucleon a$_{Sym}$   &(MeV) &30.68& 36.10 \\
Effective mass ratio m$^{*}$/m & &0.635& 0.60 \\
\hline
\end{tabular}
\end{table}

\end{document}